\documentclass[twoside,12pt]{article}
\usepackage{epsfig,latexsym}

\def\Journal#1#2#3#4{{#1} {#2} (#4) #3 }

\def\NPA{{\em Nucl. Phys.} A}

\def\PRO{{\em Prog. Theor. Phys.}}
\def\NPB{{\em Nucl. Phys.} B}

\def\PLB{{\em Phys. Lett.} B}

\def\PRL{\em Phys. Rev. Lett.}
\def\PREV{\em Phys. Rev.}
\def\PREP{\em Phys. Rep.}
\def\PRA{{\em Phys. Rev.} A}
\def\PRD{{\em Phys. Rev.} D}
\def\PRC{{\em Phys. Rev.} C}
\def\PRB{{\em Phys. Rev.} B}

\def\ANNP{\em Ann. Phys. (N.Y.)}
\def\RMP{{\em Rev. Mod. Phys.}}

\def\INT{{\em Int. J. Mod. Phys.} E}

\newcommand{\be}{\begin{equation}}
\newcommand{\ee}{\end{equation}}
\newcommand{\bea}{\begin{eqnarray}}
\newcommand{\eea}{\end{eqnarray}}

\newcommand{\ub}[1]{\underline{#1}}
\newcommand{\ob}[1]{\overline{#1}}
\newcommand{\Pminus}{{\cal P}^-}
\newcommand{\veck}{\vec{k}_\perp}

\def\senk#1{\vec{#1}_\perp}
\def\psibar{\overline{\psi}}
\def\g{\gamma}

\begin{document}

\title{ \vspace{1cm} Nonperturbative light-front Hamiltonian methods}

\author{J.R.\ Hiller\\
\\
Department of Physics and Astronomy, \\
University of Minnesota-Duluth, \\
Duluth, MN 55812 USA}

\maketitle

\begin{abstract} 

We examine the current state-of-the-art in nonperturbative calculations
done with Hamiltonians constructed in light-front quantization of 
various field theories.  The language of light-front quantization is
introduced, and important (numerical) techniques, such as Pauli--Villars
regularization, discrete light-cone quantization, basis light-front
quantization, the light-front coupled-cluster method, the renormalization
group procedure for effective particles, sector-dependent
renormalization, and the Lanczos diagonalization method, are surveyed.
Specific applications are discussed for quenched scalar Yukawa theory,
$\phi^4$ theory, ordinary Yukawa theory, supersymmetric Yang--Mills
theory, quantum electrodynamics, and quantum chromodynamics.  The 
content should serve as an introduction to these methods for anyone
interested in doing such calculations and as a rallying point for
those who wish to solve quantum chromodynamics in terms of wave 
functions rather than random samplings of Euclidean field
configurations.

\end{abstract}

\section{Introduction} \label{sec:Intro}

After many years in gestation, light-front 
quantization~\cite{Dirac,LFreview1,LFreview2,LFreview3,LFreview4}
is now poised as a viable tool for the nonperturbative solution of
quantum chromodynamics (QCD)~\cite{WhitePaper}.  This will establish
an approach complementary to lattice gauge theory~\cite{lattice},
one where wave functions return to their usual
central role.  Observables can then be computed
as expectation values.  In addition, the method
is formulated in Minkowski space-time, rather
than the Euclidean space-time of lattice theory,
making time-like quantities more readily accessible.
In comparison with equal-time quantization, use
of the light-front affords boost-invariant wave functions
without spurious vacuum contributions.

The purpose here is not to review the historical
development of light-front methods; this is done
quite nicely elsewhere~\cite{LFreview1}.  The
purpose is instead to summarize the state of
the art in nonperturbative light-front calculations, in particular
those aspects applicable to QCD, and thereby provide
the impetus and the foundation for the massive
computational effort required to complete the task.
The effort {\em is} massive but then so was the development
of lattice gauge theory. 

Other methods are also candidates for calculations in
QCD.  Among them are Dyson--Schwinger equations~\cite{DSE}, 
which is also a Euclidean method; the truncated conformal
space approach~\cite{TCSA}; and 
the transverse lattice~\cite{TransLattice}, which combines 
a lattice in transverse coordinates with two-dimensional
light-front quantization for longitudinal space and time.
These are quite adequately addressed elsewhere.

Perturbative calculations can also benefit from a light-front
approach; however, these are also outside the scope of the
present review.  Instead, the recent review by Cruz-Santiago, 
Kotko, and Sta\'{s}to~\cite{CruzSantiago} provides an excellent
introduction to light-front calculations of scattering amplitudes.

The focus here is on light-front Hamiltonian methods for
nonperturbative bound-state problems.  The methods
are, at least loosely, based on Fock-state expansions of
the eigenstates.  The Fock states are eigenstates of
momentum, particle number, and fundamental quantum
numbers associated with any symmetries or charges.
The wave functions appear as the coefficients of the 
Fock states in the expansion; they are functions of (relative) 
momenta\footnote{Unlike equal-time coordinates, light-front 
coordinates admit a separation of external and relative momenta.}
and are indexed by the particle count and quantum numbers.
The Hamiltonian eigenvalue problem is then transformed
into a coupled set of integral equations for these
wave functions, with the invariant mass of the eigenstate
as the eigenvalue.  As such, the approach lends itself
well to numerical solution by discretization~\cite{PauliBrodsky,LFreview1}
and by basis-function expansions~\cite{Wilsonremarks,BLFQ}.

The light-front Fock vacuum is an eigenstate of the full
Hamiltonian, including interactions, provided that zero
modes are excluded~\cite{LFreview1}.  The solution of the 
eigenvalue problem for the light-front Hamiltonian can then 
focus on the massive
states, unlike equal-time quantization where the vacuum
state itself must be computed as well.  This is a
significant advantage for light-front quantization.
As is the added characteristic that vacuum contributions
are absent from the Fock-state expansions of the
massive states.  The Fock-state wave functions can
then be interpreted as defining the massive state
itself.

However, this is a much weaker statement about the
vacuum than to claim that the light-front Fock
vacuum is the physical vacuum.  The latter is not
empty, and any physics that, in equal-time quantization
flows from the structure of the physical vacuum is
typically difficult to reproduce in light-front
quantization.  One exception to this is a light-front
derivation of the Casimir effect~\cite{LFCasimir},
but quantities such as critical couplings and
exponents in $\phi^4$ theory remain elusive.

To solve the infinite system of equations for the 
masses and wave functions requires some form
of truncation.  This is usually done as a truncation in
Fock space to maximum numbers of particle types.
However, such a truncation causes uncanceled divergences
because cancellations between contributions to a 
particular process frequently require contributions
from (disallowed) intermediate states with additional particles.
Two solutions to this difficulty have been proposed.  One
is sector-dependent 
renormalization~\cite{SecDep-Wilson,HillerBrodsky,Karmanov,KarmanovYukawa}, 
where bare parameters
of the Lagrangian are allowed to depend on the Fock sector
or sectors on which the particular interaction term acts.
The uncanceled divergences are absorbed into renormalization 
of the couplings. This can lead to inconsistencies in the interpretation of
the wave functions~\cite{SecDep}.  The other solution
is the light-front coupled-cluster (LFCC) method~\cite{LFCC},
in which the truncation is in the way that higher Fock-state
wave functions are related to the lower wave functions, with no
Fock-space truncation.  In either case, the bare parameters
are fixed by fitting observables.

For theories beyond two dimensions, the integral operators of
the integral equations are associated with divergences even 
without Fock-space truncation.  These are the usual divergences
of quantum field theory, and they require regularization.
Various schemes have been proposed, particularly momentum
cutoffs, in the transverse momenta for UV divergences and in the
longitudinal momenta for IR divergences.  Modifications of
this include use of a cutoff on the invariant mass of the Fock state
and on the change in the invariant mass across each interaction event.
Such cutoffs violate Lorentz and gauge invariance and require
counterterms for the restoration of the symmetries.

An alternative that avoids breaking these symmetries, and
which has proven quite useful, is Pauli--Villars (PV)
regularization~\cite{PauliVillars}.  In the present
context of nonperturbative calculations, this is implemented
by inclusion of massive PV fields with negative metric
in the Lagrangian and, consequently, the Fock space.  Modification 
of loops in individual diagrams, as is frequently done in 
perturbation theory, is not an option here.\footnote{Similarly,
dimensional regularization~\protect\cite{dimreg} is also not an option,
because the integrals to be modified are only implicit in the
nonperturbative action of the Hamiltonian.}  

The regulating PV fields are removed in the limit of infinite mass.
This opens a third possibility for coping with uncanceled divergences
in Fock-space truncation.  One can seek plateaus in the PV-mass
dependence and remain at finite values for one or more of the
regulating masses~\cite{OnePhotonQED}.

The introduction of PV particles to the Lagrangian leads to a 
non-Hermitian Hamiltonian and a loss of unitarity.  These effects 
are caused by the negative metric assigned to some or all of the 
PV fields, in order to arrange the minus signs needed to achieve 
the necessary subtractions.  This introduces unphysical features 
that are to be minimized by keeping the PV masses large, if not 
taken all the way to infinity.  In practice, matrix representations 
of the Hamiltonian in numerical calculations are then also
non-Hermitian,\footnote{As discussed in Sec.~\ref{sec:Lanczos}, 
a special form of the Lanczos diagonalization algorithm has been
developed to handle such matrices.} and subsequently there are 
unphysical, negatively normed eigenvectors, as well as negatively 
normed contributions to the Fock-state expansions of physical
eigenvectors.  Numerical results must be carefully vetted for 
spurious eigenvectors.  Also, variational methods are of limited 
utility, because the lowest states in the spectrum are frequently 
unphysical.

As compensation for the computational load and memory requirements
associated with the additional Fock states containing PV particles, 
the PV interactions can be arranged to cancel the instantaneous
fermion interactions~\cite{LFreview1}.  These interactions are 
characteristic of light-front quantization, where part of a Dirac 
fermion field is constrained rather than dynamical.  When the 
constrained components are eliminated from the Lagrangian, additional 
interactions are induced for the remaining dynamical components.  They 
are four-point interactions and as such they significantly reduce the 
sparseness of any matrix representation of the Hamiltonian and greatly 
increase the time required for computation of nonzero matrix elements.  
Given that sparseness is very important for numerical calculations,
because the matrices are much too large to be stored in full and must 
be stored in compressed form, a matrix representation that includes 
the PV Fock states can be an advantage because it is more sparse
even though larger.

The physics of the instantaneous interactions is, however, not missing.
When PV regularization is properly introduced, these interactions
are factorized into two three-point interactions that involve an 
intermediate PV fermion.  The precise form of the original four-point 
interactions is recovered in the infinite PV-mass limit.  This is 
critical because, as is known from perturbation theory, the
instantaneous fermion interactions play important roles in the 
cancellation of singularities and restoration of covariance~\cite{BoxYukawa}.

Light-front Hamiltonian methods for gauge theories necessarily require 
a choice of gauge.  The traditional choice is light-cone gauge~\cite{LFreview1}, 
which has two advantages: a directly soluble constraint equation for Dirac fermions 
and no need for unphysical degrees of freedom such as ghosts.  Unfortunately, 
working with a single fixed gauge makes impossible any check of gauge invariance
and blocks the use of BRST invariance~\cite{BRST} for any attempt at a proof of 
renormalizability.  The broken symmetry also makes a calculation vulnerable to 
dependence on its regularization parameters, and any results will be suspect.
Although not a serious problem for calculations in QED, any attempt at a 
non-Abelian theory, such as QCD, is at a serious disadvantage without gauge
invariance.  Both lattice QCD and perturbative QCD are done in ways that 
respect gauge invariance as much as possible, and light-front calculations must 
do the same.

A major theme of this review is that the use of PV regularization allows the choice of
a family of covariant gauges.  Gauge invariance within this family can be checked
by varying the gauge fixing parameter.  This was first done for PV-regulated 
QED~\cite{ArbGauge}, and, although PV regularization was traditionally considered
inapplicable to non-Abelian theories, a new formulation has been constructed
for PV-regulated Yang--Mills theories to include a BRST invariance with ghost 
and anti-ghost fields~~\cite{BRSTPVQCD}.  As emphasized above, the presence of such 
symmetries provides an important check on any calculation; conversely, the violation 
of such symmetries frequently leads to a strong dependence on the regularization 
parameters.  Hence, the preservation of symmetries is more important than the 
extra effort associated with additional degrees of freedom.  If instead, 
avoidance of ghosts was paramount over gauge invariance, most covariant 
perturbative QCD calculations would be done in Coulomb gauge, which is not the case.

Regularization can also be provided by the numerical approximation
to the integral equations for the wave functions.  For example, a 
basis-function expansion~\cite{BLFQ} for the wave functions is truncated, 
to establish a finite matrix representation of the original 
integral equations.  The truncation provides a regularization
which is removed in the limit of infinite basis size.
This, however, entangles the renormalization process with
the numerics, making control of the numerical approximation
more difficult, and may lead to a net increase in difficulty,
even though the numerical regularization itself may be simple.

The work on PV regularization has emphasized the philosophy
that the regularization and numerics should be kept separate.
The numerical approximation is of a finite theory, with 
relatively clear goals for numerical convergence.  The
renormalization of the bare parameters is investigated 
only in the continuum limit.

The Fock space need not be constructed directly in terms
of bare particles, but can instead be made from effective
particles, as is done specifically in the renormalization
group procedure for effective particles (RGPEP)~\cite{RGPEP}.
More generically, one can carry out renormalization-group
analyses~\cite{Perry:1993mn,Wilson:1994fk,Similarity} to 
construct effective Hamiltonians that may be better suited
for use in nonperturbative calculations.

The remainder of this review is broken into three main
sections and a brief summary of questions to pursue.  Section~\ref{sec:LFquant}
presents the basic ideas and notation of light-front 
quantization. These include the coordinates; the free-field
quantizations for scalars, fermions, and vector bosons;
Fock-state expansions; and the calculation of observables. 
This is followed by Sec.~\ref{sec:Methods}
on the various methods used for light-front calculations,
from the original discretized light-cone quantization 
(DLCQ)~\cite{PauliBrodsky},
including its supersymmetric extension, SDLCQ~\cite{SDLCQ},
to function expansion methods~\cite{BLFQ,GenSymPolys}, 
the LFCC method~\cite{LFCC}, and the effective particle
approach~\cite{RGPEP}.  The third main section,
Sec.~\ref{sec:Applications}, is reserved for illustrative
applications to various field theories: quenched scalar Yukawa theory,
$\phi^4$ theory in two dimensions, Yukawa theory, 
supersymmetric Yang--Mills theory, QED, and QCD.

\section{Light-Front Quantization} \label{sec:LFquant}

This section sets the basic notation and provides
the foundation for the remaining sections.  As such, it
does not contain anything new but does keep the review
self-contained.

\subsection{\it Light-Front Coordinates \label{sec:LFcoords}}

\begin{figure}[tb]
\begin{center}
\epsfig{file=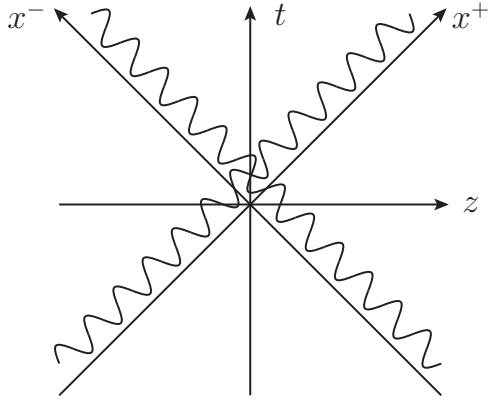,scale=0.75}
\caption{Light-front coordinates.  The light wave along the $x^-$ axis
is everywhere at one light-front time $x^+$.
\label{fig:litecone}}
\end{center}
\end{figure}

Light-front coordinates~\cite{Dirac} are defined by
\be
x^\pm \equiv x^0 \pm z,
\ee 
with $x^+$ as the light-front time and
with the transverse coordinates $\senk{x}\equiv (x,y)$ unchanged.
Figure~\ref{fig:litecone} depicts the relationship between $x^\pm$ and
the equal-time coordinates $z$ and $t$.  In two dimensions, the origin
of the name light-cone coordinates, which was used more frequently in 
the past, is readily apparent.  In more than two dimensions, light fronts
in the $x^+$ direction are tangential to the light cone; hence the
name light-front coordinates, which is now the more common usage.
Some work is done with definitions of $x^\pm$ that differ by a factor 
of $\sqrt{2}$.  This shifts factors of 2 and 1/2 in what follows.

Spatial four-vectors are written as $x^\mu = (x^+,x^-,\senk{x})$
and light-front three-vectors as $\ub{x}=(x^-,\senk{x})$.  The
conjugate four-momentum is $p^\mu=(p^-,p^+,\senk{p})$, with
$p^-\equiv E-p_z$ the light-front energy and $p^+\equiv E+p_z$
the longitudinal light-front momentum.  The light-front three-momentum
is denoted $\ub{p}=(p^+,\senk{p})$.  The dot product of a coordinate
vector and a momentum vector is $p\cdot x=\frac12(p^-x^++p^+x^-)-\senk{p}\cdot\senk{x}$,
hence the designations of $p^-$ and $p^+$ as light-front energy and
momentum, conjugate to $x^+$ and $x^-$, respectively.  The four-vector
dot product implies the following spacetime metric
\begin{equation}
g_{\mu\nu}=\left(\begin{array}{cccc} 0&1/2&0&0\\
		   1/2&0&0&0\\
		   0&0&-1&0\\
		   0&0&0&-1  \end{array}\right) .
\end{equation}
A dot product for light-front three-vectors is 
defined as $\ub{p}\cdot\ub{x}=\frac12 p^+x^--\senk{p}\cdot\senk{x}$.
The mass-shell condition $p^2=m^2$ becomes $p^-=(m^2+p_\perp^2)/p^+$.

Spatial derivatives are defined by
\begin{equation}
\partial_\pm \equiv \frac{\partial}{\partial x^\pm}
            =\frac12\left(\frac{\partial}{\partial t}\pm\frac{\partial}{\partial z}\right),\;\;
\senk{\partial}\equiv\senk{\nabla}=\left(\frac{\partial}{\partial x},\frac{\partial}{\partial y}\right) .
\end{equation} 
The factor of $\frac12$ comes from the inversion, $t=\frac12(x^++x^-)$, $z=\frac12(x^+-x^-)$.

For a system of $n$ particles, with $p_i$ the momentum of the $i$th particle
and $M$ the invariant mass, we define the total momentum $P$ with 
components $P^+=\sum_i p_i^+$, $\senk{P}=\sum_i\senk{p_i}$, and 
$P^-=(M^2+P_\perp^2)/P^+$.  Notice, however, that, except for free particles,
$P^-$ is not equal to $\sum_i p_i^-$; the momentum is not on the light-front energy shell.

It is convenient to define relative momenta for the constituents, as
a longitudinal momentum fraction $x_i\equiv p_i^+/P^+$ and a relative
transverse momentum $\senk{k_i}\equiv \senk{p_i}-x_i\senk{P}$.
Clearly, the $x_i$ sum to unity and the $\senk{k_i}$ sum to zero.
Also, if the total transverse momentum $\senk{P}$ is zero, the 
relative transverse momenta are just the transverse momenta, making
this a convenient and frequent choice of frame.

Any two frames of reference can be connected by a combination of
longitudinal boosts along $z$ and light-front transverse boosts
that combine a transverse boost and a rotation.  A longitudinal
boost in equal-time coordinates is defined as
\be
E'=\g(E-\beta p_z),\;\;
p'_z=\g(p_z-\beta E),\;\;
\senk{p'}=\senk{p},
\ee
with $\beta$ the relative velocity and $\g=1/\sqrt{1-\beta^2}$.
In terms of light-front coordinates, this boost is
\be
p^{\prime -}=E'-p'_z=\sqrt{\frac{1+\beta}{1-\beta}} p^-,\;\; 
p^{\prime +}=E'+p'_z=\sqrt{\frac{1-\beta}{1+\beta}}p^+,\;\;
\senk{p'}=\senk{p},
\ee
The light-front transverse boost is
\be
p^{\prime -}=p^-+2\senk{p}\cdot\senk{\beta}+\beta^2p^+,\;\;
p^{\prime +}=p^+,\;\;
\senk{p'}=\senk{p}+p^+\senk{\beta},
\ee
where $\senk{\beta}=(\beta_x,\beta_y)$ is again the relative
velocity.  The relative momenta of constituents, $x_i$ and $\senk{k_i}$
are invariant with respect to these boosts.  Therefore, wave functions
constructed in terms of these variables will themselves be
invariant.  This separation of internal (relative) momenta 
and the external momentum is responsible for much of the utility
of light-front quantization.

One disadvantage of light-front coordinates is the loss of explicit
rotational symmetry.  Except for rotations about the $z$ axis,
rotations are associated with dynamical operators that include
the interaction.  This means that Fock states cannot be eigenstates
of total angular momentum; the rotation operator changes the 
particle count.  For some discussion of the restoration of
rotational invariance, see \cite{BurkardtLangnau}.

\subsection{\it Free Fields \label{sec:FreeFields}}

\subsubsection{\it scalar field \label{sec:scalars}}

The Lagrangian for a free neutral scalar field $\phi$ of mass $m$ is
\be
{\cal L}=\frac12(\partial_\mu\phi)^2-\frac12 m^2\phi^2
=\frac12\partial_-\phi\partial_+\phi-\frac12(\senk{\partial}\phi)^2-\frac12 m^2\phi^2.
\ee
From this Lagrangian, we obtain the field equation as the Klein--Gordon equation,
$\partial^\mu\partial_\mu\phi+m^2\phi=0$, and the
conjugate momentum $\pi\equiv \delta{\cal L}/\delta(\partial_+\phi)=\frac12\partial_-\phi$.
Therefore, the Hamiltonian density for translations in light-front time $x^+$ is
\be
{\cal H}\equiv \pi\partial_+\phi-{\cal L}=\frac12(\senk{\partial}\phi)^2+\frac12 m^2\phi^2.
\ee
The light-front Hamiltonian $\Pminus$ is just the integral of this density (normal ordered)
over the light-front spatial volume at $x^+=0$: $\Pminus=\int d\ub{x} :{\cal H}(x^+=0):$,
with $d\ub{x}=dx^- d^2x_\perp$.

The field equation is solved by the Fourier decomposition\footnote{A common
alternative notation is to normalize $\phi$ with $16\pi^3p^+$ in place
of $\sqrt{16\pi^3 p^+}$ and with a compensating nonzero commutation relation of
$[a(\ub{p}),a^\dagger(\ub{p'})]=16 \pi^3 p^+ \delta(\ub{p}-\ub{p'})$.}
\be
\phi(x)=\int_{p^+\geq0}\frac{dp^+}{\sqrt{4\pi p^+}}\frac{d^2p_\perp}{2\pi}
        \left[a(\ub{p})e^{-ip\cdot x}+a^\dagger(\ub{p})e^{ip\cdot x}\right],
\ee
with $p^2=m^2$.  The range of integration is restricted to $p^+\geq0$
because $p^+=\sqrt{m^2+p_z^2+\senk{p}^2}+p_z$ is always nonnegative.
The $p^+$ and $\senk{p}$ integration factors are written separately,
to make easy identification of the parts applicable to 1+1 and 
3+1 dimensional theories.
The creation and annihilation operators satisfy commutation relations
\be \label{eq:scalarcommutators}
{[}a(\ub{p}),a(\ub{p'})]=0,\;\;
{[}a^\dagger(\ub{p}),a^\dagger(\ub{p'})]=0,\;\;
{[}a(\ub{p}),a^\dagger(\ub{p'})]=\delta(\ub{p}-\ub{p'})
             \equiv\delta(p^+-p^{\prime +})\delta(\senk{p}-\senk{p'}).
\ee
With use of the identity $\int dx^- \exp[i(k^+-q^+)x^-/2]=4\pi\delta(k^+-q^+)$,
the light-front Hamiltonian can be reduced to
\be
\Pminus=\int d\ub{p}\frac{m^2+\senk{p}^2}{p^+}a^\dagger(\ub{p})a(\ub{p}).
\ee
Clearly, this operator sums up the $p^-$ contributions identified by
the number operator $a^\dagger(\ub{p})a(\ub{p})$ and therefore
represents the light-front kinetic energy of the free field.

The case of the charged scalar is a slight generalization of
the neutral case.  The Lagrangian is
\be
{\cal L}=\partial_\mu\phi^*\partial^\mu\phi-m^2|\phi|^2.
\ee
The field equation is again the Klein--Gordon equation.
The Hamiltonian density is
\be
{\cal H}=|\senk{\partial}\phi|^2+ m^2|\phi|^2.
\ee
The field equation is solved by the Fourier decomposition
\be
\phi(x)=\int \frac{dp^+d^2p_\perp}{\sqrt{16\pi^3 p^+}}
        \left[c_+(\ub{p})e^{-ip\cdot x}+c_-^\dagger(\ub{p})e^{ip\cdot x}\right].
\ee
The nonzero commutation relations of the creation and annihilation operators are
\be \label{eq:chargedcommutator}
{[}c_\pm(\ub{p}),c_\pm^\dagger(\ub{p'})]=\delta(\ub{p}-\ub{p'}).
\ee
The normal-ordered light-front Hamiltonian can be reduced to
\be
\Pminus=\int d\ub{p}\frac{m^2+\senk{p}^2}{p^+}
     \left[c_+^\dagger(\ub{p})c_+(\ub{p})+c_-^\dagger(\ub{p})c_-(\ub{p})\right].
\ee

\subsubsection{\it fermion field}

The Lagrangian for a free fermion field is
\be
{\cal L}=\psibar(i\g\cdot\partial-m)\psi,
\ee
where the $\g^\mu=(\g^0,\vec\g)=(\beta,\beta\vec\alpha)$ are the
Dirac matrices.  One useful representation of these matrices is
\be
\beta=\g^0=\left(\begin{array}{cc} I & 0 \\ 0 & -I \end{array}\right), \;\;
\vec\alpha=\left(\begin{array}{cc} 0 & \vec\sigma \\ \vec\sigma & 0 \end{array}\right), \;\;
\vec\g=\left(\begin{array}{cc} 0 & \vec\sigma \\ -\vec\sigma & 0 \end{array}\right),
\ee
with $I$ the $2\times2$ identity matrix and $\vec\sigma$ the Pauli matrices.

The field equation is the Dirac equation
\be
(i\g\cdot\partial-m)\psi=0
\ee
The light-front gamma matrices are $\g^\pm\equiv\g^0\pm\g^3=(\g^\mp)^\dagger$.  They
obey the usual anticommutation relation $\{\g^\mu,\g^\nu\}=2g^{\mu\nu}$,
with $g^{\mu\nu}$ the light-front metric. The Dirac equation can then be
written as
\be
(i\frac12\g^0\g^+\partial_++i\frac12\g^0\g^-\partial_-
+i\senk{\alpha}\cdot\senk{\partial}-\beta m)\psi=0.
\ee

With use of the projections $\Lambda_\pm\equiv\frac12 \g^0\g^\pm$, which
satisfy
\be
\Lambda_\pm^2=\Lambda_\pm,\;\;
\Lambda_\pm\Lambda_\mp=0,\;\;
\Lambda_++\Lambda_-=1 ,
\ee 
the Dirac equation separates into a dynamical equation for $\psi_+\equiv\Lambda_+\psi$
\be
i\partial_+\psi_+ -(-i\senk{\alpha}\cdot\senk{\partial}+\beta m)\psi_-=0
\ee
and a constraint equation for $\psi_-\equiv\Lambda_-\psi$
\be
i\partial_-\psi_- -(-i\senk{\alpha}\cdot\senk{\partial}+\beta m)\psi_+=0.
\ee

Plane-wave solutions are readily found.  Let $\psi_\pm(x)=w_\pm(p)e^{-ip\cdot x}$,
where $\Lambda_\pm w_\pm=w_\pm$.  Then the dynamical and constraint equations
become
\be
p^+w_--(\senk{\alpha}\cdot\senk{p}+\beta m)w_+=0,\;\;
p^-w_+-(\senk{\alpha}\cdot\senk{p}+\beta m)w_-=0.
\ee
The constraint equation is solved, to yield $w_-=\frac{1}{p^+}(\senk{\alpha}\cdot\senk{p}+\beta m)w_+$.
Substitution of this solution into the dynamical equation, combined with reduction of 
Dirac matrix products, leaves
\be
p^-w_+-\frac{m^2+\senk{p}^2}{p^+}w_+=0
\ee
which is immediately satisfied by $p$ on the mass shell.  The only condition
that remains to be satisfied is $\Lambda_+w_+=w_+$.  In other words, $w_+$
must be an eigenvector of the projection with eigenvalue 1.

In the Dirac representation of the $\g$-matrices, the projection matrix
has the form
\be
\Lambda_+=\frac12\left(\begin{array}{cccc}1 & 0 & 1 & 0 \\ 
                                          0 & 1 & 0 & -1 \\
		                                      1 & 0 & 1 & 0  \\ 
		                                      0 & -1 & 0 & 1 \end{array}\right).
\ee 
This matrix has two eigenvectors, both with eigenvalue $+1$,
\be
\chi_{+\frac12}=\frac{1}{\sqrt2}\left(\begin{array}{c} 1\\0\\1\\0 \end{array}\right), \;\;
\chi_{-\frac12}=\frac{1}{\sqrt2}\left(\begin{array}{c} 0\\1\\0\\-1 \end{array}\right) .
\ee 
These can be used as a spinor basis for the Fourier expansion of the field $\psi_+$.

The solution of the Dirac equation for the dynamical field can then be written as
\be
\psi_+(x)=\int\frac{d\ub{p}}{\sqrt{16\pi^3}}\sum_{s=\pm1/2}
     \left[b_s(\ub{p})\chi_s e^{-ip\cdot x}
           +d_s^\dagger(\ub{p})\chi_{-s} e^{ip\cdot x}\right].
\ee
The nonzero anticommutators are
\be
\{b_s(\ub{p}),b_{s'}^\dagger(\ub{p}')\}=\delta_{ss'}\delta(\ub{p}-\ub{p'}), \;\;
\{d_s(\ub{p}),d_{s'}^\dagger(\ub{p}')\}=\delta_{ss'}\delta(\ub{p}-\ub{p'}).
\ee
The expansion for the constrained field $\psi_-$ is obtained by
applying $(\senk{\alpha}\cdot\senk{p}+\beta m)/p^+$ to the expansion
for $\psi_+$.  The complete fermion field $\psi=\psi_++\psi_-$ can
then be simplified to
\be
\psi(x)=\int\frac{d\ub{p}}{\sqrt{16\pi^3 p^+}}\sum_{s=\pm1/2}
     \left[b_s(\ub{p})u_s(\ub{p})e^{-ip\cdot x}
           +d^\dagger_s(\ub{p})v_s(\ub{p}) e^{ip\cdot x}\right],
\ee
with
\be
u_s(\ub{p})\equiv\frac{1}{\sqrt{p^+}}\left[p^++\senk{\alpha}\cdot\senk{p}+\beta m\right]\chi_s, \;\;
v_s(\ub{p})\equiv\frac{1}{\sqrt{p^+}}\left[p^++\senk{\alpha}\cdot\senk{p}-\beta m\right]\chi_{-s}.
\ee
The index $s$ is the light-front helicity, sometimes loosely called the spin (projection) and
not to be confused with the ordinary (Jacob--Wick) helicity, the latter being the projection
onto the equal-time three-momentum.  The light-front helicity is left invariant by boosts
that the leave the light front itself invariant.

In terms of the dynamical field, the Lagrangian is
\be
{\cal L}=i\psi_+^\dagger\partial_+\psi_+
   -\psi_+^\dagger(i\senk{\alpha}\cdot\senk{\partial}-\beta m)
         \frac{1}{i\partial_-}(i\senk{\alpha}\cdot\senk{\partial}-\beta m)\psi_+,
\ee
where $\frac{1}{i\partial_-}$ appears as part of a formal solution of the
constraint equation that eliminates $\psi_-$.  The Hamiltonian density is
\be
{\cal H}=\psi_+^\dagger(i\senk{\alpha}\cdot\senk{\partial}-\beta m)
         \frac{1}{i\partial_-}(i\senk{\alpha}\cdot\senk{\partial}-\beta m)\psi_+,
\ee
and the normal-ordered light-front Hamiltonian is
\be
\Pminus=\int  d\ub{p}\frac{m^2+\senk{p}^2}{p^+}\sum_s
       \left[b_s^\dagger(\ub{p})b_s(\ub{p})+d_s^\dagger(\ub{p})d_s(\ub{p})\right].
\ee

\subsubsection{\it vector field}

The Lagrangian of a free massive vector field is
\be
{\cal L}=-\frac14 F^2+\frac12\mu A^2-\frac12\zeta(\partial\cdot A)^2,
\ee
where $\zeta$ is the gauge-fixing parameter and 
$F^{\mu\nu}=\partial^\mu A^\nu-\partial^\nu A^\mu$ is the field-strength
tensor.  The Lorentz gauge condition $\partial\cdot A=0$
is to be satisfied by a projection of states onto a physical subspace.

An alternative gauge choice~\cite{LFreview1} is light-cone gauge $A^+=0$.  
The $A^-$ component is then a constrained field and only the transverse
components $\senk{A}$ are dynamical.  Thus, the quantization has the
advantage of requiring only physical fields with positive metric.
However, there is the disadvantage that there is no gauge parameter with
respect to which one might check gauge invariance (in the massless limit).

The Euler--Lagrange field equation for the Lorentz-gauge Lagrangian is
\be
(\Box +\mu^2)A_\mu-(1-\zeta)\partial_\mu(\partial\cdot A)=0.
\ee
The solution as a Fourier expansion can be constructed by
the light-front analog~\cite{ArbGauge} of a method due to 
Stueckelberg~\cite{Stueckelberg}.  With the introduction of
a four-momentum $\tilde k$ associated with a different
mass $\tilde\mu\equiv\mu/\sqrt{\zeta}$, defined by
\be
\ub{\tilde k}=\ub{k}, \;\; \tilde k^-=(k_\perp^2+\tilde\mu^2)/k^+,
\ee
the expansion can be written as
\be
A_\mu(x)=\int\frac{d\ub{k}}{\sqrt{16\pi^3 k^+}}\left\{\sum_{\lambda=1}^3
   e_\mu^{(\lambda)}(\ub{k})\left[ a_\lambda(\ub{k})e^{-ik\cdot x}
            + a_\lambda^\dagger(\ub{k})e^{ik\cdot x}\right]
+e_\mu^{(0)}(\ub{k})\left[ a_0(\ub{k})e^{-i\tilde k\cdot x}
            + a_0^\dagger(\ub{k})e^{i\tilde k\cdot x}\right]\right\},
\ee
with polarization vectors defined by
\be
e^{(1,2)}(\ub{k})=(0,2 \hat e_{1,2}\cdot \vec{k}_\perp/k^+,\hat e_{1,2}), \;\;
e^{(3)}(\ub{k})=((k_\perp^2-\mu^2)/k^+,k^+,\vec k_\perp)/\mu, \;\;
e^{(0)}(\ub{k})=\tilde k/\mu. 
\ee
The $\hat e_{1,2}$ are transverse unit vectors.
These polarizations satisfy $k\cdot e^{(\lambda)}=0$ 
and $e^{(\lambda)}\cdot e^{(\lambda')}=-\delta_{\lambda\lambda'}$
for $\lambda,\lambda'=1,2,3$.
The first term in $A_\mu$ satisfies both $(\Box +\mu^2)A_\mu=0$ and
$\partial\cdot A=0$ separately.  The $\lambda=0$ term violates each, 
but the violations cancel in the sum, which is the field equation.
The nonzero commutators are
\be
[a_\lambda(\ub{k}),a_{\lambda'}^\dagger(\ub{k'})]
     =\epsilon^\lambda \delta_{\lambda\lambda'}\delta(\ub{k}-\ub{k'}),
\ee
with $\epsilon^\lambda=(-1,1,1,1)$ the metric of each field component.
 
The light-front Hamiltonian density is
\be
{\cal H}=\frac12\sum_{\mu=0}^3 \epsilon^\mu
       \left[(\partial_\perp A^\mu)^2+\mu^2 (A^\mu)^2\right] 
  +\frac12(1-\zeta)(\partial\cdot A)(\partial\cdot A
                                -2\partial_-A^- -2\partial_\perp\cdot\vec A_\perp).
\ee
The first term is obviously the Feynman-gauge ($\zeta=1$) piece.
The light-front Hamiltonian for the free massive vector field is then found to be
\be
 \Pminus=\int d\ub{k} \sum_\lambda \epsilon^\lambda \frac{k_\perp^2+\mu_\lambda^2}{k^+}
                a_\lambda^\dagger(\ub{k})a_\lambda(\ub{k}),
\ee
with $\mu_\lambda=\mu$ for $\lambda=1,2,3$, but $\mu_0=\tilde\mu$.
Thus, the Hamiltonian for the free vector field takes the usual
form except that the mass of the fourth polarization is
different and gauge dependent and that the metric of this
polarization is opposite that of the other polarizations.
In Feynman gauge, this reduces to the usual Gupta--Bleuler
quantization~\cite{GuptaBleuler} with $\tilde\mu=\mu$.

\subsection{\it Wave Functions \label{sec:WaveFns}}

The light-front Hamiltonian eigenvalue problem is
\be
\Pminus|\psi(\ub{P})\rangle=\frac{M^2+P_\perp^2}{P^+}|\psi(\ub{P})\rangle, \;\;
\ub{\cal P}|\psi(\ub{P})\rangle=\ub{P}|\psi(\ub{P})\rangle.
\ee
The second equation is solved explicitly by expanding the eigenstate $|\psi(\ub{P})\rangle$
in Fock states with $n$ constituents with relative momenta $x_i$ and $\senk{k_i}$
\be
|x_i,\senk{k_i},\ub{P},n\rangle=\frac{1}{\sqrt{n!}}\prod_{i=1}^n 
                a^\dagger(x_iP^+,\senk{k_i}+x_i\senk{P})|0\rangle.
\ee
Here the construction is for scalar constituents of a single type, 
but the form is easily generalized. 

The empty Fock vacuum $|0\rangle$ is an eigenstate of 
$\Pminus$,\footnote{For an in-depth discussion of the light-front vacuum,
see \protect\cite{HerrmannPolyzou}.}
even in the presence of interactions, with the possible exception of
contributions from modes of zero $p^+$.  This is because the
longitudinal light-front momentum $p^+$ cannot be negative, and
no interaction can create particles from the vacuum without violating
momentum conservation.  For practical calculations, this is an
advantage of light-front quantization, in that massive eigenstates
can be computed without first computing the vacuum state, and 
any Fock-state expansion for a massive state does not have spurious
vacuum contributions.  The latter aspect makes the Fock-state
wave functions unambiguous as probability amplitudes for 
constituent momentum distributions.  This does leave open
the question of the connection with the equal-time vacuum
and properties such as symmetry breaking that are usually
associated with the vacuum~\cite{HerrmannPolyzou,Martinovic}; however, these
considerations are beyond the scope of this review.

The Fock-state expansion for the eigenstate is
\be \label{eq:FSexpansion}
|\psi(\ub{P})\rangle=\sum_n (P^+)^{\frac{n-1}{2}}\int\prod_i^n dx_i d^2k_{i\perp}
       \delta(1-\sum_i^n x_i)\delta(\sum_i^n \senk{k_i})\psi_n(x_i,\senk{k_i})|x_i,\senk{k_i},\ub{P},n\rangle.
\ee
The $\psi_n$ are the Fock-state momentum-space wave functions for $n$ constituents.
Substitution into the first equation of the eigenvalue problem yields a coupled
set of integral equations for these wave functions, with the total mass $M$ as the
eigenvalue.  These equations and the wave functions are independent of the
total momentum $\ub{P}$.  

The normalization of the wave functions is determined
by the requirement $\langle\psi(\ub{P'})|\psi(\ub{P})\rangle=\delta(\ub{P'}-\ub{P})$.
Once the contractions of the creation and annihilation operators are carried out,
this reduces to
\be
1=\sum_n \int\prod_i^n dx_i d^2k_{i\perp}
       \delta(1-\sum_i^n x_i)\delta(\sum_i^n \senk{k_i})|\psi_n(x_i,\senk{k_i})|^2.
\ee
The individual terms in the sum over $n$ provide the probabilities for each Fock sector.
The wave functions and their normalization are independent of $P^+$ because the
factor of $(P^+)^{\frac{n-1}{2}}$ in (\ref{eq:FSexpansion}) has been arranged to
cancel the factors of $1/P^+$ that come from the contractions
\be
[a(x'_{i'}P^{\prime+},\senk{k'_{i'}}+x'_{i'}\senk{P'}),a^\dagger(x_iP^+,\senk{k_i}+x_i\senk{P})]
=\delta(x'_{i'}P^{\prime+}-x_iP^+)\delta(\senk{k'_{i'}}+x'_{i'}\senk{P'}-\senk{k_i}-x_i\senk{P})
\ee
when integrated with respect to $x_i$.

For the free scalar, the equations decouple as simply
\be
\left[\sum_i \frac{m^2+k_{i\perp}^2}{x_i}\right]\psi_n(x_i,\senk{k_i})=M^2\psi_n(x_i,\senk{k_i}).
\ee
This happens because 
\bea
\sum_i\frac{m^2+(\senk{k_i}+x_i\senk{P})^2}{x_iP^+}
  &=&\frac{1}{P^+}\sum_i\left[\frac{m^2+k_{i\perp}^2+2x_i\senk{k_i}\cdot\senk{P}+x_i^2P_\perp^2}{x_i}\right]  \\
  &=&\frac{1}{P^+}\left[\sum_i\frac{m^2+k_{i\perp}^2}{x_i}+2\left(\sum_i\senk{k_i}\right)\cdot\senk{P}
                       +\left(\sum_i x_i\right)P_\perp^2\right] \nonumber \\
  &=&\frac{1}{P^+}\left[\sum_i\frac{m^2+k_{i\perp}^2}{x_i}+P_\perp^2\right]. \nonumber
\eea
The last step follows from the momentum-conservation constraints $\sum_i\senk{k_i}=0$ and $\sum_i x_i=1$.

To make a connection with nonrelativistic quantum mechanics, consider
the invariant mass $\sum_i (m^2+k_{i\perp}^2)/x_i$ in the center of
mass frame, where $P^+=M$ and $\senk{P}=0$.  In this frame,
$x_i=(\sqrt{m^2+\vec{k}_i^2}+k_{iz})/M$ and
\bea
\sum_i \frac{m^2+k_{i\perp}^2}{x_i}=
  &=& M\sum_i \frac{m^2+\vec{k}_i^2-k_{iz}^2}{\sqrt{m^2+\vec{k}_i^2}+k_{iz}}
    = mM\sum_i\frac{1+\vec{k}_i^2/m^2-k_{iz}^2/m^2}{\sqrt{1+\vec{k}_i^2/m^2}-k_{iz}/m} \nonumber \\
   &\simeq & mM\left(n-\sum_i\frac{k_{iz}}{m}+\sum_i\frac{\vec{k}_i^2}{2m^2}\right)
   =M\left(nm+\sum_i\frac{\vec{k}_i^2}{2m}\right).
\eea
Thus, the equation for the $n$-body wave function $\psi_n$ is approximately
\be
\sum_i\frac{\vec{k}_i^2}{2m}\psi_n=(M-nm)\psi_n.
\ee

\subsection{\it Observables \label{sec:Observables}}

As in ordinary quantum mechanics, physical observables can 
be computed from matrix elements of
appropriate operators with respect to chosen eigenstates.
For example, the anomalous magnetic moment of the electron
can be computed from the spin-flip matrix element of 
the electromagnetic current~\cite{BrodskyDrell}.  This is most
conveniently done in the Drell--Yan frame~\cite{DrellYan}, 
where the photon transfers zero longitudinal momentum.  The electron
itself is represented by a Fock-state expansion that includes dressing
by photons and electron-positron pairs, as obtained from solving the
Hamiltonian eigenvalue problem in QED.  For the extension to
generalized parton distributions~\cite{GPDreviews}, see the recent work of 
Chakrabarti {\em et al.}~\cite{GPDs}.

In general, the transition
amplitude for absorption of a photon of momentum $q$ by a dressed
electron is given by~\cite{BRS}\footnote{Factors of $16\pi^3$ and $P^+$
are different from the expressions in \protect\cite{BRS} and 
\protect\cite{BrodskyDrell} because here states are normalized by 
$\langle\psi(\ub{P'})|\psi(\ub{P})\rangle=\delta(\ub{P'}-\ub{P})$
rather than $\langle\psi(\ub{P'})|\psi(\ub{P})\rangle=16\pi^3P^+\delta(\ub{P'}-\ub{P})$.}
\be \label{eq:Jplus}
16\pi^3\langle\psi^{\sigma}(\ub{P}+\ub{q})|J^+(0)|\psi^\pm(\ub{P})\rangle
=\bar{u}_\sigma(\ub{P}+\ub{q})\left[F_1(q^2)
             +i\frac{\sigma^{\mu\nu}q_\nu}{2M}F_2(q^2)\right]u_\pm(\ub{p}),
\ee
where $F_1$ and $F_2$ are the usual Dirac and Pauli form factors
and $|\psi^\sigma(\ub{P})\rangle$ is the dressed electron state 
with light-front helicity $\sigma$.
With $\senk{P}=0$, $q^+=0$, and $q^-=2q\cdot P/P^+$, the form factors can
be obtained from~\cite{BrodskyDrell}
\be
F_1(q^2)=\frac{16\pi^3}{2}\langle \psi^\sigma(\ub{P}+\ub{q})| J^+(0) |\psi^\sigma(\ub{P})\rangle
\ee
and
\be
-\left(\frac{q_x-iq_y}{2M}\right) F_2(q^2)=
        \frac{16\pi^3}{4\sigma}\langle \psi^\sigma(\ub{P}+\ub{q})| J^+(0) |\psi^{-\sigma}(\ub{P})\rangle.
\ee
Note that the factor of 1/2 comes from the normalization of the helicity spinors: 
$\bar{u}\gamma^+ u=2p^+$.
The plus component is used because, unlike the other components, it is not renormalized
when the Fock space is truncated~\cite{BRS,ChiralLimit}.  The truncation destroys
covariance, and the calculation of the other components requires great care~\cite{ChoiJi}.

The normal-ordered current operator $J^+$ is
\be
J^+(0)=:\psi(0)^\dagger\g^0\g^+\psi(0):
      =2:\psi(0)^\dagger\Lambda_+\psi(0):
      =2:\psi_+(0)^\dagger\psi_+(0):,
\ee
which, on use of the mode expansion for $\psi_+$, simplifies to
\bea
J^+(0)&=&\int\frac{d\ub{p'}}{\sqrt{16\pi^3}}\sum_{s'=\pm1/2}
     \int\frac{d\ub{p}}{\sqrt{16\pi^3}}\sum_{s=\pm1/2}
     :\left[b_{s'}^\dagger(\ub{p'})\chi_{s'}^\dagger +d_{s'}(\ub{p'})\chi_{-s}^\dagger \right]
     \left[b_s(\ub{p})\chi_s +d_s^\dagger(\ub{p})\chi_{-s} \right]: \\
     &=&\int\frac{d\ub{p}d\ub{p'}}{16\pi^3}\sum_{s=\pm1/2}
       \left[b_s^\dagger(\ub{p'})b_s(\ub{p})+b_s^\dagger(\ub{p'})d_{-s}^\dagger(\ub{p})
            -d_s^\dagger(\ub{p})d_s(\ub{p'})+d_{-s}(\ub{p'})b_s(\ub{p})\right].
\eea
The last step required use of the orthonormality of $\chi_s$.  With the current
in this form, the utility of the $q^+=0$ frame becomes apparent; with no change
in the longitudinal momentum, the pair creation and annihilation terms do not
contribute and the current is diagonal in particle number.\footnote{However,
there can be zero-mode contributions~\protect\cite{deMelo}.}
     
Substitution of the final expression for the current and of the Fock-state expansion 
for the electron eigenstate into the matrix elements for the form factors leads 
to~\cite{BrodskyDrell}
\be
F_1(q^2)=\sum_n\sum_je_j\int\delta(1-\sum_i x_i)\prod_i dx_i 
     \delta(\sum_i \senk{k_i}) \prod_i d^2k_{i\perp }
   \psi_n^{\sigma *}(x_i,\vec {k}_{i\perp}^{\,\prime})
   \psi_n^\sigma(x_i,\vec {k}_{i\perp})   
\ee
and
\be
-\left(\frac{q_x-iq_y}{2M}\right) F_2(q^2)=\sum_n\sum_je_j\int\delta(1-\sum_i x_i)\prod_i dx_i
    \delta(\sum_i \senk{k_i})\prod_i d^2k_{i\perp}
   \psi_n^{1/2*}(x_i,\vec {k}_{i\perp}^{\,\prime})
   \psi_n^{-1/2}(x_i,\vec {k}_{i\perp}^{\,\prime}).
\ee
Here $\psi_n^\sigma$ is the $n$-body Fock-state wave function for the 
eigenstate with light-front helicity $\sigma$, $e_j$ is the fractional charge of the 
struck constituent, and $\senk{k'_i}$ is
\be
\senk{k'_i}=\left\{\begin{array}{ll} \senk{k_i}-x_i\senk{q}, & i\neq j \\
                                     \senk{k_j}+(1-x_j)\senk{q}, & i=j. \end{array}\right.
\ee
As can be easily seen, the normalization of the eigenstate
is equivalent to $F_1(0)=1$.  

The anomalous moment is $a_e=F_2(0)$, which
requires the taking of the limit to zero momentum transfer. As 
shown in \cite{BrodskyDrell}, this limit can be expressed as
\bea \label{eq:ae}
a_e&=&\mp M \sum_n \sum_j e_j \int \delta(1-\sum_i x_i)\delta(\sum_i\senk{k_i})
       \left(\prod_i dx_id^2k_{i\perp}\right) \\
   && \times \psi_n^{\pm *}(x_i,\senk{k_i})
       \left[\sum_{i\neq j} x_i\left(\frac{\partial}{\partial k_{ix}}
                                    \pm i\frac{\partial}{\partial k_{iy}}\right)\right]
        \psi_n^\mp(x_i,\senk{k_i}). \nonumber
\eea
Thus, the anomalous moment can be computed given the solution to
the coupled equations for the wave functions.

The Dirac form factor, or more specifically its slope at zero momentum
transfer, can be used to measure the average radius of the eigenstate
as $R=\sqrt{-6F'_1(0)}$.  The slope is obtained from the derivative
of the expression for $F_1(q^2)$, which can be simplified to
\be \label{eq:Fprime}
F'_1(0)=\sum_n\sum_j \frac{e_j}{2}\int\delta(1-\sum_i x_i)\prod_i dx_i 
    \delta(\sum_i \senk{k_i}) \prod_i d^2k_{i\perp } \sum_{i\neq j}
   |x_i\senk{\nabla_i}\psi_n^\sigma(x_i,\vec {k}_{i\perp})|^2.
\ee

The finite temperature properties of a theory
can computed from the partition function $Z=e^{-E/T}$.
One does not use the light-front analog $e^{-P^-/T_{\rm LF}}$
because it does not correspond to a heat bath at rest~\cite{Elser}.
Other examples of where this choice matters can be found in the 
variational analysis of $\phi^4$ theory~\cite{phi4variational}
and the light-front derivation of the Casimir effect~\cite{LFCasimir,Almeida,Lenz}.
This is not to say that light-front quantization cannot be used;
the physics should be the same in any coordinate system.

To compute the partition function, one needs the spectrum
of the theory, which is what nonperturbative
light-front methods can yield.  Each mass eigenstate contributes 
according to its ordinary energy $E$.  For bosonic states of mass $M_n$
in one space dimension this yields a free-energy contribution of~\cite{FiniteTemp}
\be
F_B=\frac{VT}{\pi}\sum_{n=1}^\infty
\int_{M_n}^\infty dp_0
\frac{p_0}{\sqrt{p_0^2-M_n^2}}
\ln \left( 1- e^{- p_0/T }\right)
\ee
in a volume $V$.
For fermions the contribution is
\be
F_F= -\frac{VT}{\pi}\sum_{n=1}^\infty
   \int_{M_n}^\infty dp_0
\frac{p_0}{\sqrt{p_0^2-M_n^2}}
 \ln \left( 1+ e^{- p_{0}/T }\right).
\ee
In supersymmetric theories, the bosonic and fermionic mass spectra
are the same, and we can readily combine these expressions to
obtain the total free energy
\be
F(T,V)=-(K-1)\pi VT^2
-\frac{2VT}{\pi}\sum_{n=1}^{\infty}{\sum_{l=0}^{\infty}}M_{n}
        \frac{K_{1}\left((2l+1)\frac{M_{n}}{T}\right)}{(2l+1)} .
\ee
Here the logarithms have been expanded, the integral over $p_0$ performed,
and the contribution of $K-1$ zero-mass states separated explicitly.
The sum over $l$ is well approximated by the first few terms.
The sum over $n$ can be represented by an integral over a density of states
$\int \rho(M) dM $.
The density can be approximated by a continuous function that is fit to
the numerical spectrum, and the integral $\int dM$ computed
by standard numerical techniques.

Additional work on finite-temperature physics on the light front 
has been done by Beyer and Strauss.  They studied both two-dimensional
QED~\cite{Strauss:2008zx} and the Nambu-Jona-Lasinio model~\cite{Strauss:2009uj}.

\section{Methods of Calculation} \label{sec:Methods}

\subsection{\it Discretized Light-Cone Quantization \label{sec:DLCQ}}

One very systematic approach to solving a quantum field theory
nonperturbatively is that of discretized light-cone quantization 
(DLCQ)~\cite{PauliBrodsky,MaskawaYamawaki}.  It has had particular
utility in two dimensions.  This includes
calculations of eigenstates in supersymmetric Yang--Mills
theories~\cite{SuperYangMills} and 
$\phi^4$ theory~\cite{RozowskyThorn,phi4phase,Chakrabarti:2003tc,Chakrabarti:2003ha,Chakrabarti:2005zy},
as well as the early applications to Yukawa theory~\cite{PauliBrodsky},
$\phi^3$ and $\phi^4$ theories~\cite{VaryHari-phi3,VaryHari-phi4,VaryHari-coherent},
QED~\cite{QED1+1}, and QCD~\cite{QCD1+1}.

Although DLCQ is in a sense a trapezoidal approximation to the coupled
integral equations for the wave functions, it is based on quantization
in a discrete basis obtained by placing the system in a light-front box
\begin{equation}
-L<x^-<L\,,\;\; -L_\perp<x,y<L_\perp.
\end{equation}
For bosons, periodic boundary conditions are used and for fermions,
antiperiodic, leading to discrete momenta
\begin{equation}
p^+\rightarrow\frac{\pi}{L}n\,, \;\;
\senk{p}\rightarrow
     (\frac{\pi}{L_\perp}n_x,\frac{\pi}{L_\perp}n_y)\,,
\end{equation}
with $n$ even for bosons and odd for fermions.
Integrals are then replaced by discrete sums obtained as 
trapezoidal approximations on the grid of momentum values.
For a generic integral, this takes the form
\begin{equation} \label{eq:rawDLCQ}
\int dp^+ \int d^2p_\perp f(p^+,\senk{p})\simeq
   \frac{2\pi}{L}\left(\frac{\pi}{L_\perp}\right)^2
   \sum_n\sum_{n_x,n_y=-N_\perp}^{N_\perp}
   f(n\pi/L,\senk{n}\pi/L_\perp).
\end{equation}
The sum on $n$ is restricted by the integer {\em resolution}~\cite{PauliBrodsky}
\begin{equation} \label{eq:resolution}
K\equiv\frac{L}{\pi}P^+,
\end{equation}
with $K$ even for a boson eigenstate and odd for a fermion.
The index $n$ can be no larger than this because all longitudinal
momenta are positive, and the maximum individual momentum can then
be no more than the total.  The sums on $n_x$ and $n_y$ have
been truncated at $\pm N_\perp$, with $N_\perp$ typically
determined by a cutoff on the transverse momentum, either
directly or as a cutoff on the invariant mass. 

The longitudinal momentum fractions $x_i$ become
ratios of integers $n_i/K$.  
Because the $n_i$ are all positive, DLCQ
automatically limits the number of particles to be no more than $\sim\!\!K/2$.
An explicit truncation in particle number, the light-front equivalent
of the Tamm--Dancoff approximation~\cite{TammDancoff}, can also
be made. 

The limit $L\rightarrow\infty$ can be exchanged for the limit
$K\rightarrow\infty$.  This is because
the combination of momentum components that defines $\Pminus$ is
simply proportional to $L$, so that the combination $P^+\Pminus$,
which has eigenvalues in the form $M^2+P_\perp^2$,
is independent of $L$.  As $K$ is increased, the longitudinal
momentum is sampled at higher resolution.

The mode expansion for the quantum field is also approximated
by a discrete sum.  For example, the neutral scalar field becomes
\be
\phi(x^+=0)=\frac{\pi}{L_\perp} \sum_{\ub{n}} \frac{1}{\sqrt{8\pi^3 n}}
        \left[a(\ub{n}) e^{-i\pi n x^-/2L+i\pi \senk{n}\cdot\senk{x}/L_\perp}
              +a^\dagger(\ub{n}) e^{i\pi n x^-/2L-i\pi \senk{n}\cdot\senk{x}/L_\perp}\right],
\ee
with $\ub{n}=(n,n_x,n_y)$ and the creation operator for discrete momenta defined by
\begin{equation}
a^\dagger(\ub{n})=
    \frac{\pi}{L_\perp}\sqrt{\frac{2\pi}{L}}a^\dagger(\ub{p}).
\end{equation}
It then satisfies a simple commutation relation
\begin{equation}
\left[a(\ub{n}),a^\dagger(\ub{n}')\right]=\delta_{\ub{n'},\ub{n}}
          \equiv\delta_{n'n}\delta_{n'_xn_x}\delta_{n'_yn_y},
\end{equation}
which follows from the continuum commutation relation and the discrete
delta-function representation
\begin{equation}
\delta(\ub{p}-\ub{p}')=
  \frac{L}{2\pi}\left(\frac{L_\perp}{\pi}\right)^2
               \delta_{\ub{n'},\ub{n}}\,.
\end{equation}

The discrete approximation of the eigenstate, with $P^+=K \pi/L$, $\senk{P}=0$, 
and $\ub{K}\equiv(K,\senk{0})$, is then
\be
|\psi(\ub{K})\rangle=\sum_n \prod_i^n \sum_{\ub{n_i}} \delta_{\ub{K},\sum_i\ub{n_i}}
                       \psi_n(\ub{n_i})|\ub{n_i},n\rangle,
\ee
where the discrete Fock states are
\be
|\ub{n_i},n\rangle=\frac{1}{\sqrt{n!}}\prod_{i=1}^n a^\dagger(\ub{n_i})|0\rangle
\ee
and the discrete wave functions are related to the continuum wave functions by
\be
\psi_n(\ub{n_i})=\left(\frac{K}{2}\frac{\pi^2}{L_\perp^2}\right)^{(n-1)/2}
                        \psi_n(n_i/K,\senk{n}\pi/L_\perp).
\ee
The discrete eigenstate is normalized as $\langle\psi(\ub{K})|\psi(\ub{K})\rangle=1$,
and the wave functions as
\be
1=\sum_n \prod_i^n \sum_{\ub{n_i}} \delta_{\ub{K},\sum_i\ub{n_i}} |\psi_n(\ub{n_i})|^2.
\ee
Although the Fock basis is a natural way to write the eigenstate, a more
convenient basis for a numerical calculation is the number basis, which
eliminates summations over states that differ only by rearrangement of bosons
of the same type.

There are zero modes, modes with zero longitudinal momentum.
In DLCQ these are not dynamical but instead constrained by the
spatial average of the Euler--Lagrange field 
equation~\cite{MaskawaYamawaki,Heinzl,Robertson}.  These zero modes 
are usually either neglected or excluded by the choice of antiperiodic
boundary conditions.  This neglect does, however, slow convergence
of the numerical solution, because contributions of order $1/K$
have been dropped; these are to be compared with the nominal
$1/K^2$ errors associated with the trapezoidal approximation.
For theories with symmetry breaking, the neglect can have 
serious consequences for the understanding of vacuum 
effects~\cite{Heinzl,Robertson,Hornbostel,phi4Pinsky,Grange}.
When included, zero modes generate effective
interactions in the DLCQ Hamiltonian~\cite{Wivoda,Maeno,ZeroModes}.
These effective interactions are typically due to end-point
corrections where, although the wave function goes to zero
as $x_i$ goes to zero, it does so slowly enough that the 
integral has a nonzero contribution which is missed by
DLCQ's neglect of the $x_i=0$ points in its trapezoidal
approximation.  They can be computed by solving the
constraint equation, either nonperturbatively~\cite{phi4Pinsky}
or as an expansion in powers of $1/K$~\cite{ZeroModes}.
There can also be quantum corrections to the 
constraint equation, such as contributions
from zero-mode loops~\cite{Hellerman,ZeroModeLoop}.

If the transverse cutoff, such as an invariant-mass cutoff, creates
a domain of integration that is not commensurate with the transverse
DLCQ grid, there are errors generated if the basic DLCQ approximation
is used.  There is a truncation error, where the edge of the domain
is not properly included, and there can be a loss of rotational
symmetry.  These can make the dependence on $K$ and $N_\perp$ very
erratic and delay numerical convergence.  However, these difficulties
can be overcome by improvements on the trapezoidal approximation
at the edge of the integration~\cite{bhm1}.  This idea also
opens the possibility of using integration schemes that are
more accurate than a trapezoidal rule for the entire domain;
Simpson's rule can be particularly helpful.  

In general, any quadrature scheme that uses equally spaced points can be
introduced.  These will place unequal weighting factors inside the discrete
sums.  For the trapezoidal rule, the weighting factors are all the
same (except for the neglected end points) and did not need to be
considered beyond an overall factor of $1/K$.  The unequal weights
will destroy the symmetry of the matrix representing the action
of $\Pminus$; however, this symmetry can be restored by a 
simple rescaling.  An eigenvalue problem of the form $\sum_j A_{ij}w_ju_j=\xi u_i$
can be rewritten as 
\begin{equation}
\sum_j\sqrt{w_i w_j}A_{ij} \sqrt{w_j}u_j=\xi\sqrt{w_i} u_i,
\end{equation}
with $\sqrt{w_i w_j}A_{ij}$ the new symmetric matrix.

The term `DLCQ' is sometimes extended to include quadratures that use
unequally spaced points to approximate the coupled integral equations.
This is at odds with the full intent of the DLCQ method, which discretizes
before quantization, a process that would not admit unequally
spaced points without spoiling momentum conservation for processes
with more than two particles.  Thus, the interaction terms of
the DLCQ Hamiltonian could not be resolved into products of
discrete creation and annihilation operators.

Nevertheless, quadratures with unequally spaced points can be a powerful
tool~\cite{PauliTrittmann,LebedLamm,YukawaTwoBoson,TwoPhotonQED}, 
even though their utility is limited to two-body equations.  This is because
the integral equations truncated at three-body contributions can usually
be reduced to an effective equation in the two-body sector, sometimes without
approximation and certainly when interactions are ignored in the three-body
sector.  The one-body and three-body wave functions are simply eliminated
in favor of expressions relating them to the two-body wave function,
which are then substituted into the original two-body equation.  The
use of unequally spaced quadratures for truncations beyond three 
constituents is best done by first introducing basis-function expansions,
as discussed below in Sec.~\ref{sec:FnExpansion}.

Quadratures with unequally spaced points can be particularly
important when PV regularization is used, because the structure
of the integrands in the effective equations can be such as
to require very high resolution near the endpoints, inversely
proportional to the PV mass squared~\cite{YukawaTwoBoson}.  
In DLCQ such resolution 
would necessitate extremely large values of $K$, making the
calculation intractable.

\subsection{\it Supersymmetric Discretized Light-Cone Quantization \label{sec:SDLCQ}}

The supersymmetric form of DLCQ (SDLCQ)~\cite{SDLCQ} is 
specifically designed to maintain supersymmetry in the discrete approximation.
Ordinary DLCQ violates supersymmetry by terms that do not survive
the continuum limit~\cite{Filippov}.  The SDLCQ construction discretizes the
supercharge $Q^-$ and {\em defines} the Hamiltonian ${\cal P}^-$
by the superalgebra relation ${\cal P}^-=\{Q^-,Q^-\}/2\sqrt{2}$.
The range of transverse momentum is limited by a simple cutoff in the momentum value.
The effects of zero modes cancel between bosonic and fermionic 
contributions, which enter with opposite signs~\cite{SDLCQzeromodes}.

The work done with SDLCQ typically uses the slightly different definition
of light-front coordinates, with division by $\sqrt{2}$.
The time coordinate is $x^+=(t+z)/\sqrt{2}$, and the space
coordinate is $x^-\equiv (t-z)/\sqrt{2}$.  The conjugate
variables are the light-front energy $p^-=(E-p_z)/\sqrt{2}$
and momentum $p^+\equiv (E+p_z)/\sqrt{2}$.  The mass-shell 
condition $p^2=m^2$ then yields $p^-=\frac{m^2 +p_\perp^2}{2p^+}$;
notice the factor of 2 in the denominator.

For example, consider supersymmetric QCD (SQCD) with a Chern--Simons (CS)
term in the large-$N_c$ approximation~\cite{SQCD-CS}.  The action is
\bea \label{eq:SQCDaction}
S&=&\int d^3x\mbox{Tr}\left\{-\frac{1}{4}F_{\mu\nu}F^{\mu\nu}
+D_\mu \xi^\dagger D^\mu \xi
+i{\bar\Psi} D_\mu\Gamma^\mu\Psi\right.
\nonumber\\
&&\left.-g\left[{\bar\Psi}\Lambda\xi
+\xi^\dagger{\bar\Lambda}\Psi\right]
+\frac{i}{2}{\bar\Lambda}\Gamma^\mu D_\mu \Lambda
+\frac{\kappa}{2}\epsilon^{\mu\nu\lambda}
  \left[A_{\mu}\partial_{\nu}A_{\lambda}
           +\frac{2i}{3}gA_\mu A_\nu A_\lambda \right]
+\kappa\bar{\Lambda}\Lambda
\right\}.
\eea
The adjoint fields are the gauge boson $A_\mu$ (gluons)
and a Majorana fermion $\Lambda$ (gluinos);
the fundamental fields are the Dirac fermion $\Psi$ (quarks)
and a complex scalar $\xi$ (squarks).
The CS coupling $\kappa$ induces a mass for the adjoint fields
without breaking the supersymmetry; this inhibits formation
of the long strings characteristic of super Yang--Mills theory.
The covariant derivatives are
\be
D_\mu\Lambda=\partial_\mu\Lambda+ig[A_\mu,\Lambda]\,,\;\;
D_\mu\xi=\partial_\mu\xi+igA_\mu\xi,  \;\;
D_\mu\Psi=\partial_\mu\Psi+igA_\mu\Psi.
\ee

The action is invariant under the following supersymmetry
transformations, which are parameterized by a two-component Majorana fermion
$\varepsilon$:
\be
\delta A_\mu=\frac{i}{2}{\bar\varepsilon}\Gamma_\mu\Lambda,\;\;
\delta\Lambda=\frac{1}{4}F_{\mu\nu}\Gamma^{\mu\nu}\varepsilon,\;\;
\delta\xi=\frac{i}{2}{\bar\varepsilon}\Psi,\;\;
\delta\Psi=-\frac{1}{2}\Gamma^\mu\varepsilon D_\mu\xi.
\ee
The supercharge associated with the corresponding Noether current is
\bea\label{sucharge}
{\bar\varepsilon}Q&=&\int dx^-dx^2\left(
\frac{i}{4}{\bar\varepsilon}\Gamma^{\alpha\beta}
\Gamma^+\mbox{tr}\left(\Lambda F_{\alpha\beta}\right)+
\frac{i}{2}D_-\xi^\dagger\
{\bar\varepsilon}\Psi+\frac{i}{2}\xi^\dagger{\bar\varepsilon}\Gamma^{+\nu}
D_\nu\Psi\right.\nonumber\\
&&-\left.\frac{i}{2}{\bar\Psi}\varepsilon D^+\xi+\frac{i}{2}D_\nu
{\bar\Psi}\Gamma^{+\nu}\varepsilon\xi\right)\,.
\eea

In order that the Majorana fermion $\Lambda$ can be chosen real,
the following representation is used for the Dirac matrices in
three dimensions:
\be
\g^0=\sigma_2, \;\; \g^1=i\sigma_1, \;\; \g^2=i\sigma_3 ,
\ee
The fermionic spinor fields and the supercharge in terms of components are
\be
\Lambda=\left(\lambda,{\tilde\lambda}\right)^T\,,\qquad
\Psi=\left(\psi,{\tilde\psi}\right)^T\,,\qquad
Q=\left(Q^+,Q^-\right)^T\,.
\ee
The superalgebra has the form
\be\label{sualg}
\{Q^+,Q^+\}=2\sqrt{2}P^+\,,\qquad \{Q^-,Q^-\}=2\sqrt{2}P^-\,,\qquad
\{Q^+,Q^-\}=-4P_\perp\,.
\ee
The supercharge $Q^-$ is then discretized, and $\Pminus$ is
constructed from the superalgebra relation.

The eigenstates of $\Pminus$ are of two types: meson-like states
\be
{\bar f}^\dagger_{i_1}(\ub{k}_1) a^\dagger_{i_1i_2}(\ub{k}_2)\dots
    b^\dagger_{i_ni_{n+1}}(\ub{k}_{n-1})\dots 
         f^\dagger_{i_p}(\ub{k}_n)|0\rangle ,
\ee
where $f^\dagger=$ creates a quark or squark, $a^\dagger$ creates
a gluon, and $b^\dagger$ creates a gluino; and glueball states
\be
{\rm Tr}[a^\dagger_{i_1i_2}(k_1)\dots b^\dagger_{i_ni_{n+1}}(k_{n})]|0\rangle .
\ee
Because of the supersymmetry, either could be a boson or a fermion.
In the large-$N_c$ limit, there is no mixing
between these states, and they are composed of single traces.  This
simplifies the calculation, particularly with respect to the
size of the matrices that need to be diagonalized.  In general,
states could be formed from multiple traces, such as
\be
{\rm Tr}[a^\dagger_{i_1i_2}(k_1)\dots b^\dagger_{i_ni_{n+1}}(k_{n})]
{\rm Tr}[a^\dagger_{j_1j_2}(p_1)\dots b^\dagger_{j_nj_{n+1}}(p_{n})]
\cdots|0\rangle,
\ee
which would lead to much larger SDLCQ matrix representations.

In addition to calculations of spectra, the SDLCQ approach can
be used to calculate matrix elements, including correlators.
Consider the (1+1)-dimensional stress-energy correlation function
\be
F(x^-,x^+)  \equiv\langle T^{++}(x)T^{++}(0)\rangle,
\ee
where $T^{\mu\nu}$ is the stress-energy tensor.
For the string theory corresponding 
to two-dimensional ${\cal N}$=(8,8) SYM theory, 
$F$ can be calculated on the string-theory side
in a weak-coupling super-gravity approximation.
Its behavior for intermediate separations $r\equiv\sqrt{2x^+x^-}$ 
is~\cite{1+1correlator,subsequent}
\be
F(x^-,x^+)=\frac {N_c^{\frac 32}}{g_{\rm YM}r^5}.
\ee
In the SDLCQ approximation, this can be computed from SYM theory
and compared.

With the total momentum $P^+=P_-$ fixed,
the Fourier transform can be expressed
in a spectral decomposed form as~\cite{1+1correlator}
\bea
\tilde F(P_-,x^+)
    &=&\frac 1{2L}\langle T^{++}(P_-,x^+) T^{++}(-P_-,0)\rangle \\
   &=&\sum_i \frac 1{2L}\langle 0|T^{++}(P_-,0)|i\rangle e^{-iP^i_+x^+}
       \langle i|T^{++}(-P_-,0)|0\rangle. \nonumber
\eea
The position-space form is recovered by the inverse transform,
with respect to $P_-=K\pi/L$. 
The continuation to Euclidean space is made by 
taking $r$ to be real.  This yields 
\be
F(x^-,x^+)=\sum_i \Big|\frac L{\pi}\langle 0|T^{++}(K)|i\rangle\Big|^2
  \left(\frac {x^+}{x^-}\right)^2 \frac{M_i^4 K_4(M_i \sqrt{2x^+x^-})}{8\pi^2K^3}.
\ee

The stress-energy operator $T^{++}$ is 
\be
T^{++}(x^-,x^+) ={\rm Tr} \left[(\partial_- X^I)^2
+\frac12 (iu^{\alpha}\partial_-u^{\alpha}-i(\partial_-u^{\alpha})u^{\alpha})\right].
\ee
In terms of the discretized creation operators, this becomes 
\be
T^{++}(-K)|0\rangle=\frac {\pi}{2L}\sum_{k=1}^{K-1} 
    \left[-\sqrt{k(K-k)} a^{\dagger}_{Iij}(K-k)a^{\dagger }_{Iji}(k)+
\left(\frac K2-k\right)
   b^{\dagger }_{\alpha ij}   (K-k)b^{\dagger }_{\alpha ji}(k)\right]|0\rangle,
\ee
with sums over $i$ and $j$ implied.
Thus $(L/\pi)\langle 0|T^{++}(K)|i\rangle$ is independent of $L$. 
Also, only one symmetry sector contributes.

The correlator behaves like $1/r^4$ at small $r$, as can be seen 
by taking the limit to obtain
\be \label{eq:smallr}
\left(\frac {x^-}{x^+}\right)^2F(x^-,x^+)
   \sim\frac{N_c^2(2n_b+n_f)}{4\pi^2r^4}(1-1/K).
\ee
To simplify the appearance of this behavior, $F$ can be rescaled
by defining
\be \label{eq:rescaledF}
f\equiv \langle T^{++}(x)T^{++}(0)\rangle 
      \left(\frac{x^-}{x^+}\right)^2
   \frac{4\pi^2r^4}{N_c^2(2n_b+n_f)}.
\ee
Then $f$ is just $(1-1/K)$ for small $r$.

The function $f$ can be computed numerically for small matrix
representations by obtaining the entire spectrum
and for large representations by using Lanczos iterations.
The Lanczos technique~\cite{Lanczos} (see Sec.~\ref{sec:Lanczos}) 
generates an approximate tridiagonal representation of the Hamiltonian 
which captures the important contributions after only a few iterations 
and which is easily diagonalized to compute the sum over eigenstates.
For the correlator, this sum is weighted by the square of the
projection $\langle i|T^{++}(-K)|0\rangle$.  The Lanczos diagonalization
algorithm  will naturally generate the states with
nonzero projection if $T^{++}(-K)|0\rangle$ is used as the initial
vector for the iterations.\footnote{Such an approach is related
to applications of the Lanczos algorithm to computation of
matrix elements of resolvents~\protect\cite{Haydock}.}

\subsection{\it Lanczos Algorithm \label{sec:Lanczos}}

The matrix approximations to the coupled integral equations
for the wave functions are typically quite large and sparse.
Standard diagonalization algorithms do not apply, because
they require storage of the entire matrix, zeros and all,
which is well beyond the capacity of current memory technology.
However, there exist alternative algorithms that rely upon 
only an ability to multiply the matrix with a vector,
something which can be done with relative ease for a
matrix stored in a compressed form that strips away the
zeros~\cite{SparseMatrix}.  These algorithms generate
better and better approximations to eigenvalues and
eigenvectors by iteration, typically converging first
to the largest and smallest eigenvalues.  The best
known of these algorithms is the Lanczos algorithm~\cite{Lanczos},
which actually predates the current standard algorithms.\footnote{The
Lanczos algorithm was temporarily abandoned due to stability issues.}

The basic Lanczos algorithm for the matrix eigenvalue problem
$A\vec\psi_\lambda=\lambda\vec\psi_\lambda$, with $A$ Hermitian,
is as follows.  Set $\vec u_1$
to some normalized initial guess and set $b_0=0$.  Then construct a sequence
of normalized vectors $\vec u_n$ according to the iteration
\be
b_n \vec u_{n+1}=A\vec u_n-a_n \vec u_n-b_{n-1}\vec u_{n-1},
\ee
with $a_n=\vec u_n^*\cdot A \vec u_n$ and $b_n$ chosen to normalize
$\vec u_{n+1}$.  The $\vec u_n$ form an
orthonormal basis with respect to which $A$ is tridiagonal
\begin{equation}
A\rightarrow T\equiv\left(\begin{array}{llllll}
              a_1 & b_1 & 0 & 0 & 0 & \ldots \\
              b_1 & a_2 & b_2 & 0 & 0 & \ldots \\
                0 & b_2 & a_3 & b_3 & 0 & \ldots \\
                0 & 0 & b_3 & .  & . & \ldots \\
                0 & 0 & 0 & . & . & \ldots \\
                . & . & . & . & . & \ldots \end{array} \right)\,. 
\end{equation}
The real symmetric matrix $T$ is easily diagonalized by ordinary means.  The
eigenvalues $\lambda_k$ of $T$ approximate the eigenvalues of $A$, and the
eigenvectors $\vec{c}_k$ of $T$ can be used to construct approximate
eigenvectors of $A$
\be
\vec \psi_{\lambda_k}=\sum_n c_{kn} \vec u_n.
\ee
The only use of $A$ is in the multiplication of $A$ times $\vec u_n$.
Generating a complete basis by iteration can yield
the exact answer; however, doing many fewer iterations, even 20,
can be sufficient to capture the extreme eigenvalues.
If $\vec u_{n+1}$ is zero, the process terminates naturally,
with $T$ an exact representation of $A$ in the subspace spanned by the
eigenvectors with nonzero projection on $\vec u_1$.

A slightly altered form of the algorithm minimizes the storage
requirements
\bea
\vec v_{n+1}&=&A \vec u_n-b_{n-1}\vec u_{n-1}, \;\;
a_n=\vec v_{n+1}^* \cdot \vec u_n, \;\;
\vec v_{n+1}^{\,\prime} = \vec v_{n+1}-a_n \vec u_n, \\
b_n&=&\sqrt{\vec v_{n+1}^{\,\prime *}\cdot\vec v_{n+1}^{\,\prime}}, \;\;
\vec u_{n+1}=\vec v_{n+1}^{\,\prime}/b_n.
\eea
In this form, the vectors $\vec u_{n-1}$, $\vec v_{n+1}$,
$\vec v_{n+1}^{\,\prime}$, and $\vec u_{n+1}$ can all be stored in 
the same array.  Therefore, storage is required for only two
vectors at a time, this sequence of overwritten vectors 
and $\vec u_n$.  To be able to
construct the eigenvectors of $A$, the vectors $\vec u_n$
do need to be saved for all $n$, written temporarily to disk 
and retrieved later, or the Lanczos algorithm can
be run a second time, after the diagonalization of $T$,
to accumulate the desired eigenvectors as each $\vec u_n$
is regenerated.  During the second pass of the algorithm,
the $a_n$ and $b_n$ are already known and do not need to be 
recalculated.

As simple as this all seems, there are limitations.
Because all of the vectors $\vec u_n$ are generated by
applying powers of $A$ to $\vec u_1$, only those eigenvectors
with nonzero projections on $\vec u_1$ should appear.
Depending on the application, this may actually be an
advantage; however, for a generic diagonalization, it may
be necessary to generate the initial guess with random
components and/or run the algorithm more than once with
different initial vectors.

Another limitation, which can be quite severe, is that
round-off errors will eventually destroy the orthogonality
of the Lanczos vectors $\vec u_n$.  This will allow
additional copies of the eigenvectors of $A$ to creep into the
calculation.  The eigenvalues of $T$ then include multiple
copies of eigenvalues of $A$, a false degeneracy.

Various strategies have been developed to overcome this limitation.
One is to re-orthogonalize the vectors as they are generated;
however, this consumes time and storage.  Another is to
simply accept the copies; the eigenvalues are not wrong, but
their degeneracy is unknown.  This is not as bad as it sounds,
since a correct estimate of any degeneracy is difficult
in any case, because any symmetry in the initial vector 
will suppress degenerate eigenvectors with different symmetry,
and multiple initial vectors will be needed to determine the
degeneracy.  A third approach is to restart the algorithm
after a few iterations, using the best found estimate of
the target eigenvector as the initial guess for the next
set of iterations.  A fourth strategy is to continue the
iterations without re-orthogonalization but then detect
and ignore the `ghost' copies.
The ghost copies can be detected by comparing the
eigenvalues of the matrix obtained from $T$ by deleting the first
row and first column~\cite{Cullum}; any eigenvalue that 
appears in both lists is spurious.

Convergence of the algorithm can be monitored by measuring 
directly the convergence of the desired eigenvalue and by 
checking an estimate of the error in the
eigenvalue, given by~\cite{Cullum} $|b_n c_{kn}|$,
where $n$ is the number of Lanczos iterations and $k$ is the index of
the desired eigenvalue of $T$.  If the
error estimate begins to grow, the iterations should be
restarted from the last best approximation to the eigenvector.

When PV regularization is used, the matrix representation has an
indefinite metric.  This could be handled with the biorthogonal
version of the Lanczos algorithm~\cite{Biorthogonal}; however,
a specialized form is much more efficient.
Let $\eta$ represent the metric signature, so that numerical dot 
products are written as ${\vec \phi}^{\,\prime *}\cdot\eta{\vec \phi}$.
The Hamiltonian matrix $A$ is not Hermitian but is self-adjoint with 
respect to this metric~\cite{Pauli} $\eta^{-1}A^\dagger\eta=A$.  
The Lanczos algorithm for the diagonalization of $H$ then takes the 
form~\cite{DLCQYukawa}
\bea 
\alpha_j&=&\nu_j{\vec q}_j^{\,*}\cdot\eta H{\vec q}_j, \;\;
   {\vec r}_j=H{\vec q}_j-\gamma_{j-1}{\vec q}_{j-1}
                                          -\alpha_j{\vec q}_j,\;\; 
   \beta_j=+\sqrt{|{\vec r}_j^{\,*}\cdot\eta{\vec r}_j|}, \;\;
 \\
{\vec q}_{j+1}&=&{\vec r}_j/\beta_j, \;\;
\nu_{j+1}=\mbox{sign}({\vec r}_j^{\,*}\cdot\eta{\vec r}_j),\;\;
   \nu_1=\mbox{sign}({\vec q}_1^{\,*}\cdot\eta{\vec q}_1), \;\;
   \gamma_j=\nu_{j+1}\nu_j\beta_j,
\eea
where ${\vec q}_1$ is chosen as a normalized initial guess and $\gamma_0=0$.
Just as for the ordinary Lanczos algorithm, 
the original matrix $A$ acquires a tridiagonal matrix 
representation $T$ with respect to the basis formed by the 
vectors ${\vec q}_j$:
\begin{equation}
A\rightarrow T\equiv\left(\begin{array}{llllll}
              \alpha_1 & \beta_1 & 0 & 0 & 0 & \ldots \\
              \gamma_1 & \alpha_2 & \beta_2 & 0 & 0 & \ldots \\
                0 & \gamma_2 & \alpha_3 & \beta_3 & 0 & \ldots \\
                0 & 0 & \gamma_3 & .  & . & \ldots \\
                0 & 0 & 0 & . & . & \ldots \\
                . & . & . & . & . & \ldots \end{array} \right)\,. 
\end{equation}
By construction, the elements of $T$ are real.  The new matrix is not symmetric
but is self-adjoint, with respect to an induced metric $\nu=\{\nu_1,\nu_2,\ldots\}$.
The eigenvalues of $T$ approximate some of the eigenvalues of $A$, even after
only a few iterations.  Approximate eigenvectors of $A$ are constructed
from the right eigenvectors ${\vec c}_k$ of $T$
as ${\vec \phi}_k=\sum_j c_{kj}{\vec q}_j$.
The process will fail if $\beta_j$ is zero for nonzero $\vec{r}_j$, which can
happen in principle, given the indefinite metric, but does not seem to 
happen in practice~\cite{Cullum}.

A useful extension of the Lanczos algorithm is a method for the
estimation of densities of states without first computing the
complete spectrum~\cite{FiniteTemp}.  For two-dimensional theories, the density 
can be written as the following trace over the
evolution operator $e^{-i\Pminus x^+}$:
\be
\rho(M^2)=\frac{1}{4\pi P^+}\int_{-\infty}^\infty e^{iM^2x^+/2P^+}
           {\rm Tr}e^{-iP^-x^+} dx^+.
\ee
The trace can be approximated by an average over a
random sample of vectors~\cite{AlbenHams}
\be
\rho(M^2)\simeq\frac1S\sum_{s=1}^S \rho_s(M^2),
\ee
with $\rho_s$ a local density for a single vector $|s\rangle$,
defined by
\be
\rho_s(M^2)=\frac{1}{4\pi P^+}\int_{-\infty}^\infty e^{iM^2x^+/2P^+}
           \langle s|e^{-i\Pminus x^+}|s\rangle dx^+.
\ee
The sample vectors $|s\rangle$ can be chosen as random 
phase vectors~\cite{Iitaka}; the coefficient
of each Fock state in the basis is a random number of 
modulus one.
  
The matrix element $\langle s|e^{-i\Pminus x^+}|s\rangle$
can be approximated by Lanczos iterations~\cite{JaklicAichhorn}.
Let $D$ be the length of $|s\rangle$, and define
$|u_1\rangle=\frac{1}{\sqrt{D}}|s\rangle$ as the
initial Lanczos vector. The matrix element
$\langle u_1|e^{-i\Pminus x^+}|u_1\rangle$ can be
approximated by the $(1,1)$ element of the 
exponentiation of the Lanczos tridiagonalization
of $\Pminus$.   

Let $P_s^-$ be the tridiagonal Lanczos matrix and $\vec{c}_j^{\,s}$
its eigenvectors, so that
\be
P_s^-\vec{c}_j^{\,s}=\frac{M_{sj}^2}{2P^+}\vec{c}_j^{\,s}.
\ee
The matrix can then be factorized as
$P_s^-=U\Lambda U^{-1}$, with $U_{ij}=(c_j^s)_i$
and $\Lambda_{ij}\equiv\delta_{ij}\frac{M_{sj}^2}{2P^+}$.  
The $(1,1)$ element is given by
\be
\left(e^{-iP_s^-x^+}\right)_{11}=\sum_j|(c_j^s)_1|^2e^{-iM_{sj}^2 x^+/2P^+}.
\ee

The local density can now be estimated by
\be
\rho_s(M^2)\simeq \sum_j w_{sj} \delta(M^2-M_{sj}^2),
\ee
where $w_{sj}\equiv D|(c_j^s)_1|^2$ is the weight of each Lanczos eigenvalue.
Only the extreme Lanczos eigenvalues are good approximations to
eigenvalues of the original $\Pminus$; however, the other Lanczos eigenvalues 
provide a smeared representation of the full spectrum.  
For construction of the full density of states, twenty sample
local densities can be sufficient; however, the number of 
Lanczos iterations needs to be on the order of 1000 per sample~\cite{FiniteTemp}.

\subsection{\it Function Expansions \label{sec:FnExpansion}}

\subsubsection{\it generic approach} \label{sec:generic}

To avoid the restriction to equally spaced quadrature points,
as imposed by momentum conservation, without limiting a
calculation to two-body equations, the Fock-state wave functions 
$\psi_n(x_i,\senk{k_i})$ can
be expanded in a set of basis functions $f_k^{(n)}(x_i,\senk{k_i})$
\be
\psi_n(x_i,\senk{k_i})=\sum_k c_k^{(n)} f_k^{(n)}(x_i,\senk{k_i}).
\ee
The overlap integrals
\be
B_{kl}^{(n)}\equiv \int \prod_i dx_i d^2k_{i\perp} \delta(1-\sum_i x_i)\delta(\sum_i\senk{k_i})
     f_k^{(n)*}(x_i,\senk{k_i})f_l^{(n)}(x_i,\senk{k_i})
\ee
and the matrix elements of the basis functions for the kinetic energy 
\be
T_{kl}^{(n)}\equiv \int \prod_i dx_i d^2k_{i\perp} \delta(1-\sum_i x_i)\delta(\sum_i\senk{k_i})
     f_k^{(n)*}(x_i,\senk{k_i})\left[\sum_i\frac{m^2+k_{i\perp}^2}{x_i}\right]f_l^{(n)}(x_i,\senk{k_i})
\ee
and the interaction\footnote{Here the interaction matrix element is written in a generic form.
In general, with the change in particle number, one basis function will depend on 
fewer momenta that are sums of individual momenta.} between Fock sectors $n$ and $m$
\be
V_{kl}^{(n,m)}\equiv \int \prod_i dx_i d^2k_{i\perp} \delta(1-\sum_i x_i)\delta(\sum_i\senk{k_i})
     f_k^{(n)*}(x_i,\senk{k_i})V(x_i,\senk{k_i})f_l^{(m)}(x_i,\senk{k_i}),
\ee
can then be computed
in various ways, perhaps analytically or at least numerically
with whatever quadrature is appropriate.  For an orthonormal
basis, the overlap integrals form just the identity matrix,
{\em i.e.} $B_{kl}^{(n)}=\delta_{kl}$;
though preferred, this may not be the most convenient choice.
A choice of basis where the kinetic energy matrix $T$ is diagonal
may also be possible and certainly useful.

If the basis is introduced for the mode expansion of the
quantum fields, this defines a new discretized quantization
that is discrete with respect to the sum over basis states.
There will then be creation and annihilation operators 
associated with each basis function.  DLCQ is of this
type, with periodic plane waves as the basis set.  The
two approaches can be combined, with function expansions
used for transverse momenta and DLCQ for the longitudinal
momenta.  An example of this is discussed in the next
subsection.

In general, given a complete set of orthonormal functions
$f_{nlm}(\ub{p})$, discrete creation operators can be
defined for neutral scalars by 
\be
a_{nlm}^\dagger=\int d\ub{p} f_{nlm}(\ub{p}) a^\dagger(\ub{p}).
\ee
The original creation operator is then expanded as
\be
a^\dagger(\ub{p})=\sum_{nlm}f_{nlm}^*(\ub{p})a_{nlm}^\dagger.
\ee
The nonzero commutator of the discrete operators is
\be
[a_{nlm},a_{n'l'm'}^\dagger]=\int d\ub{p} d\ub{p'} 
  f_{nlm}^*(\ub{p}) f_{n'l'm'}(\ub{p'})[a(\ub{p}),a^\dagger(\ub{p'})]
  =\int d\ub{p} f_{nlm}^*(\ub{p})f_{n'l'm'}(\ub{p})
  =\delta_{nn'}\delta_{ll'}\delta_{mm'},
\ee
which, given the assumed completeness of the basis, guarantees that
\be
[a(\ub{p}),a^\dagger(\ub{p'})]=\sum_{nlm} f_{nlm}^*(\ub{p'}) f_{nlm}(\ub{p})
   =\delta(\ub{p}-\ub{p'}).
\ee
The field operator is then simply
\be
\phi(x)=\sum_{nlm}\left[\tilde{f}_{nlm}(x)a_{nlm}+\tilde{f}_{nlm}^*(x)a_{nlm}^\dagger\right],
\ee
with 
\be
\tilde{f}_{nlm}(x)\equiv \int \frac{d\ub{p}}{\sqrt{16\pi^3 p^+}}e^{-ip\cdot x}f_{nlm}(\ub{p}).
\ee
The extension to other types of fields is straightforward.  The discrete expansions can then
be used to construct $\Pminus$ and Fock-state expansions in terms of the discrete operators,
which will lead to a discrete matrix representation for the eigenvalue problem.  However,
the longitudinal momentum ${\cal P}^+$ will no longer be diagonal; therefore, such 
discretizations are most useful when other quantum numbers, such as angular momentum, 
are of particular importance.

The key approximation made in the use of function expansions, besides 
truncations in Fock space, is a truncation of the basis set.  Convergence
as the basis set is increased must then be studied.
If the basis-set truncation provides the regularization
as well as the finiteness of the matrix representation,
then the convergence is more than just numerical
convergence and must include some form of renormalization.

The matrix representation of the eigenvalue problem will take the form
\be
\sum_l T_{kl}^{(n)}c_l^{(n)}+\sum_{m,l}V_{kl}^{(n,m)}c_l^{(m)}
    =\frac{M^2+P_\perp^2}{P^+}\sum_l B_{kl}^{(n)}c_l^{n}.
\ee
Obviously, this is a generalized eigenvalue problem, written
more compactly as $H\vec{c}=\lambda B\vec{c}$, which
can be solved in various ways.  Usually $B$ is factorized
in terms of lower and upper triangular matrices $L$ and $U$
and the problem converted to an ordinary one, 
$H'\vec{c}^{\,\prime}=\lambda \vec{c}^{\,\prime}$, 
with $B=LU$, $H'=L^{-1}HU^{-1}$, and $\vec{c}=U^{-1}\vec{c}^{\,\prime}$.
The upper and lower triangular matrices are easily inverted
implicitly through the solution of the associated linear
equations $U\vec x=\vec y$ and $L\vec x=\vec y$ by 
backward or forward substitution.  An alternative
factorization, which is more robust, follows from the
singular value decomposition $B=UDU^T$, where the columns of the 
unitary matrix $U$ are the eigenvectors of $B$ and 
$D$ is a diagonal matrix of the eigenvalues of $B$.
Now the definitions of $H'$ and $\vec{c}^{\,\prime}$ are
$H'=D^{-1/2}U^T HUD^{-1/2}$ and $\vec c^{\,\prime}=D^{1/2}U^T\vec c$.
There is also a Lanczos algorithm for the generalized 
eigenvalue problem that avoids the factorization~\cite{GenLanczos}.

\subsubsection{\it basis light-front quantization} \label{sec:BLFQ}

The basis light-front quantization (BLFQ)
approach~\cite{BLFQ,BLFQelectron,BLFQelectron2,BLFQpositronium} is a hybrid method
which uses discretization in the longitudinal direction combined
with products of single-particle basis functions in the transverse.
It is an adaptation of {\em ab initio} no-core methods developed
for problems in nuclear structure~\cite{no-core}.  The use of
single-particle functions sacrifices strict conservation of 
transverse momentum for the flexibility of easily formed products
that satisfy symmetries of the many-body wave functions.  The
desired transverse momentum eigenstates are then identified
by a Lagrangian multiplier method that shifts eigenstates with
excited center of mass motion to high energies, as is commonly 
done in nuclear many-body calculations~\cite{no-core}.

The transverse basis functions are two-dimensional oscillator
functions, as given in \cite{BLFQpositronium}
\be
\Psi_{nm}(\vec q_\perp)=\frac{1}{b}\sqrt{\frac{4\pi n!}{(n+|m|)!}}
    e^{im\phi}\rho^{|m|}e^{-\rho^2}L_n^{|m|}(\rho^2),
\ee
with $\vec q_\perp=\vec p_\perp/\sqrt{x}$, $\rho=|\vec q_\perp|/b$, 
$\phi=\tan^{-1}(q_y/q_x)$, and $b=\sqrt{P^+\Omega}$.  Here $xP^+$ and
$\vec p_\perp$ are the longitudinal and transverse momenta of the 
individual particle, the $L_n^m$ are associated Laguerre polynomials,
and $\Omega$ is the oscillator angular frequency.  The single
particle energies are $E_{nm}=(2n+|m|+1)\Omega$, with $n$ and $m$
the radial and azimuthal quantum numbers.  The basis is truncated
by limiting the total of single-particle energies with the 
constraint
\be
\sum_i (2n_i+|m_i|+1)\leq N_{\rm max}
\ee
Convergence in the $N_{\rm max}\rightarrow\infty$ limit is then to
be studied.

The particular choice of coordinates is made in order that
the wave functions factorize into center-of-momentum and 
internal components~\cite{BLFQcoordinates}.  The c.m.\ motion
is removed from the lower part of the spectrum by a
Lagrange-multiplier technique~\cite{LagrangeMultiplier}.

The basis states are combined to form an eigenstate of the total
angular momentum projection $M_J$.  This is guaranteed by forming
only products for which $M_J=\sum_i(m_i+s_i)$, where $s_i$ is
the fermion spin projection.  However, the product states are
not eigenstates of the total angular momentum $J$; diagonalization
of the Hamiltonian yields states with a range of total $J\geq M_J$.

The choice of oscillator basis functions is convenient for several
reasons.  It is the natural basis for states trapped in cavities
maintained by magnetic fields.  The coordinate-space wave functions
can be obtained exactly by Fourier transform and take the same
functional form.  Matrix elements of the Hamiltonian are well
converged in the ultraviolet; thus the regularization 
is provided by the basis functions and the truncation of the 
basis set, rather than by use of a PV regularization or a cutoff.
Perhaps most important, transverse
oscillator functions have a close connection to the successful
AdS/QCD-based quark models~\cite{AdSQCD}.

Of course, the convergence of matrix elements does not 
guarantee finiteness in the continuum limit.  As the number
of basis functions is taken to infinity, there can and will
be divergences in general.  For example, in QED the wave functions
of the dressed electron are known to fall off too slowly to be
normalizable~\cite{OnePhotonQED}, but in the BLFQ approach
the approximate wave functions are normalizable; the normalization
becomes infinite only in the limit of infinite $N_{\rm max}$.
This then requires some care in the regularization of the 
original Hamiltonian and in the process of taking the continuum
limit.  For QCD, where quarks are confined, this is less of
a concern, and harmonic oscillator functions have a long 
history of utility.

The BLFQ method has been extended to include time-dependent
processes~\cite{BLFQtime}.  The light-front time evolution of
a state is determined by
\be
i\frac{\partial}{\partial x^+}|\psi(x^+)\rangle=\frac12\Pminus(x^+)|\psi(x^+)\rangle.
\ee
The light-front Hamiltonian is split as $\Pminus=\Pminus_0+V$, to isolate the interaction
of interest $V$ (perhaps with an external field).  In the interaction
picture, the time evolution is then
\be
i\frac{\partial}{\partial x^+}|\psi(x^+)\rangle_I=\frac12 V_I(x^+)|\psi(x^+)\rangle_I,
\ee
with the formal solution
\be
|\psi(x^+)\rangle_I={\cal T}_+e^{-\frac{i}{2}\int_0^{x^+} dx^{\prime+}V_I(x^{\prime+})}|\psi(0)\rangle_I.
\ee
Here ${\cal T}_+$ is the light-front time-ordering operator.
The initial state is expanded in terms of eigenstates $|n\rangle$ of $\Pminus_0$
with coefficients chosen to match the particular physical situation.  The
eigenstates are approximated by time-independent BLFQ in a truncated Fock space,
and the time evolution is approximated by a sequence of discrete time steps
$x_i^+=i\delta x^+$ as
\be
{\cal T}_+e^{-\frac{i}{2}\int_0^{x^+} dx^{\prime+}V_I(x^{\prime+})}\simeq
\prod_{i=1}^n \left[1-\frac{i}{2}V_I(x_i^+)\delta x^+\right].
\ee
This approach has been applied to photon emission from an electron
in a background laser field~\cite{BLFQtime}
and should be useful for the analysis of particle production in the
chromodynamic fields of high-energy heavy-ion collisions.

\subsubsection{\it symmetric multivariate polynomials} \label{sec:sympolys}

For two-dimensional theories, there exists a basis of
symmetric multivariate polynomials~\cite{GenSymPolys} $P_{ki}^{(n)}(x_i)$.
The subscript $k$ is the order, and $i$ differentiates the various 
possibilities at that order.  They are 
fully symmetric with respect to interchange of the $n$ momenta $x_i$
and yet respect the momentum conservation constraint $\sum_i x_i=1$.
For $n=2$ constituents there is only one possibility
at each order, but for $n>2$ there can be more than one.
For example, for three constituents there are two sixth-order
polynomials, $P^{(3)}_{61}=(x_1x_2x_3)^2$ and $P^{(3)}_{62}=(x_1x_2+x_1x_3+x_2x_3)^3$.
If not for the momentum-conservation constraint, there would of
course be even more possibilities.  For example,
$P^{(3)}_{2}=x_1x_2+x_1x_3+x_2x_3$ is equivalent to $x_1^2+x_2^2+x_3^2$, 
up to a constant, when $x_3$ is replaced by $1-x_1-x_2$.

The linearly independent symmetric polynomials can be written as
products of powers of simpler polynomials.
Define $C_m(x_i)$ as a multivariate polynomial of order $m$
that is a sum of 
simple monomials $\prod_j^n x_j^{m_j}$, where $m_j$ is 0 or 1,
$\sum_j^n m_j=m$, and the sum over the monomials ranges over all
possible choices for the $m_j$, making each $C_m$ fully symmetric.
As examples of the $C_m$, consider the general case of 
$n$ longitudinal momentum variables.  Then $C_2$ is 
just $\sum_j^n \left(x_j\sum_{k>j}^n x_k\right)$,
$C_{n-1}$ is $\sum_j^n \prod_{k\neq j} x_k$, and
$C_n$ is $\prod_k^n x_k$.  In particular, for n=3,
$C_2=x_1x_2+x_1x_3+x_2x_3$ and $C_3=x_1x_2x_3$.

The full polynomial of order $k$ is then built as
\be
P_{ki}^{(n)}=C_2^{i_2} C_3^{i_3}\cdots C_n^{i_n},
\ee
with $i=(i_2,i_3,\ldots,i_n)$, as restricted by $k=\sum_j j i_j$.
Thus, each way of decomposing $k$ into a sum of $n$ integers $i_j>1$
provides a different polynomial of the order $k$.  That this
captures all such linearly independent polynomials can be
shown by a simple counting argument~\cite{GenSymPolys}.
The absence of a first-order polynomial is a direct
consequence of momentum conservation, since the linear
fully symmetric multivariate polynomial is $\sum_i x_i=1$.

The polynomials in this form are not orthonormal.  A
Gram--Schmidt process or a factorization of the overlap
matrix will produce the orthonormal combinations~\cite{GenSymPolys}.
However, if these combinations cannot be computed in
exact arithmetic, round-off errors can spoil the
orthogonality and make computation of matrix elements
actually less reliable when large orders are reached.

For antisymmetric polynomials, potentially useful for fermion 
wave functions, there is no known closed form.  However,
the constraints of momentum conservation and antisymmetry
can be used to determine a set of linear equations for
the polynomial coefficients~\cite{GenSymPolys}.  These
would allow construction of the polynomials order by
order.

\subsection{\it Regularization \label{sec:Regularization}}

\subsubsection{\it general considerations \label{sec:general}}

For all but two-dimensional theories, there are infinities
in the integrals of the equations for the wave functions.
These need to be regulated in some controlled way, such
that when the regulators are removed, the theory is
predictive.  In other words, the regularization must
provide for renormalizability.  For nonperturbative
calculations, renormalization is done by fixing
bare parameters with fits to data.  For model theories,
the `data' would be values of observables that would
have a physical interpretation for a real-world theory.
Such an observable might be a mass, a mass ratio,
an average radius, or a magnetic moment.

Obviously, the dependence on the regulator needs to
disappear, and regularizations with a strong dependence
require some form of adjustment, which is typically
the addition of counterterms to the Hamiltonian.
These terms are removed as the regulator is removed
but cancel the worst of the regulator dependence
before it is removed.  As is known from perturbation
theory, a regularization that breaks some symmetry
of the theory is likely to induce a strong dependence
on the regulator, and counterterms should be chosen
to restore the symmetry.  A nonperturbative example
of this can be found in the restoration of chiral
symmetry in QED~\cite{ChiralLimit,TwoPhotonQED}.

A distinction needs to be made between cutoffs that are
made for a numerical calculation, to make the calculation
finite in size, and cutoffs that are made for a regularization.
If the numerical approximation is made to a finite theory,
one that is already regulated, then the removal of
any (additional) cutoff made for numerical reasons
is strictly a matter for numerical convergence; any
renormalization is done after the numerical cutoff is
removed and numerical convergence has been achieved.
If, however, the numerical cutoff is also the regulator,
then investigation of numerical convergence must be
combined with the renormalization.  Given the additional
complications, a preferred approach is to apply numerical
approximations to an already regulated theory.

The simplest ultra-violet (UV) regulator is a transverse
momentum cutoff.  However, this breaks Lorentz invariance
as well as gauge invariance.  A cutoff on the invariant
mass of the Fock state, $\sum_i (m_i^2+k_{i\perp}^2)/x_i$,
is better, but still not ideal.  Of course, dimensional
regularization~\cite{dimreg} has been a well-received method,
particularly because Lorentz and gauge symmetries are
preserved, but it is tied to modifications of integrals for which
the integrand is known (as they arise in perturbation
theory); here the integrands involve unknown wave functions
and unknown pole structures.  

A much more workable method, which also preserves Lorentz
and gauge symmetries,
is Pauli--Villars (PV) regularization~\cite{PauliVillars},
though it is not used in the way it frequently is in perturbation
theory.  Instead of modifying loops in the individual
integrals associated with each Feynman diagram, the
heavy PV fields are added to the Lagrangian.  This is
equivalent to the modification of propagators in individual integrals,
because the additional terms in the Lagrangian generate
diagrams where each field is replaced by its PV 
partners.\footnote{This is equivalent to higher covariant derivatives
in the kinetic energy~\protect\cite{higherderiv}.}
All that is needed is for at least some of the PV partners
to have a negative metric, so that the contraction associated
with a line in a diagram will have the opposite sign and
cause a subtraction between diagrams.  The required number of 
PV fields and their metrics are determined by the number
of subtractions needed and any need for symmetry 
restoration~\cite{ChiralLimit}.\footnote{Supersymmetry 
provides this kind of regularization quite naturally
and has been exploited in SDLCQ calculations~\protect\cite{SDLCQ}; 
however, the superpartners are of the same mass and cannot correspond
to known physics until the supersymmetry is broken, a nontrivial task.}
Details of PV regularization for Abelian and non-Abelian gauge
theories are given in the next two subsections.

\subsubsection{\it quantum electrodynamics \label{sec:PVQED}}

To see how this form of PV regularization works, consider QED.  The 
basic Lagrangian is
\be
{\cal L}= -\frac14 F^{\mu \nu} F_{\mu \nu} 
      \bar{\psi} (i \gamma^\mu \partial_\mu - m) \psi
  - e \bar{\psi}\gamma^\mu \psi A_\mu .
\ee
The nondynamical part of the fermion field must satisfy the
constraint equation
\be
i\partial_-\psi_-+e A_-\psi_-
  =(i\gamma^0\gamma^\perp)
     \left[\partial_\perp \psi_+-ie A_\perp\psi_+\right]  - m \gamma^0\psi_+ .
\ee
Unlike the free case, $\psi_-$ is coupled to the photon field.  The 
coupling to $A_\perp$ induces instantaneous-fermion interactions in
the light-front Hamiltonian~\cite{LFreview1}, where a fermion is
coupled to two photons with an intermediate `instantaneous' fermion
in between the photon couplings.  The presence of $A_-=A^+$ makes 
explicit inversion impossible; hence, the nominal choice of light-cone
gauge, where $A_-=0$.  In light-cone gauge, the $A^-$ component is
also nondynamical, and the solution of its constraint equation
generates instantaneous-photon interactions where fermions
exchange an `instantaneous' photon~\cite{LFreview1}.

A PV-regulated QED Lagrangian takes the form
\bea  \label{eq:Lagrangian}
{\cal L} &=&  \sum_k r_k \left[-\frac14 F_k^{\mu \nu} F_{k,\mu \nu} 
         +\frac12 \mu_k^2 A_k^\mu A_{k\mu} 
         -\frac12 \zeta \left(\partial^\mu A_{k\mu}\right)^2\right] \\
&& + \sum_i s_i \bar{\psi_i} (i \gamma^\mu \partial_\mu - m_i) \psi_i 
  - e \sum_{ijk} \beta_i\bar\psi_i\gamma^\mu \beta_j \psi_j \xi_k A_{k\mu} , \nonumber
\eea
with $F_{k,\mu \nu} = \partial_\mu A_{k\nu}-\partial_\nu A_{k\mu}$.
The PV indices $i$, $j$, and $k$ each take the value of zero for a physical
field.  The metric signatures of the PV photons and PV fermions are $r_k$
and $s_i$, which are equal to $\pm1$.  The photon mass is $\mu_k$, with $\mu_0$ as an
infrared regulator to be taken to zero.  A gauge-fixing term has been included,
with gauge parameter $\zeta$.  The interaction term involves the following
combinations:
\be \label{eq:NullFields}
  \psi =  \sum_i \beta_i\psi_i, \;\;
  A_\mu  = \sum_k \xi_k A_{k\mu},
\ee
which include the coupling coefficients $\beta_i$ and $\xi_k$, 
to be chosen to enforce the necessary subtractions.
To keep $e$ as the charge
of the physical fermion, the physical coefficients and
metric signatures are set to unity, $\beta_0=1$, $\xi_0=1$, $s_0=1$, $r_0=1$.

The meaning of the metric is in the (anti)commutation relations
for the creation and annihilation operators for the photon
and fermion fields.  The nonzero relations become
\be  \label{eq:commutationrelations}
[a_{k\lambda}(\ub{k}),a_{k'\lambda'}^\dagger(\ub{k}')]
     =r_k\delta_{kk'}\epsilon^\lambda \delta_{\lambda\lambda'}\delta(\ub{k}-\ub{k}')
\ee
and
\be
\{b_{is}(\ub{k}),b_{i's'}^\dagger(\ub{k}')\}
   =s_i\delta_{ii'}\delta_{ss'}\delta(\ub{k}-\ub{k}'), \;\;
\{d_{is}(\ub{k}),d_{i's'}^\dagger(\ub{k}')\}
   =s_i\delta_{ii'}\delta_{ss'}\delta(\ub{k}-\ub{k}').
\ee
The factors of $r_k$ and $s_i$ carry the metric.

The interactions between the field combinations $\psi$ and $A_\mu$, defined
in (\ref{eq:NullFields}), are what provide
the PV subtractions that regulate any loop.  For a loop with 
one photon contraction and one fermion contraction, the interaction
vertex implies that the contraction of the $k$th photon 
field yields the metric signature $r_k$ and contraction of the $i$th 
fermion field, $s_i$.  The coupling coefficients from the vertices 
are $\xi_k$ and $\beta_i$.  The loop contribution then contains the
factors $r_k \xi_k^2$ and $s_i\beta_i^2$.  By imposing the constraints
\be
\sum_k r_k\xi_k^2=0, \;\;
\sum_i s_i\beta_i^2=0,
\ee
the loop contribution, when summed over $k$ and $i$, will contain
two subtractions.  This extends to more complicated loops with
overlapping divergences, because each internal line is associated
with a subtraction.  

These constraints make the combined fields $A_\mu$ and $\psi$ null.
The creation operators for the combined fields are 
$a_\lambda^\dagger(\ub{k})\equiv\sum_k \xi_k a_{k\lambda}(\ub{k})$,
$b_s^\dagger(\ub{p})\equiv\sum_i \beta_i b_{is}(\ub{p})$,
and $d_s^\dagger(\ub{p})\equiv\sum_i \beta_i d_{is}(\ub{p})$.
They commute with the annihilation operators, hence the designation
as null.  More generally, if there are two types of
vertices of the same general form but different coupling coefficients
$\xi_k$, $\xi'_k$, $\beta_i$, and $\beta'_i$, the loop contribution 
then contains the factors $r_k \xi_k \xi'_k$ and $s_i \beta_i \beta'_i$,
and the two subtractions are attained if the field combinations in
the two vertices are mutually null, in the sense that
\be
\sum_k r_k\xi'_k \xi_k=0, \;\;
\sum_i s_i\beta'_i \beta_i=0.
\ee

If there is more than one PV field, the associated coupling coefficient
can be chosen by some additional constraint, perhaps to restore a
symmetry.  For example, restoration of chiral symmetry in the
limit of zero fermion mass requires a second PV photon~\cite{ChiralLimit}
and restoration of a zero mass for the photon eigenstate requires
a second PV fermion~\cite{VacPol}.
  
Given the interaction between the fermion and vector fields, the 
constraint equation for the nondynamical components of the fermion
field is coupled to the vector field.  The constraint is
\be \label{eq:FermionConstraint}
is_i\partial_-\psi_{i-}+e A_-\beta_i\sum_j\beta_j\psi_{j-} 
  =(i\gamma^0\gamma^\perp)
     \left[s_i\partial_\perp \psi_{i+}-ie A_\perp\beta_i\sum_j\beta_j\psi_{j+}\right] 
      -s_i m_i \gamma^0\psi_{i+} . \nonumber
\ee
As discussed above, light-cone gauge ($A^+=A_-=0$) is ordinarily chosen, to make the
constraint explicitly invertible.  However, the interaction Lagrangian has
been arranged in just such a way that the $A$-dependent terms
can be canceled between the constraints for individual fields~\cite{OnePhotonQED}.
Multiplication by $(-1)^i\sqrt{\beta_i}$ and a sum over $i$ yields
\be
i\partial_-\psi_-
  =(i\gamma^0\gamma^\perp)
     \partial_\perp \psi_+
      - \gamma^0\sum_i\beta_im_i\psi_{i+},
\ee
as the constraint for the null fermion field that appears
in the interaction Lagrangian.  This constraint is the
same as the free-fermion constraint, in any gauge, and
the interaction Hamiltonian can be constructed from the
free-field solution.\footnote{The analogous cancellation occurs
in Yukawa theory~\protect\cite{YukawaOneBoson}, where the 
individual fermion constraint equations contain
couplings to the scalar field that cancel for the
null fermion field.}

Without this cancellation of $A$-dependent terms, the 
constraint would generate the four-point interactions
between fermion and photon fields, the 
instantaneous-fermion interactions~\cite{LFreview1}
discussed above.
The addition of the PV-fermion fields has, in effect,
factorized these interactions into type-changing
photon emission and absorption three-point vertices.
The instantaneous interactions are recovered in the
limit of infinite PV fermion masses, because in the 
contraction of two three-point vertices the light-front
energy denominator with an intermediate PV fermion cancels the PV-mass
factors in the emission and absorption vertices and
the contraction survives the infinite-PV-mass limit.
The absence of instantaneous fermion and instantaneous photon contributions is
important for numerical calculations, where such four-point interactions can
greatly increase the computational load and matrix storage requirements;
this is partial compensation for the increase in basis size brought by the
PV fields.

The PV-regulated light-front QED Hamiltonian is then~\cite{OnePhotonQED,VacPol}
\bea \label{eq:QEDP-}
{\cal P}^-&=&
   \sum_{i,s}s_i\int d\ub{p}
      \frac{m_i^2+p_\perp^2}{p^+}
          b_{i,s}^\dagger(\ub{p}) b_{i,s}(\ub{p}) 
    +\sum_{i,s}s_i\int d\ub{p}
      \frac{m_i^2+p_\perp^2}{p^+}
          d_{i,s}^\dagger(\ub{p}) d_{i,s}(\ub{p}) \\
   && +\sum_{k,\mu}r_k\int d\ub{k}
          \frac{\mu_l^2+k_\perp^2}{k^+}\epsilon^\mu
             a_{k\mu}^\dagger(\ub{k}) a_{k\mu}(\ub{k})
          \nonumber \\
   && +\sum_{i,j,k,s,\mu}\beta_i\beta_j\xi_k \int d\ub{p} d\ub{q}\left\{
      b_{i,s}^\dagger(\ub{p}) \left[ b_{j,s}(\ub{q})
       V^\mu_{ij,2s}(\ub{p},\ub{q})\right.\right.\nonumber \\
      &&\left.\left.\rule{1.75in}{0in}
+ b_{j,-s}(\ub{q})
      U^\mu_{ij,-2s}(\ub{p},\ub{q})\right] 
            a_{k\mu}^\dagger(\ub{q}-\ub{p})  \right. \nonumber \\
&& +     b_{i,s}^\dagger(\ub{p}) \left[ d_{j,s}^\dagger(\ub{q})
       \bar V^\mu_{ij,2s}(\ub{p},\ub{q}) 
+ d_{j,-s}^\dagger(\ub{q})
      \bar U^\mu_{ij,-2s}(\ub{p},\ub{q})\right] 
            a_{k\mu}(\ub{q}+\ub{p})
                      \nonumber \\
&& -   \left.   d_{i,s}^\dagger(\ub{p}) \left[ d_{j,s}(\ub{q})
       \tilde V^\mu_{ij,2s}(\ub{p},\ub{q}) 
+ d_{j,-s}(\ub{q})
      \tilde U^\mu_{ij,-2s}(\ub{p},\ub{q})\right] 
            a_{k\mu}^\dagger(\ub{q}-\ub{p})
                    + H.c.\right\}.  \nonumber
\eea
The vertex functions $V$ and $U$ are those given in \cite{OnePhotonQED}:
\bea \label{eq:vertices}
    V^0_{ij\pm}(\ub{p},\ub{q}) &=& \frac{e_0}{\sqrt{16 \pi^3 }}
                   \frac{ \vec{p}_\perp\cdot\vec{q}_\perp
                      \pm i\vec{p}_\perp\times\vec{q}_\perp
                       + m_i m_j + p^+q^+}{p^+q^+\sqrt{q^+-p^+}} , \\
    V^3_{ij\pm}(\ub{p},\ub{q}) &=& \frac{-e_0}{\sqrt{16 \pi^3}}
                        \frac{ \vec{p}_\perp\cdot\vec{q}_\perp
                      \pm i\vec{p}_\perp\times\vec{q}_\perp
                       + m_i m_j - p^+q^+ }{p^+q^+\sqrt{q^+-p^+}} , \nonumber\\
    V^1_{ij\pm}(\ub{p},\ub{q}) &=& \frac{e_0}{\sqrt{16 \pi^3}}
       \frac{ p^+(q^1\pm i q^2)+q^+(p^1\mp ip^2)}{p^+q^+\sqrt{q^+-p^+}} , \nonumber\\
    V^2_{ij\pm}(\ub{p},\ub{q}) &=& \frac{e_0}{\sqrt{16 \pi^3}}
       \frac{ p^+(q^2\mp i q^1)+q^+(p^2\pm ip^1)}{p^+q^+\sqrt{q^+-p^+}} , \nonumber\\
    U^0_{ij\pm}(\ub{p},\ub{q}) &=& \frac{\mp e_0}{\sqrt{16 \pi^3}}
       \frac{m_j(p^1\pm ip^2)-m_i(q^1\pm iq^2)}{p^+q^+\sqrt{q^+-p^+}} , \nonumber\\
    U^3_{ij\pm}(\ub{p},\ub{q}) &=& \frac{\pm e_0}{\sqrt{16 \pi^3}}
       \frac{m_j(p^1\pm ip^2)-m_i(q^1\pm iq^2)}{p^+q^+\sqrt{q^+-p^+}} , \nonumber\\
    U^1_{ij\pm}(\ub{p},\ub{q}) &=& \frac{\pm e_0}{\sqrt{16 \pi^3}}
                            \frac{m_iq^+-m_jp^+ }{p^+q^+\sqrt{q^+-p^+}} , \nonumber\\
    U^2_{ij\pm}(\ub{p},\ub{q}) &=& \frac{i e_0}{\sqrt{16 \pi^3}}
                     \frac{m_iq^+-m_jp^+ }{p^+q^+\sqrt{q^+-p^+}} . \nonumber
\eea
The other four vertex functions are related to these by~\cite{VacPol}
\bea
\label{eq:BarVertexFunctions}
\bar V_{ij,2s}^\mu(\ub{p},\ub{q})&=&\sqrt{\frac{q^+-p^+}{q^++p^+}}
    \left.V_{ij,2s}^\mu(\ub{p},\ub{q})\right|_{m_j\rightarrow -m_j}, \\
\bar U_{ij,2s}^\mu(\ub{p},\ub{q})&=&\sqrt{\frac{q^+-p^+}{q^++p^+}}
    \left.U_{ij,2s}^\mu(\ub{p},\ub{q})\right|_{m_j\rightarrow -m_j}, \nonumber \\
\tilde V_{ij,2s}^\mu(\ub{p},\ub{q})&=&\sqrt{\frac{p^+-q^+}{q^+-p^+}}
    \left.V_{ij,2s}^\mu(\ub{q},\ub{p})\right|_{m_j\rightarrow -m_j,\,m_i\rightarrow -m_i}, \\
\tilde U_{ij,2s}^\mu(\ub{p},\ub{q})&=&\sqrt{\frac{p^+-q^+}{q^+-p^+}}
    \left.U_{ij,2s}^\mu(\ub{q},\ub{p})\right|_{m_j\rightarrow -m_j,\,m_i\rightarrow -m_i}.
    \nonumber
\eea
The factors of $r_k$ and $s_i$ guarantee that the kinetic-energy terms 
have the correct signatures.  For example, when the number operator 
$b_{is}^\dagger(\ub{p})b_{is}(\ub{p})$ acts on a Fock state and 
contracts with $b_{js'}^\dagger(\ub{p'})$, the result is 
$s_j b_{js'}^\dagger(\ub{p'})\delta_{ij}\delta(\ub{p}-\ub{p'})$;
the leading factor of $s_j$ is canceled by the $s_i$ in the
kinetic energy term, yielding a positive kinetic-energy contribution
for the PV constituent, independent of whether $s_i$ is positive or
negative.

\subsubsection{\it non-Abelian gauge theories \label{sec:PVQCD}}

In order to do nonperturbative calculations with light-front Hamiltonian
methods in a non-Abelian gauge theory such as QCD, we must have a 
regularization for which renormalizability can be
shown.  Proofs of renormalizability for non-Abelian gauge theories\footnote{For
Abelian gauge theories, renormalizability can be shown even when a mass term
is added~\protect\cite{MassiveVector}.} typically
rely on BRST invariance~\cite{BRST}, which is the remnant of gauge invariance
after the gauge is fixed.  The underlying gauge invariance requires massless
gauge particles, unless some form of spontaneous symmetry breaking is invoked.
If massive Pauli--Villars particles are to be used as the regulators, then
their mass and the modification of interactions to include PV-index-changing 
currents, both break ordinary gauge invariance. 

However, this breaking can be resolved with two modifications~\cite{BRSTPVQCD}.  
One is a generalization of the ordinary gauge transformation
to include mixing of fields with different PV indices; this re-establishes the
gauge invariance for massless PV gluons and mass-degenerate PV quarks.  The
other is the introduction of masses for the PV particles
through the addition of interactions with auxiliary scalars, in a
non-Abelian extension of a method due to Stueckelberg~\cite{Stueckelberg2,Marnelius:1997rx}.
A particular gauge-fixing term is also part of the method, and the appropriate
Faddeev--Popov ghost terms~\cite{FaddeevPopov} can then be computed.
The ghost terms restore the BRST invariance.  These steps, while not specifically
light-front in character, yield a Lagrangian from which a light-front Hamiltonian 
can be constructed in an arbitrary covariant gauge~\cite{BRSTPVQCD}.

A PV-regulated Lagrangian for a non-Abelian gauge theory can be built 
from four terms
\be
{\cal L}={\cal L}_{\rm massless}+{\cal L}_{\rm gluon}
           +{\cal L}_{\rm quark}+{\cal L}_{\rm ghost}.
\ee
The first, ${\cal L}_{\rm massless}$, is a gauge-invariant Lagrangian
for massless gluons and quarks;
${\cal L}_{\rm gluon}$ is the mass and gauge-fixing term
for gluons and auxiliary scalars;
${\cal L}_{\rm quark}$ is the mass term for quarks; and
${\cal L}_{\rm ghost}$ is the Faddeev-Popov ghost term.
The starting point is the first term, given by 
\be
{\cal L}_{\rm massless}=-\frac14\sum_k r_k F_{ak}^{\mu\nu} F_{ak\mu\nu}
  +\sum_i s_i \bar\psi_i i\gamma^\mu\partial_\mu \psi_i  
   +g\sum_{ijk}\beta_i \beta_j \xi_k \bar\psi_i\gamma^\mu T_a A_{ak\mu}\psi_j,
\ee
where the field tensor is
\be
F_{ak}^{\mu\nu}=\partial^\mu A_{ak}^\nu-\partial^\nu A_{ak}^\mu
      -r_k \xi_k gf_{abc}\sum_{lm}\xi_l\xi_m A_{bl}^\mu A_{cm}^\nu.
\ee
The indices are $k$ for (PV) gluons and $i$, $j$ for (PV) quarks;
each takes the value of zero for a physical field.
The metric signatures are $r_k$ and $s_i$ are just $\pm1$
({\em e.g.} $(-1)^k$ and $(-1)^i$).
The extended gauge transformations are
\be
A_{ak}^\mu \longrightarrow  A_{ak}^\mu+\partial^\mu\Lambda_{ak}
+r_k\xi_k g f_{abc}\Lambda_b A_c^\mu,
\ee
\be
\psi_i \longrightarrow \psi_i+ig s_i\beta_i T_a\Lambda_a \psi,
\ee
with $\Lambda_a\equiv\sum_k\xi_k\Lambda_{ak}$ and $[T_a,T_b]=if_{abc}T_c$. 

The subtractions needed for regularization are provided by the 
null field combinations
\be
A_a^\mu\equiv \sum_k \xi_k A_{ak}^\mu, \;\;
\psi\equiv \sum_i \beta_i \psi_i,
\ee
with $\sum_k r_k \xi_k^2=0$ and $\sum_i s_i \beta_i^2=0$.
These make $A_a^\mu$ Abelian and $\psi$ gauge invariant.
With these definitions, this first term of the Lagrangian can be written as
\be \label{eq:Lbasenull}
{\cal L}_{\rm massless}=-\frac14\sum_k r_k(\partial^\mu A_{ak}^\nu-\partial^\nu A_{ak}^\mu)^2
                  +gf_{abc} \partial^\mu A_a^\nu A_{b\mu} A_{c\nu} 
                   +\sum_is_i\bar\psi_i i\gamma^\mu\partial_\mu\psi_i
                  +g\bar\psi\gamma^\mu T_a A_{a\mu}\psi.
\ee
The Lagrangian includes kinetic energy terms for fields with metrics $r_k$ and $s_i$,
and the interaction terms involve only null fields.

The four-gluon interaction is implicit in the infinite-PV-mass limit
through a contraction of two three-gluon interactions, with the contraction
being a PV gluon.  This is reminiscent of a method used to simplify
color factors in perturbation theory, by introducing an auxiliary field
to two three-gluon interactions~\cite{fourgluontrick}.

The gluon mass and gauge-fixing term is constructed from a gauge-invariant
piece and a gauge-fixing piece, each with couplings to an auxiliary set of
PV scalar fields $\phi_{ak}$, in a non-Abelian extension of a Stueckelberg 
mechanism~\cite{Stueckelberg,Stueckelberg2,Marnelius:1997rx}
\be
{\cal L}_{\rm gluon}=\frac12\sum_k r_k \left(\mu_k A_{ak}^\mu-\partial^\mu\phi_{ak}\right)^2 
   -\frac{\zeta}{2}\sum_k r_k \left(\partial_\mu A_{ak}^\mu +\frac{\mu_k}{\zeta}\phi_{ak}\right)^2.
\ee
The $\phi_{ak}$ obey the gauge transformation
\be
\phi_{ak} \longrightarrow \phi_{ak}+\mu_k \Lambda_{ak}
    +\mu_k r_k \xi_k g f_{abc} \int^x \!\!\!\! dx'_\mu \Lambda_b(x') A_{c}^\mu(x').
\ee
The line integral is needed in order to allow the derivative to transform as
\be
\partial^\mu\phi_{ak} \longrightarrow \partial^\mu\phi_{ak}+\mu_k \partial^\mu\Lambda_{ak}
    +\mu_k r_k \xi_k g f_{abc} \Lambda_b A_{c}^\mu.
\ee
When the two terms of this part of the Lagrangian are combined, the cross terms 
form a total derivative that can be neglected.  The remaining terms leave
\be \label{eq:Lgreduced}
{\cal L}_{\rm gluon}=\frac12\sum_k r_k \mu_k^2 \left(A_{ak}^\mu\right)^2
           -\frac{\zeta}{2}\sum_k r_k \left(\partial_\mu A_{ak}^\mu\right)^2 
   +\frac12\sum_k r_k\left[\left(\partial_\mu\phi_{ak}\right)^2
                                  -\frac{\mu_k^2}{\zeta}\phi_{ak}^2\right],
\ee
where the gluon acquires the mass $\mu_k$ and the scalar, a mass of
$\mu_k/\sqrt\zeta$.  The scalar field also inherits the metric $r_k$.

The quark mass term also involves a coupling to the auxiliary scalar:
\be
{\cal L}_{\rm quark}=-\sum_i s_i m_i (\bar\psi_i+ig\frac{s_i\beta_i}{\mu_{\rm PV}}\widetilde\phi_a \bar\psi T_a)
                 (\psi_i-ig\frac{s_i\beta_i}{\mu_{\rm PV}}\widetilde\phi_a T_a\psi),
\ee
with $\mu_{\rm PV}\equiv \max_k \mu_k$.  The combination
\be
\widetilde\phi_a\equiv\sum_k \xi_k\frac{\mu_{\rm PV}}{\mu_k}\phi_{ak}
\ee
must also be null, which is enforced by the additional constraint 
$\sum_k r_k \frac{\xi_k^2}{\mu_k^2}=0$.
The gauge transformation of the combination is Abelian:
\be
\widetilde\phi_a \longrightarrow \widetilde\phi_a+\mu_{\rm PV}\Lambda_a.
\ee

In order that all couplings be null, the combination
\be
\widetilde\psi=\sum_i \beta_i \frac{m_i}{m_{\rm PV}}\psi_i,
\ee
with $m_{\rm PV}\equiv\max_i m_i$, can be made null by imposition
of the constraint $\sum_i s_i m_i^2 \beta_i^2=0$ and
made mutually null with $\psi$ by imposition of 
$\sum_i s_i m_i \beta_i^2=0$.  This second constraint
also eliminates the quartic coupling term.  With these
definitions and constraints, the quark mass term becomes
\be \label{eq:Lq}
{\cal L}_{\rm quark}=-\sum_i s_i m_i \bar\psi_i\psi_i
            -ig \frac{m_{\rm PV}}{\mu_{\rm PV}}\left[\bar\psi T_a \widetilde\phi_a\widetilde\psi
                                             -\bar{\widetilde\psi} T_a \widetilde\phi_a\psi \right]
\ee

The ghost term~\cite{FaddeevPopov} is obtained from a standard 
construction~\cite{BailinLove}
\be \label{eq:LFP}
{\cal L}_{\rm ghost}=\sum_k r_k \partial_\mu \bar c_{ak}\partial^\mu c_{ak}
                  -\sum_k r_k \frac{\mu_k^2}{\zeta} \bar c_{ak} c_{ak}
            +g f_{abc}\left[\partial_\mu \bar c_a c_b A_c^\mu
                  -\frac{\mu_{\rm PV}^2}{\zeta}\bar{\widetilde c}_a
                       \int^x\!\!\!\! dx'_\mu c_b(x') A_c^\mu(x')\right],
\ee
for ghosts $c_{ak}$ and anti-ghosts $\bar c_{ak}$,
with null combinations defined as
\be
c_a\equiv\sum_k\xi_k c_{ak}, \;\;
\bar c_a\equiv\sum_k\xi_k \bar c_{ak}, \;\;
\bar{\widetilde c}_a\equiv\sum_k \xi_k\frac{\mu_k^2}{\mu_{\rm PV}^2}\bar c_{ak}.
\ee
For these to be (mutually) null, the additional constraints
$\sum_k r_k \mu_k^2 \xi_k^2=0$ and 
$\sum_k r_k \mu_k^4 \xi_k^2=0$ must be required.

To summarize the constraints, there are four for the adjoint fields: 
\be
\sum_k r_k \xi_k^2=0, \;\;
\sum_k r_k \frac{\xi_k^2}{\mu_k^2}=0, \;\;
\sum_k r_k \mu_k^2 \xi_k^2=0, \;\;
\sum_k r_k \mu_k^4 \xi_k^2=0,
\ee
and three for the quark fields: 
\be
\sum_i s_i \beta_i^2=0,  \;\;
\sum_i s_i m_i^2 \beta_i^2=0, \;\;
\sum_i s_i m_i \beta_i^2=0.
\ee
If the PV masses are to be chosen independently, the constraints
require four PV gluons, four PV ghosts and antighosts, five PV 
scalars, and three PV quarks.  For pure Yang--Mills theory, the 
number of PV scalars is reduced to four, because the $k=0$ fields 
can be dropped and the physical gluon mass $\mu_0$ set to zero.
The eliminated fields $\phi_{a0}$ are needed only to split the
masses of the PV quarks.  In either case, the number of PV fields
implies that the computational load will necessarily be large.

The BRST transformations are
\bea
\delta A_{ak}^\mu&=&\epsilon\partial^\mu c_{ak}
                      +\epsilon r_k\xi_k g f_{abc}c_b A_c^\mu, \\
\delta \psi_i &=& i\epsilon g s_i \beta_i T_a c_a \psi, \;\;
\delta \bar\psi_i = -i\epsilon g s_i \beta_i \bar\psi T_a c_a, \\
\delta \phi_{ak}&=&\epsilon\mu_k c_{ak} +\epsilon r_k\xi_k \mu_k g f_{abc}
                             \int^x\!\!\!\! dx'_\mu c_b(x') A_c^\mu(x'), \\
\delta \partial^\mu\phi_{ak}&=&\epsilon\mu_k \partial^\mu c_{ak}
                      +\epsilon r_k\xi_k \mu_k g f_{abc} c_b A_c^\mu, \\
\delta \bar c_{ak} & =& -\zeta\epsilon\left(\partial_\mu A_{ak}^\mu+\frac{\mu_k}{\zeta}\phi_{ak}\right), \;\;
\delta c_{ak} = \frac12 \epsilon r_k \xi_k g f_{abc}c_b c_c,
\eea
with $\epsilon$ a real Grassmann constant and $\epsilon^2=0$.
For the various null combinations, the transformations are
\be
\delta A_a^\mu = \epsilon\partial^\mu c_a, \;\;
\delta \widetilde\phi_a= \epsilon\mu_{\rm PV}c_a, \;\;
\delta c_a = 0, \;\;
\delta \psi =0, \;\;
\delta \widetilde\psi = 0.
\ee
The full Lagrangian is invariant with respect to these transformations.

The construction of the light-front Hamiltonian from this Lagrangian
follows the pattern laid out for QED.  Calculations would then be
done for a range of PV masses and the limit of infinite PV mass
studied.  This would all be done for a series of values for
the gauge-fixing parameter, in order to investigate the gauge-independence
of computed observables.

\subsection{\it Sector Dependent Renormalization \label{sec:SecDep}}

In the usual approach to renormalization of a quantum field
theory, one seeks to give meaning to the bare parameters of
the Lagrangian in relation to physical observables.  For
nonperturbative calculations in a truncated Fock space,
there can be some utility in allowing these bare parameters
to be Fock-sector dependent; this is known as sector-dependent
renormalization.  This was originally proposed by Perry, Harindranath,
and Wilson~\cite{SecDep-Wilson} and applied to QED
by Hiller and Brodsky~\cite{HillerBrodsky}.  More recent work
with this approach has been by Karmanov, Mathiot, and
Smirnov~\cite{Karmanov}.
 
\begin{figure}[tb]
\begin{center}
\epsfig{file=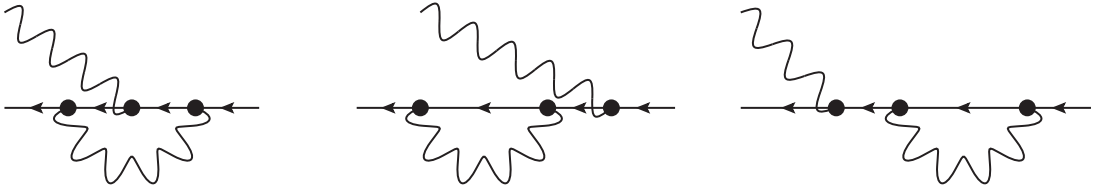,scale=0.75}
\caption{Contributions to the Ward identity in QED.  Only the
last survives a truncation to one photon and one fermion.
\label{fig:WardID}}
\end{center}
\end{figure}

The simplest way to motivate sector-dependent bare parameters
is to consider what happens
to self-energy contributions as the edge of the truncation
is approached.  Clearly, the contributions are reduced as
more and more potential intermediate states are forbidden
by the truncation.  In particular, for constituents in
the highest Fock sector, there are no self-energy 
contributions, which suggests that each bare mass
should be equal to the physical mass.  Similarly,
in a transition between the highest Fock sector and
a sector just below, with one less constituent, there
can be no loop corrections and the only self-energy 
corrections are on the side of the lower sector; this
is illustrated in Fig.~\ref{fig:WardID}.  This would
suggest that the bare coupling associated with the
transition should be renormalized differently than
the bare coupling for a transition of the same
type between lower sectors, where loop corrections
and different self-energy corrections are possible.
The calculation of the sector-dependent bare parameters
can be done systematically~\cite{HillerBrodsky,Karmanov} by 
an iterative procedure, beginning with the most severe
truncation and working up to the desired truncation;
at each step, the bare parameters of all but the lowest
sector are held fixed at the values obtained in the
previous iteration.

If sector-dependent renormalization is not done, there
are uncanceled divergences that arise because the
truncation has eliminated contributions that would
otherwise cancel against contributions that are kept.
The distinction between which are kept and which not
is the number of constituents required by intermediate states.
This prevents complete removal of the regulators and
requires a strategy of seeking a range of regulation
parameters ({\em e.g.}, the PV masses) within which
the physical observables are insensitive to the
regularization~\cite{OnePhotonQED}.  This minimizes the
net effect of two sources of error: the truncation
itself, which creates the sensitivity to the regulators,
and the presence of PV fields with finite masses.

Sector-dependent bare parameters are chosen to absorb the
uncanceled divergences.  However, this leads to some unexpected
consequences.  In QED, there can be coupling
renormalization without vacuum polarization; normally
the Ward identity prevents this, but a Fock-space
truncation violates the Ward identity, as illustrated
by Fig.~\ref{fig:WardID}.  There are also difficulties
in removal of regulators, in that Fock-sector probabilities
move outside the range of zero to one and some bare
couplings can become imaginary.  For a consistent theory,
the regulators may not be completely removed, just as
for the standard approach to renormalization; there
may be no uncanceled divergences but Fock-sector
probabilities and couplings may take on unphysical values.
Any calculation that uses the sector-dependent approach
must check for this behavior.

As discussed below, the difficulties are associated 
with the sector-dependent coupling renormalization.  This suggests
a middle ground where only bare masses are sector dependent.
Such an arrangement can be useful in avoiding the 
difficulties of sector-dependent renormalization 
and yet making the invariant mass of higher Fock states
more physically reasonable.  The scale of the
invariant mass for the higher Fock states can be
important because Fock states with an invariant mass
much larger than the eigenstate mass will not make
a significant contribution to the eigenstate.  In
calculations where the eigenstate/physical mass is much smaller
than the bare mass and where higher Fock states should
be included, sector-dependent masses will allow a
better approximation.

Of course, in the limit that the Fock-space truncation is
removed, both the standard renormalization and the
sector-dependent renormalization should yield the same
result.  For the standard case, the limit simply removes
the uncanceled divergences.  For the sector-dependent
case, the bare parameters become sector independent.

For a detailed comparison of the two renormalization methods,
consider the dressed-electron state in QED with a truncation
to at most one photon~\cite{SecDep}.  The Fock-state expansion is
\be \label{eq:FockExpansion}
|\psi^\pm(\ub{P})\rangle=\sum_i z_i b_{i\pm}^\dagger(\ub{P})|0\rangle
  +\sum_{ijs\mu}\int d\ub{k} C_{ijs}^{\mu\pm}(\ub{k})b_{is}^\dagger(\ub{P}-\ub{k})
                                       a_{j\mu}^\dagger(\ub{k})|0\rangle.
\ee
A second PV fermion flavor plays no role in this
sector, so that sums over the PV fermion indices can be limited
to 0 and 1, with $\beta_1=1$.
The coupled equations for the one-electron amplitude $z_i$
and the one-electron/one-photon wave function $C_{ijs}^{\mu\pm}(\ub{k})$
are, with $y=k^+/P^+$:
\be \label{eq:firstcoupledequation}
[M^2-m_i^2]z_i = \int (P^+)^2 dy d^2k_\perp 
     \sum_{j,l,\mu}\xi_l s_j r_l\epsilon^\mu 
  \left[V_{ji\pm}^{\mu*}(\ub{P}-\ub{k},\ub{P})C^{\mu\pm}_{jl\pm}(\ub{k}) 
        +U_{ji\pm}^{\mu*}(\ub{P}-\ub{k},\ub{P}) C^{\mu\pm}_{jl\mp}(\ub{k})\right] ,
\ee
and
\be \label{eq:TwoBodyEqns}
\left[M^2 - \frac{m_j^2 + k_\perp^2}{1-y} - \frac{\mu_l^2 + k_\perp^2}{y}\right]
  C^{\mu\pm}_{jl\pm}(\ub{k})
    =\xi_l\sum_{i'} s_{i'} z_{i'} P^+ V_{ji'\pm}^\mu(\ub{P}-\ub{k},\ub{P}),
\ee
\be \label{eq:TwoBodyEqns2}
\left[M^2 - \frac{m_j^2 + k_\perp^2}{1-y} - \frac{\mu_l^2 + k_\perp^2}{y}\right]
C^{\mu\pm}_{jl\mp}(\ub{k}) 
    = \xi_l\sum_{i'} s_{i'} z_{i'} P^+ U_{ji'\pm}^\mu(\ub{P}-\ub{k},\ub{P}).
\ee
The indexing is arranged such that an index of $i$ or $i'$ corresponds to the 
one-electron sector and $j$ to the one-electron/one-photon sector. 
For a sector-dependent approach, the mass $m_i$ is chosen to be the bare mass $m_0$, 
and $m_j$ the physical mass, $M=m_e$.  In the standard parameterization, all are bare masses.

The coupled equations can be solved analytically~\cite{OnePhotonQED}. 
The wave functions $C_{ils}^{\mu\pm}$ are
\be \label{eq:wavefn1}
C^{\mu\pm}_{il\pm}(\ub{k}) = \xi_l
  \frac{\sum_j s_j z_j P^+ V_{ij\pm}^\mu(\ub{P}-\ub{k},\ub{P})}
    {M^2 - \frac{m_i^2 + k_\perp^2}{1-y} - \frac{\mu_l^2 + k_\perp^2}{y}} , \;\;
%
C^{\mu\pm}_{il\mp}(\ub{k}) = \xi_l
\frac{\sum_j s_j z_j P^+ U_{ij\pm}^\mu(\ub{P}-\ub{k},\ub{P})}
     {M^2 - \frac{m_i^2 + k_\perp^2}{1-y} - \frac{\mu_l^2 + k_\perp^2}{y}} ,
\ee
and the one-electron amplitudes satisfy
\be \label{eq:FeynEigen}
(M^2-m_i^2)z_i =
      2e_0^2\sum_{i'} s_{i'}z_{i'}\left[\bar{J}+m_im_{i'} \bar{I}_0
  -2(m_i+m_{i'}) \bar{I}_1 \right],
\ee
with
\begin{eqnarray} \label{eq:In}
\bar{I}_n(M^2)&\equiv&\int\frac{dy dk_\perp^2}{16\pi^2}
   \sum_{jl}\frac{s_j r_l\xi_l^2}{M^2-\frac{m_j^2+k_\perp^2}{1-y}
                                   -\frac{\mu_l^2+k_\perp^2}{y}}
   \frac{m_j^n}{y(1-y)^n}\,, \\
\bar{J}(M^2)&\equiv&\int\frac{dy dk_\perp^2}{16\pi^2}  \label{eq:J}
   \sum_{jl}\frac{s_j r_l\xi_l^2}{M^2-\frac{m_j^2+k_\perp^2}{1-y}
                                   -\frac{\mu_l^2+k_\perp^2}{y}}
   \frac{m_j^2+k_\perp^2}{y(1-y)^2} .
\end{eqnarray}
These integrals satisfy the identity~\cite{ChiralLimit} $\bar{J}=M^2 \bar{I}_0$.
The coupling coefficient $\xi_2$ is fixed by requiring that
$M=0$ when $m_0=0$, to retain the chiral symmetry~\cite{ChiralLimit}. 
However, to further simplify the calculation, the second PV photon
will be dropped and $\xi_1=1$ for the first.

The analytic solution for the remaining PV amplitude is
\be
z_1=\frac{M \pm m_0}{M \pm m_1}z_0 ,
\ee
with the bare coupling $\alpha_0=e_0^2/4\pi$ restricted to two possibilities
\be \label{eq:OneBosonEigenvalueProb}
\alpha_{0\pm}=\frac{(M\pm m_0)(M\pm m_1)}{8\pi (m_1-m_0)(2 \bar{I}_1\pm M\bar{I}_0)}.
\ee
The lower sign corresponds to the physical answer, because $m_0$ then
becomes the physical mass $M=m_e$ at zero coupling.  The amplitude
$z_0$ is determined by the normalization.
 
In this truncation, the limit $m_1\rightarrow\infty$ can be taken, to
simplify the remaining calculation, leaving only $\mu_1$ as
the regulating mass.  In this limit, $z_1$ is zero, $m_1z_1\rightarrow\pm(M-m_0)z_0$,
and
\be  \label{eq:alpha0}
\alpha_{0\pm}=\pm\frac{M(M\pm m_0)}{8\pi (2 \bar{I}_1\pm M\bar{I}_0)}.
\ee
In the sector-dependent approach, $\bar I_1$ and $\bar I_0$ are independent
of $m_0$.  This allows the solution for $\alpha_0$ to be rearranged as
an explicit expression for $m_0$
\be
m_0=\mp M + 8\pi\frac{\alpha_{0\pm}}{M}(2 \bar{I}_1\pm M\bar{I}_0).
\ee

The anomalous magnetic moment can be computed in this
one-photon truncation as
\bea
a_e&=&m_e\sum_{s\mu}\int d\ub{k}\epsilon^\mu \sum_{j=0,2}\xi_j^2
  \left(\sum_{i'=0}^1\sum_{k'=j/2}^{j/2+1}
    \frac{s_{i'}r_{k'}}{\xi_{k'}}C_{i'k's}^{\mu+}(\ub{k})\right)^* \\
  && \times y\left(\frac{\partial}{\partial k_x}+i\frac{\partial}{\partial k_y}\right)
  \left(\sum_{i=0}^1\sum_{k=j/2}^{j/2+1}
    \frac{s_i r_k}{\xi_k}C_{iks}^{\mu-}(\ub{k})\right). \nonumber
\eea
In the limit where the PV electron mass $m_1$ is infinite, this reduces to
\be
a_e=\frac{\alpha_0}{\pi}m_e^2z_0^2\int y^2 (1-y) dy dk_\perp^2
  \left(\sum_{k=0}^1 \frac{r_k}{ym_0^2+(1-y)\mu_k^2+k_\perp^2-m_e^2y(1-y)}\right)^2 .
\ee
For the sector-dependent parameterization, the product $\alpha_0 z_0^2$ is equal to
the physical coupling $\alpha$,
and the bare mass $m_0$ in the denominator is replaced by the physical mass $m_e$.

Even though we do not include the vacuum polarization 
contribution to the dressed-electron state, the
sector-dependent bare coupling is not equal to the
physical coupling.  Instead, they are related by
$e_0=e/z_0$, where $z_0$ is the amplitude for the
bare-electron Fock state computed without projection onto
the physical subspace.  

In general, the bare coupling is given by $e_0=Z_1e/\sqrt{Z_{2i}Z_{2f}}Z_3$.
Here, however, there is no vacuum polarization and $Z_3=1$.  Also, there is
no vertex correction and $Z_1=1$.  This leaves the wave-function renormalization,
which has been split~\cite{OSUqed} between initial and final contributions 
$\sqrt{Z_{2i}}$ and $\sqrt{Z_{2f}}$; the split is due to the effects of truncation
which limit the contributions to $Z_2$ differently on opposite sides of
the photon interaction.  Figure~\ref{fig:WardID} shows how this works. In
the one-photon truncation, only the self-energy loop on one fermion leg
can contribute to $Z_2$; the loop on the other leg is eliminated by the
truncation. This leaves $\sqrt{Z_{2i}Z_{2f}}=z_0$ and $\alpha=\alpha_0/z_0$.

The normalization
$\langle\psi^{\sigma'}(\ub{P}')|\psi^\sigma(\ub{P})\rangle
                =\delta(\ub{P}'-\ub{P})\delta_{\sigma'\sigma}$ is
what determines the amplitude $z_0$.
In the sector-dependent approach, this reduces to
$1=z_0^2+e_0^2z_0^2\tilde J_2$, with
\be \label{eq:tildeJ2}
\tilde J_2\equiv\frac{1}{8\pi^2}\int y\, dy dk_\perp^2
  \sum_{k=0}^1 r_k
  \frac{(y^2+2y-2)m_e^2+k_\perp^2}
       {[k_\perp^2+(1-y)\mu_{k}^2+y^2m_e^2]^2}.
\ee
With the replacement of $e_0$ by $e/z_0$, $z_0$ can be obtained as
$z_0=\sqrt{1-e^2\tilde J_2}$, which implies $e_0=e/\sqrt{1-e^2\tilde J_2}$.
For large $\mu_1$, the integral $\tilde J_2$ behaves as
$\tilde J_2\simeq \frac{1}{8\pi^2}\left(\ln\frac{\mu_1\mu_0^2}{m_e^3}+\frac98\right)$.
Thus, $e_0$ can become imaginary and Fock-sector
probabilities, which are proportional to $|z_0|^2$, can range outside $[0,1]$
as $\mu_1\rightarrow\infty$ and $\mu_0\rightarrow0$.  Therefore, consistency
imposes limits on the UV regulator $\mu_1$ and on the infrared regulator $\mu_0$.

In the standard parameterization, the bare amplitude
is determined by $1=z_0^2+e^2z_0^2 J_2$, with
\bea \label{eq:J2}
J_2&=&\frac{1}{8\pi^2}\int y\, dy dk_\perp^2
   [m_0^2-4m_0m_e(1-y)+m_e^2(1-y)^2+k_\perp^2] \\
 && \times \left(\sum_{k=0}^1 r_k
  \frac{1}
       {[k_\perp^2+(1-y)\mu_{k}^2+ym_0^2-y(1-y)m_e^2]}\right)^2 .  \nonumber
\eea
Thus the bare amplitude is $z_0=1/\sqrt{1+e^2J_2}$,
and is driven to zero as $\mu_1\rightarrow\infty$.  This causes
most expectation values also to go to zero.  Therefore,
in this case there is a limit on $\mu_1$, but $\mu_0$ can be zero.

\begin{figure}[tb]
\begin{center}
\epsfig{file=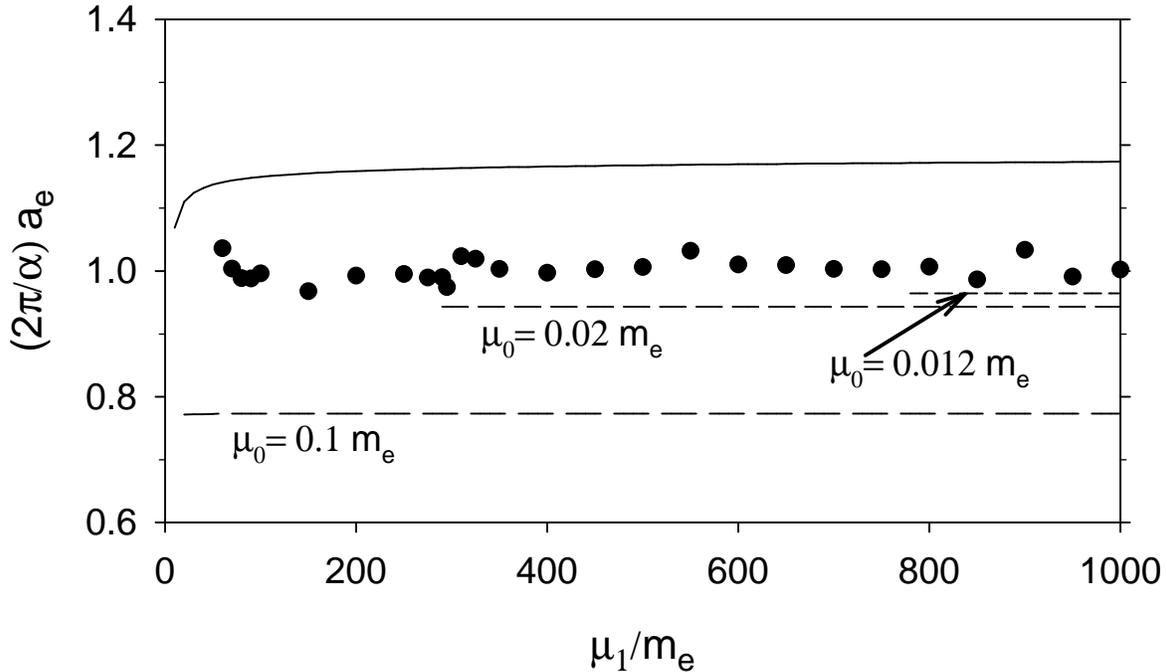,scale=1}
\caption{\label{fig:ae} The anomalous moment of the electron in
units of the Schwinger term ($\alpha/2\pi$)~\protect\cite{Schwinger} plotted versus
the PV photon mass, $\mu_1$, as shown in \protect\cite{SecDep}.  
The solid line is the 
standard-parameterization result for the one-electron/one-photon
truncation and the dashed lines are the results for
sector-dependent parameterization at three different values of
photon mass $\mu_0$; both use $m_1=\infty$.  The results for the
sector-dependent parameterization are plotted only for values
of $\mu_1$ for which the probability of the two-particle
sector remains between 0 and 1; for values of $\mu_0\leq0.01m_e$,
this requires $\mu_1>1000m_e$, which is beyond the range of
the plot.  For the case of the standard parameterization,
$\mu_0$ has its physical value of zero.
The filled circles are from a calculation with the standard parameterization
that includes the self-energy contribution from the one-electron/two-photon
sector~\protect\cite{SecDep}.  It also includes a second PV photon flavor, with its mass, 
$\mu_2$, set to $\sqrt{2}\mu_1$; the PV electron mass $m_1$ 
is equal to $2\cdot10^4\,m_e$. 
The variation is due to errors in numerical quadratures.}
\end{center}
\end{figure}

The anomalous moment in the sector-dependent case is
\be \label{eq:secdepae}
a_e=\frac{\alpha}{\pi}m_e^2\int y^2 (1-y) dy dk_\perp^2 \sum_{k=0}^1 r_k
    \left( \frac{1}{ym_e^2+(1-y)\mu_k^2+k_\perp^2-m_e^2y(1-y)}\right)^2
\ee
In the $\mu_1\rightarrow\infty$,
$\mu_0\rightarrow0$ limit, this becomes
exactly the Schwinger result~\cite{Schwinger}
\be \label{eq:Schwinger}
a_e=\frac{\alpha}{\pi}m_e^2\int
     \frac{ dy dq_\perp^2/(1-y)}{\left[\frac{m_e^2+q_\perp^2}{1-y}
     +\frac{q_\perp^2}{y}-m_e^2\right]^2} = \frac{\alpha}{2\pi}.
\ee
However, this limit cannot be taken without making the underlying
theory inconsistent, and the result for finite regulator masses
is quite different from the exact one-photon result, as can
be seen in Fig.~\ref{fig:ae}.

For both the standard and the sector-dependent parameterizations,
truncation of the Fock space results in uncanceled
divergences.  For the standard case, these divergences are 
explicit; for the sector-dependent case, they are found in
the renormalization of the coupling.  However, the sector-dependent
coupling renormalization is not the usual coupling renormalization;
instead of being driven by vacuum polarization, it is the result
of unbalanced wave-function renormalization and the breaking of
the Ward identity by the truncation.  In both parameterizations,
consistency requires a limit on one or more of the regulators,
and not all PV masses can be taken to infinity.  Any results
need to be extracted at finite PV masses.  For the
sector-dependent approach, this is complicated
by infrared divergences.

With so much difficulty caused by the Fock-space truncation,
there is a strong motivation to avoid the truncation all
together.  Unfortunately, a finite calculation in Fock space
requires a truncation of some kind.  However, this need not
be a truncation in particle number.  Other truncations are
possible, as discussed in the next section.

\subsection{\it Light-Front Coupled-Cluster Method \label{sec:LFCC}}

The LFCC method~\cite{LFCC} avoids the difficulties 
associated with an explicit Fock-space truncation
by truncating the set of coupled equations in a very different way.
Instead of truncating the number of particles, it truncates
the way in which wave functions are related to each
other; the wave functions of higher Fock states are
determined by the lower-state wave functions and the
exponentiation of an operator $T$.  Specifically,
the eigenstate is written in the form $\sqrt{Z}e^T|\phi\rangle$,
where $\sqrt{Z}$ is a normalization factor and
$|\phi\rangle$ is a state with the minimal number
of constituents.  The operator $T$ increases particle
number and conserves all relevant quantum numbers,
including the total light-front momentum.  This is in principle
exact but also still infinite, because $T$ can have an
infinite number of terms.  

The truncation made is a truncation of $T$.
The original eigenvalue problem becomes a finite-sized eigenvalue
problem for the valence state $|\phi\rangle$, combined
with auxiliary equations for the terms retained in $T$:
\be \label{eq:LFCCeqns}
P_v\ob{\Pminus}|\phi\rangle=\frac{M^2+P_\perp^2}{P^+}|\phi\rangle, \;\;\;\;
(1-P_v)\ob{\Pminus}|\phi\rangle=0.
\ee
Here $P_v$ is a projection onto the valence sector, and
$\ob{\Pminus}\equiv e^{-T}\Pminus e^T$ is the LFCC effective
Hamiltonian.  The projection $1-P_v$ is truncated to
provide just enough auxiliary equations to determine
the functions in the truncated $T$ operator.  The effective
Hamiltonian is computed from its Baker--Hausdorff expansion
$\ob{\Pminus}=\Pminus+[\Pminus,T]+\frac12[[\Pminus,T],T]+\cdots$,
which can be terminated at the point where more particles
are being created than are kept by the truncated
projection $1-P_v$.  The use of the exponential of $T$ rather
than some other function is convenient, not only because
of the Baker--Hausdorff expansion but more generally because
it is invertible; in principle, other functions could be used
and would also provide an exact representation until a
truncation is made.

The truncation of $T$ can be handled systematically.  Terms
can be classified by the number of annihilated constituents
and the net increase in particle number.  For example, in QCD
the lowest-order contributions annihilate one particle and increase
the total by one.  These are one-gluon emission from a quark,
quark pair creation from one gluon, and gluon pair creation
from one gluon.  Each involves a function of relative momentum
for the transition from one to two particles.  Higher order terms 
annihilate more particles and/or increase the total by
more than one.  These provide additional contributions to
higher-order wave functions and even to low-order wave
functions for more complicated valence states.  For example,
the wave function for the $|q\bar{q}g\rangle$ Fock state
of a meson can have a contribution from a term in
$T$ that annihilates a $q\bar{q}$ pair and creates a
pair plus a gluon, when this acts on the meson
valence state $|q\bar{q}\rangle$.

Zero modes can be included in the LFCC method~\cite{LFCCZeroModes}.
The vacuum must then also be computed as a nontrivial eigenstate,
which takes the form of a generalized coherent state of zero
modes~\cite{VaryHari-coherent,coherentstates}.
The zero modes are included by first introducing modes of 
infinitesimal longitudinal momentum and taking the limit of 
zero momentum at the end of the calculation.
Inclusion of four zero modes in the $T$ operator should
be sufficient for a calculation of the critical coupling
for dynamical symmetry breaking directly in terms of the
vacuum state.

The mathematics of the LFCC method has its origin in the
many-body coupled-cluster method~\cite{CCorigin} used in
nonrelativistic nuclear physics and quantum 
chemistry~\cite{CCreviews}.\footnote{Some applications to 
field theory of the coupled-cluster
method have been previously considered~\protect\cite{CC-QFT}, for
Fock-state expansions in equal-time quantization.  The 
focus was on the non-trivial vacuum structure, and
particle states are then built on the vacuum.  There was
some success in the analysis of $\phi^4_{1+1}$, particularly
of symmetry-breaking effects.}
The physics is, however, quite different.  The many-body
method works with a state of a large number
of particles and uses the exponentiation of $T$ to
build in correlations of excitations to higher 
single-particle states; the particle number does not change.
The LFCC method starts from a small number of constituents
in a valence state and uses $e^T$ to build states with
more particles; the method of solution of the valence-state
eigenvalue problem is left unspecified.

The computation of physical observables from matrix elements 
of operators requires some care.  Direct
computation would require an infinite sum over Fock space.
We instead borrow from the many-body coupled cluster method~\cite{CCreviews}
a construction that computes expectation values from 
right and left eigenstates\footnote{The effective Hamiltonian
is not Hermitian, because every term of $T$ must increase
the particle count and therefore $T$ itself cannot be Hermitian.}
of $\ob{\Pminus}$.  This can be extended to include
off-diagonal matrix elements and gauge projections~\cite{LFCCqed}.

Consider the expectation value for an operator $\hat{O}$:
\be
\langle\hat O\rangle=\frac{\langle\phi| e^{T^\dagger}\hat O e^T|\phi\rangle}
                      {\langle\phi| e^{T^\dagger} e^T|\phi\rangle}.
\ee
Define
$\ob{O}=e^{-T}\hat O e^T$ and 
$\langle\widetilde\psi|=\langle\phi|\frac{e^{T^\dagger}e^T}
      {\langle\phi|e^{T^\dagger} e^T|\phi\rangle}$.  The
expectation value can then be expressed as
$\langle\hat O\rangle=\langle\widetilde\psi|\ob{O}|\phi\rangle$, and
the dual vector $\langle\widetilde\psi|$ is normalized as
\be
\langle\widetilde\psi'|\phi\rangle
=\langle\phi'|\frac{e^{T^\dagger}e^T}{\langle\phi| e^{T^\dagger} e^T|\phi\rangle}|\phi\rangle
=\delta(\ub{P}'-\ub{P}).
\ee
The effective operator $\ob{O}$ can be computed from its
Baker--Hausdorff expansion,
$\ob{O}=\hat O + [\hat O,T]+\frac12[[\hat O,T],T]+\cdots$.
The dual vector $\langle\widetilde\psi|$ is a left eigenvector of $\ob{\Pminus}$,
as can be seen from
\be
\langle\widetilde\psi|\ob{\Pminus}
=\langle\phi|\frac{e^{T^\dagger}\Pminus e^T}{\langle\phi| e^{T^\dagger} e^T|\phi\rangle}
=\langle\phi|\ob{\Pminus}^\dagger \frac{e^{T^\dagger}e^T}
                            {\langle\phi| e^{T^\dagger} e^T|\phi\rangle}
=\frac{M^2+P_\perp^2}{P^+}\langle\widetilde\psi|.
\ee
Physical quantities can then be computed from the right and left LFCC eigenstates.

The LFCC method can also be extended to consider time-dependent problems,
such as the strong-field dynamics studied by Zhao {\em et al}.~\cite{BLFQtime}.
The fundamental time-evolution equation 
$\Pminus|\psi\rangle=i\frac{\partial}{\partial x^+}|\psi\rangle$
is replaced by
\be
P_v\left(\ob{\Pminus}-i\frac{\partial T}{\partial x^+}\right)|\phi\rangle
   =i\frac{\partial}{\partial x^+}|\phi\rangle\;\;
\mbox{and}\;\;
(1-P_v)\left(\ob{\Pminus}-i\frac{\partial T}{\partial x^+}\right)|\phi\rangle=0.
\ee
Because $\partial T/\partial x^+$ increases particle number, the valence
time-evolution equation reduces to
\be
P_v\ob{\Pminus}|\phi\rangle=i\frac{\partial}{\partial x^+}|\phi\rangle.
\ee
For a single-particle valence state, there is only a time-dependent
phase.  The time evolution of the $T$ operator is determined by the
auxiliary equation, subject to appropriate initial conditions.

\subsection{\it Effective Particles \label{sec:EffParticles}}

The renormalization-group procedure for effective particles 
(RGPEP)~\cite{RGPEP} has been
developed as an extension of the program for nonperturbative
QCD by Wilson {\em et al.}~\cite{Wilson:1994fk}.  It builds on
the idea of the similarity renormalization procedure proposed
by Glazek and Wilson~\cite{Similarity}.  Effective Hamiltonians
are constructed in terms of creation and annihilation operators
for effective particles, which allows for constituent masses
that differ from the current masses in the original Hamiltonian.
This facilitates a direct connection with constituent quark models
and admits a perturbative construction of the effective
Hamiltonian.

A given quantum field $\psi_0$, built from creation and
annihilation operators for bare quanta, is transformed
to a field $\psi_s={\cal U}_s \psi_0 {\cal U}_s^\dagger$,
built from creation and annihilation operators for
effective particles with scale $s$.  Kinematical quantum
numbers are unchanged; however, masses are treated as
dynamical, so that effective particles, such as gluons,
can acquire mass.  The effective fields are used to
construct an effective Hamiltonian ${\cal H}_t$,
with $t=s^4$ a convenient re-parameterization.  
The effective Hamiltonian is band diagonal, with 
bandwidth $\sim1/s$.

For a generic Hamiltonian, written in terms of annihilation operators $q_{0i}$
\be
{\cal H}_t(q_{0i})=\sum_{n=2}^\infty\sum_{i_1,i_2,\ldots,i_n}
   c_t(i_1,\ldots,i_n)q_{0i_1}^\dagger\cdots q_{0i_n},
\ee
the evolution in the scale $t$ is given by
\be
\frac{\partial}{\partial t}{\cal H}_t=[[{\cal H}_f,{\cal H}_{Pt}],{\cal H}_t],
\ee
where ${\cal H}_f\equiv\sum_i p_i^- q_{0i}^\dagger q_{0i}$ is the free part of
the Hamiltonian and
\be
{\cal H}_{Pt}\equiv\sum_{n=2}^\infty\sum_{i_1,i_2,\ldots,i_n}
   \left(\frac12\sum_{k=1}^np_{i_k}^-\right)^2
   c_t(i_1,\ldots,i_n)q_{0i_1}^\dagger\cdots q_{0i_n}.
\ee
This reduces to coupled equations for the coefficients $c_t$,
which then determine the effective Hamiltonian ${\cal H}_t(q_{ti})$
as a polynomial in the effective-particle operators $q_{ti}$.

The band-diagonal structure occurs because the evolution of
matrix elements for the interaction Hamiltonian ${\cal H}_I\equiv {\cal H}-{\cal H}_f$
is given by~\cite{RGPEP}
\be
\frac{\partial}{\partial t}\left(\sum_{mn}|{\cal H}_{Imn}|^2\right)
   =-2\sum_{km}({\cal M}_{km}^2-{\cal M}_{mk}^2)^2|{\cal H}_{Ikm}|^2\leq 0,
\ee
with ${\cal M}_{km}^2$ the invariant mass squared for the particles in the
$k$th Fock state that are connected by the interaction to the particles in
the $m$th state.  These Fock states are eigenstates of ${\cal H}_f$.
Thus, the off-diagonal matrix elements decrease in magnitude with increasing
$t=s^4$ and do so most rapidly for those states most greatly separated in
invariant mass.

Illustrations of the RGPEP in terms of simple theories, where the interaction
is a mixing of particle states, can be found in \cite{RGPEP}.  The solutions are consistent
with those obtained with ordinary light-front Fock-space methods.

\section{Applications} \label{sec:Applications}

To illustrate the use of light-front methods, some
applications to a range of quantum field theories
are considered.

\subsection{\it Quenched Scalar Yukawa Theory \label{sec:QSY}}

Also known as the (massive) Wick--Cutkosky model~\cite{WickCutkosky}, 
quenched scalar Yukawa theory involves charged and neutral scalars
coupled by a cubic interaction and quenched, to exclude pair production.
The Wick--Cutkosky model focuses on two charged scalars interacting
through the exchange of the neutral, which may or may not be massive.
Various light-front analyses have been 
done~\cite{Sawicki,JiFurnstahl,WC1+1,Wivoda,Swenson,Ji:1994zx,Cooke:2000ef,%
EarlierQSY,HwangKarmanov,Karmanov,JiTokunaga}, including both two-dimensional
and four-dimensional theories.  The most recent work  focuses on the 
construction of the eigenstate for a charged scalar dressed by
a cloud of neutrals~\cite{QSY}.

Without quenching, the theory is ill defined, having a spectrum
that extends to negative infinity, as is generally true of cubic scalar
theories~\cite{Baym,Gross}.  Within a DLCQ approximation,
this instability can be difficult to detect~\cite{Swenson} unless
constrained longitudinal zero modes are included~\cite{ZeroModes}.
However, the constraint equation for the neutral scalar can be solved 
exactly, and the effective interactions that this generates include
zero-mode exchange as well as a destabilizing doorway to
infinite numbers of charged pairs.  The exchange of zero modes
is important for obtaining the nominal
$1/K^2$ convergence of the trapezoidal quadrature rule~\cite{Wivoda}.

The Lagrangian of the model is
\be
{\cal L}=\partial_\mu\chi^*\partial^\mu\chi-m^2|\chi|^2
  +\frac12(\partial_\mu\phi)^2-\frac12 \mu^2\phi^2-g\phi|\chi|^2.
\ee
The Hamiltonian density is
\be
{\cal H}=|\senk{\partial}\chi|^2+ m^2|\chi|^2
   +\frac12(\senk{\partial}\phi)^2+\frac12 \mu^2\phi^2+g\phi|\chi|^2.
\ee
The mode expansions for the fields are, as in Sec.~\ref{sec:scalars},
\be
\phi(x)=\int \frac{dp^+d^2p_\perp}{\sqrt{16\pi^3 p^+}}
        \left[a(\ub{p})e^{-ip\cdot x}+a^\dagger(\ub{p})e^{ip\cdot x}\right]
\ee
and
\be
\chi(x)=\int \frac{dp^+d^2p_\perp}{\sqrt{16\pi^3 p^+}}
        \left[c_+(\ub{p})e^{-ip\cdot x}+c_-^\dagger(\ub{p})e^{ip\cdot x}\right].
\ee
The necessary commutators are given in (\ref{eq:scalarcommutators}) and 
(\ref{eq:chargedcommutator}).

If the $\phi$ zero mode is to be included in a DLCQ calculation, then
the mode expansion is periodic on the interval $-L<x^-<L$ and $\phi$ is
equal to this expansion plus the zero-mode contribution $\phi_0(\vec{x}_\perp)$.
When the Euler--Lagrange equation 
$\partial_-\partial_+\phi-\senk{\partial}\cdot\senk{\partial}\phi+m^2\phi+g|\chi|^2=0$
is averaged over the $x^-$ interval, only the $\phi_0$ contribution survives,
leaving the constraint
\be
m^2\phi_0-\nabla_\perp^2\phi_0=-\frac{g}{L}\int_{-L}^L dx^-|\chi|^2.
\ee
When $\phi_0$ is eliminated from the Hamiltonian density, the
$|\chi|^2$ term induces a negative term in the Hamiltonian,
proportional to $g^2|\chi|^4$.  The expectation value of this term can be made
arbitrarily negative by including a large number of charged pairs with
momentum fraction $1/K$~\cite{ZeroModes}, 
and the spectrum must therefore be unbounded from below.  This term also 
contributes an effective quartic interaction between charged scalars that can be
interpreted as zero-mode exchange~\cite{Wivoda}.

In the continuum, the light-front Hamiltonian is $\Pminus=\Pminus_0+\Pminus_{\rm int}$,
with
\be
\Pminus_0=\int d\ub{p}\frac{m^2+\senk{p}^2}{p^+}
     \left[c_+^\dagger(\ub{p})c_+(\ub{p})+c_-^\dagger(\ub{p})c_-(\ub{p})\right]
     +\int d\ub{q}\frac{\mu^2+\senk{q}^2}{q^+}a^\dagger(\ub{q})a(\ub{q}).
\ee
and
\bea
\Pminus_{\rm int}&=&g\int\frac{d\ub{p} d\ub{q}}{\sqrt{16\pi^3 p^+ q^+(p^++q^+)}}
  \left[\left(c_+^\dagger(\ub{p}+\ub{q})c_+(\ub{p})
                       +c_-^\dagger(\ub{p}+\ub{q})c_-(\ub{p})\right)a(\ub{q}) \right. \\
  && \rule{2in}{0mm} \left.
     +a^\dagger(\ub{q})\left(c_+^\dagger(\ub{p})c_+(\ub{p}+\ub{q})
                            +c_-^\dagger(\ub{p})c_-(\ub{p}+\ub{q})\right)\right] 
        \nonumber \\
 && +g\int \frac{d\ub{p}_1 d\ub{p}_2}{\sqrt{16\pi^3 p_1^+ p_2^+ (p_1^++p_2^+)}}
\left[c_+^\dagger(\ub{p}_1)c_-^\dagger(\ub{p}_2)a(\ub{p}_1+\ub{p}_2)
      +a^\dagger(\ub{p}_1+\ub{p}_2)c_+(\ub{p}_1)c_-(\ub{p}_2)\right]. \nonumber
\eea
For the quenched theory, the second term in $\Pminus_{\rm int}$ is dropped.

The Fock-state expansion for the eigenstate with charge $\pm1$ is
\be
|\psi_\pm(\ub{P})\rangle=\sum_{n=0}^\infty (P^+)^{n/2}
      \int \left(\prod_i^n dx_i d^2k_{i\perp}\right)\theta(1-\sum_i^n x_i)
           \psi_n^\pm(x_i;\senk{k_i})|x_i,\senk{k_i},\ub{P},n,\pm\rangle,
\ee
with the Fock states written as
\be
|x_i,\senk{k_i},\ub{P},n,\pm\rangle=\frac{1}{\sqrt{n!}} 
           c_\pm^\dagger((1-\sum_i^n x_i)P^+,(1-\sum_i^n x_i)\senk{P}-\sum_i^n\senk{k_i})
           \prod_i^n a^\dagger(x_iP^+,\senk{k_i}+x_i\senk{P})|0\rangle.
\ee
Substitution into the eigenvalue problem 
$\Pminus|\psi_\pm(\ub{P})\rangle=\frac{M_\pm^2+P_\perp^2}{P^+}|\psi_\pm(\ub{P})\rangle$
yields the coupled system for the wave functions $\psi_n$
\bea
\lefteqn{\left(\frac{m^2+(\sum_i \senk{k_i})^2}{1-\sum_i x_i}
        +\sum_i^n \frac{\mu^2+k_{i\perp}^2}{x_i}\right)\psi_n^\pm(x_i;\senk{k_i})} && \\
 && + \frac{g}{\sqrt{16\pi^3n}}\sum_j^n
       \frac{\psi_{n-1}^\pm(x_1,\senk{k_1};\ldots;x_{j-1},\vec{k}_{j-1\perp};
       x_{j+1},\vec{k}_{j+1\perp};\ldots;x_n,\senk{k_n})}{\sqrt{x_j(1-\sum_{i\neq j}x_i)(1-\sum_i x_i)}}
       \nonumber \\
&& + \frac{g\sqrt{n+1}}{\sqrt{16\pi^3}}  \int dy d^2q_\perp \theta(1-\sum_i^n x_i-y) 
         \frac{\psi_{n+1}^\pm(x_1,\senk{k_1};\ldots;x_n,\senk{k_n};y,\senk{q})}
               {\sqrt{y(1-\sum_i x_i-y)(1-\sum_i x_i)}}=M_\pm^2\psi_n^\pm(x_i,\senk{k_i}).
               \nonumber
\eea
This system is then solved for $M_\pm$ and the $\psi_n^\pm$.

\begin{figure}[ht]
\begin{center}
\begin{tabular}{cc}
\epsfig{file=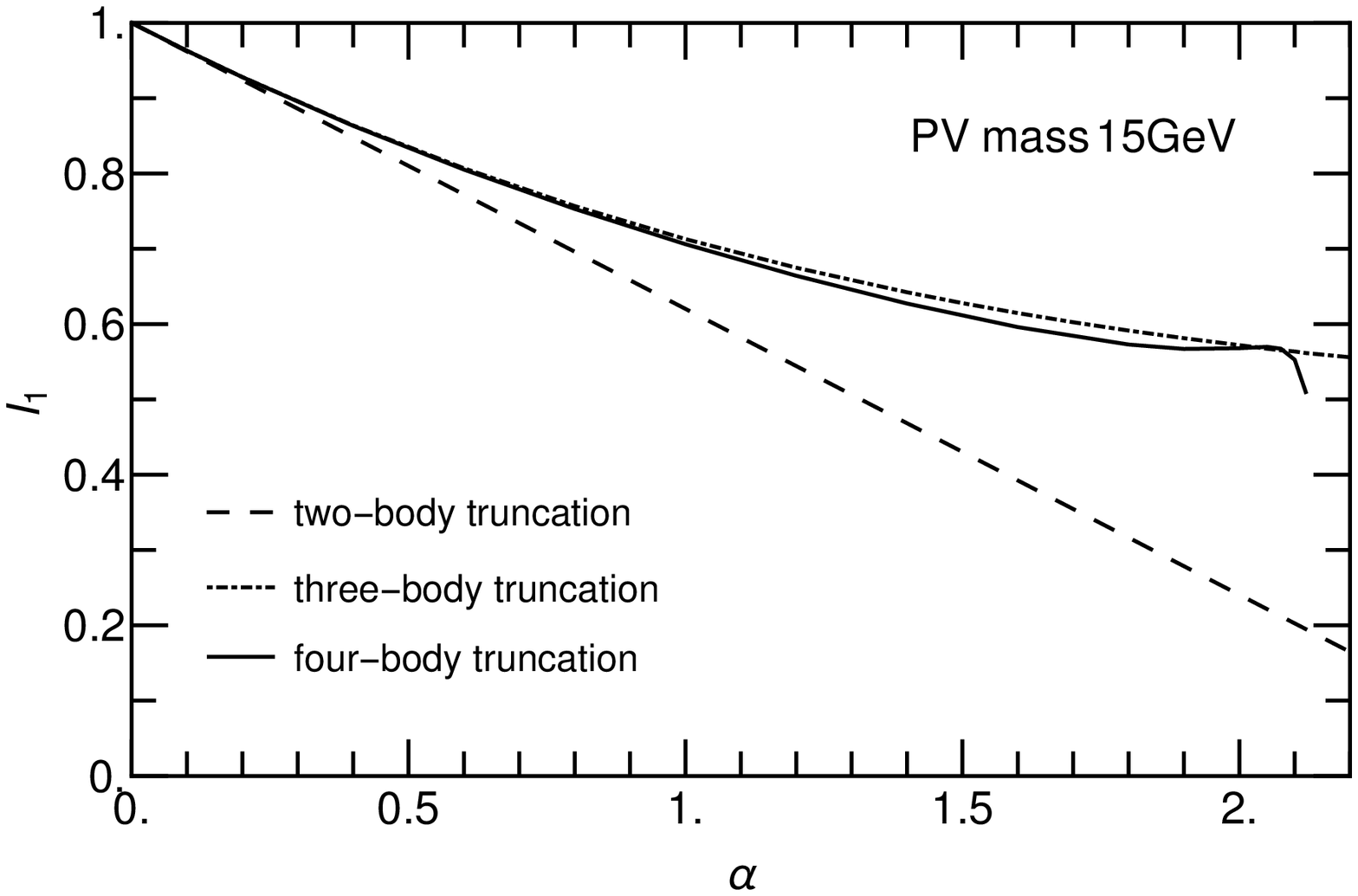,scale=0.48} & \epsfig{file=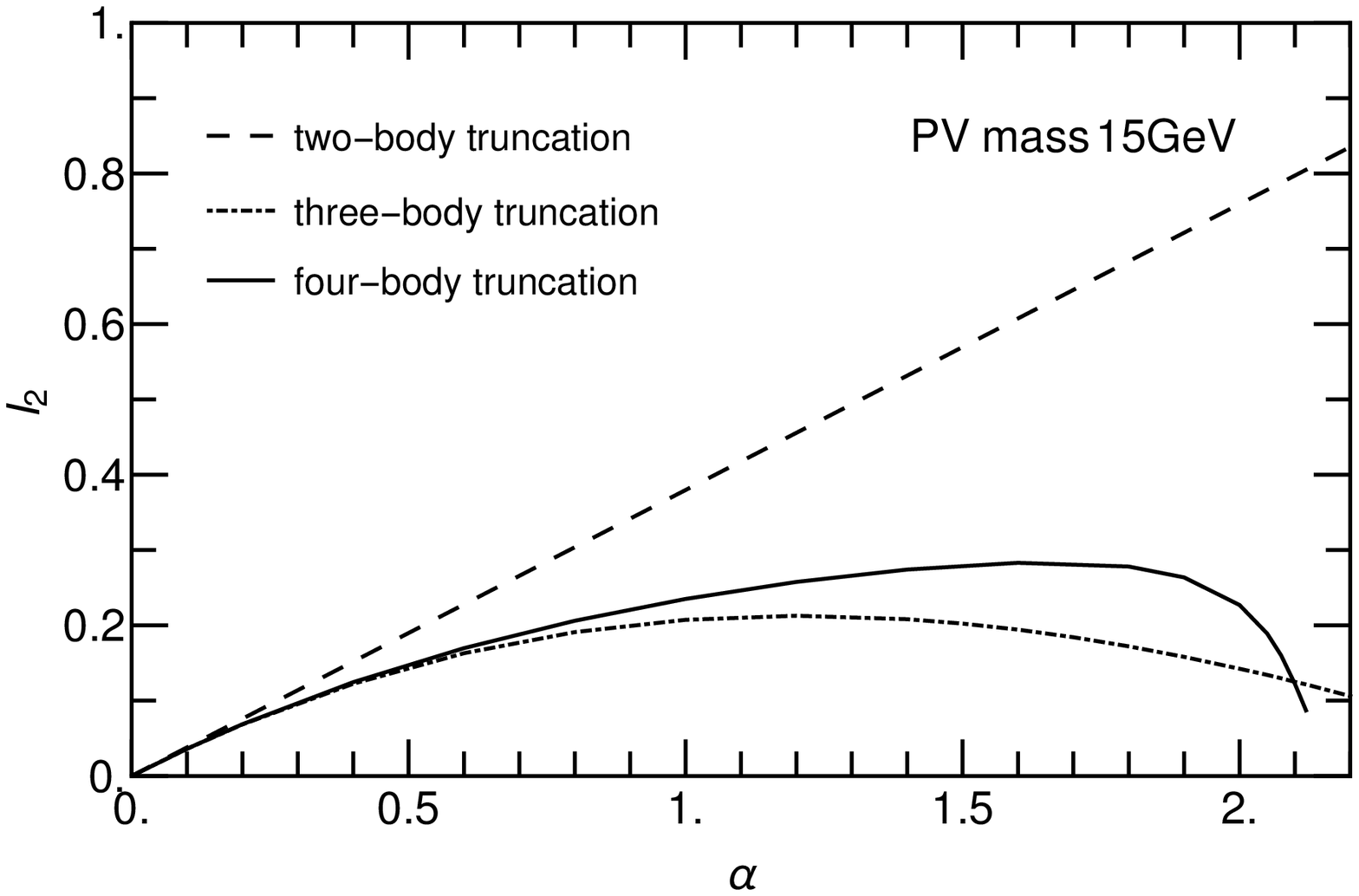,scale=0.48} \\
(a) & (b) 
\end{tabular}
\caption{\label{fig:QSYconvergence} 
Fock sector probabilities as a function of coupling strength $\alpha\equiv\frac{g^2}{16\pi m^2}$
for the (a) one-particle and (b) two-particle sectors
of the dressed charged scalar in quenched scalar Yukawa theory, as shown in
\protect\cite{QSY}, for a sequence of Fock-space truncations to two, three, and
four particles.  The regulating PV mass is set at 15 GeV; the bound-state mass
is 0.94 GeV; and the neutral scalar's mass is 0.14 GeV, chosen to mimic
a nucleon dressed by pions.
}
\end{center}
\end{figure}

Numerical solutions have been obtained by Li {\em et al.}~\cite{QSY}.  They used
PV regularization with one PV neutral scalar and sector-dependent renormalization.
Fock-space truncations up to four particles were investigated.  Earlier work~\cite{EarlierQSY}
had included only three particles (one charged and two neutral).  The formulation is
slightly different, being based on the covariant light-front dynamics scheme~\cite{LFreview2},
but equivalent.  Gauss-Legendre quadrature is used for longitudinal, transverse, and azimuthal
integrations.  Truncations to two, three, and four particles are compared, in order to
study convergence as the Fock-space truncation is relaxed; for weak to moderate coupling,
convergence is observed, as illustrated in Fig.~\ref{fig:QSYconvergence}.  The solution 
is also used to compute the form factor of the dressed charged scalar.
%

For the LFCC method, this model is probably the most straightforward
application, aside from the original exactly soluble case~\cite{LFCC}.
The charge $\pm1$ valence state is just $c_\pm^\dagger(\ub{P})|0\rangle$,
and the exact $T$ operator can be written as
\be
T=\sum_n \int \prod_i^n d\ub{q}_i d\ub{p}\, t_n(\ub{q}_1,\ldots,\ub{q}_n,\ub{p})
\left(\prod_i^n a^\dagger(\ub{q}_i)\right)c_\pm^\dagger(\ub{p})c_\pm(\ub{p}+\sum_i^n\ub{q}_i).
\ee
The action of $e^T$ on the valence state then generates all the Fock states
of the quenched theory, with a one-to-one correspondence between the functions
$t_n$ and the Fock-state wave functions $\psi_n$; each is a nonlinear
combination of the others.  Convergence with respect to truncations
of the sum over $n$ in $T$ can be studied.

\subsection{\it $\phi^4_2$ Theory \label{sec:phi4}}

Two-dimensional $\phi^4$ theory has been a focus for nonperturbative
light-front methods from almost the 
beginning~\cite{VaryHari-phi4,VaryHari-coherent,phi4phase,Chakrabarti:2003tc,%
Chakrabarti:2003ha,Chakrabarti:2005zy,Heinzl}, 
at least partly because the theory provides a relatively simple instance
of symmetry breaking and the possible importance of zero modes~\cite{RozowskyThorn,ZeroModes}.
The phase transition has been studied on the light front~\cite{phi4phase},
and there have been various attempts at the calculation of critical couplings
and even critical exponents~\cite{phi4Pinsky}.  Comparison with results
from equal-time quantization require some care if phrased in terms
of bare parameters; the renormalization of these parameters is
different in the different quantizations~\cite{SineGordon}.

The Lagrangian for $\phi^4_2$ theory is
${\cal L}=\frac12(\partial_\mu\phi)^2-\frac12\mu^2\phi^2-\frac{\lambda}{4!}\phi^4$.
Some choose to normalize the coupling differently, replacing $\lambda/4!$ with
$\lambda/4$ or even simply $\lambda$.
The light-front Hamiltonian density is
${\cal H}=\frac12 \mu^2 \phi^2+\frac{\lambda}{4!}\phi^4$.
The mode expansion for the field at zero light-front time is, as usual,
\be \label{eq:mode}
\phi=\int \frac{dp^+}{\sqrt{4\pi p^+}}
   \left\{ a(p^+)e^{-ip^+x^-/2} + a^\dagger(p^+)e^{ip^+x^-/2}\right\},
\ee
with the nonzero commutator
$[a(p^+),a^\dagger(p^{\prime +})]=\delta(p^+-p^{\prime +})$.

The light-front Hamiltonian can be divided into a kinetic piece and three
interaction pieces, each with a different number of creation and
annihilation operators,
$\Pminus=\Pminus_{11}+\Pminus_{13}+\Pminus_{31}+\Pminus_{22}$,
where
\bea \label{eq:Pminus11}
\Pminus_{11}&=&\int dp^+ \frac{\mu^2}{p^+} a^\dagger(p^+)a(p^+),  \\
\label{eq:Pminus13}
\Pminus_{13}&=&\frac{\lambda}{6}\int \frac{dp_1^+dp_2^+dp_3^+}
                              {4\pi \sqrt{p_1^+p_2^+p_3^+(p_1^++p_2^++p_3^+)}} 
     a^\dagger(p_1^++p_2^++p_3^+)a(p_1^+)a(p_2^+)a(p_3^+), \\
\label{eq:Pminus31}
\Pminus_{31}&=&\frac{\lambda}{6}\int \frac{dp_1^+dp_2^+dp_3^+}
                              {4\pi \sqrt{p_1^+p_2^+p_3^+(p_1^++p_2^++p_3^+)}} 
      a^\dagger(p_1^+)a^\dagger(p_2^+)a^\dagger(p_3^+)a(p_1^++p_2^++p_3^+), \\
\label{eq:Pminus22}
\Pminus_{22}&=&\frac{\lambda}{4}\int\frac{dp_1^+ dp_2^+}{4\pi\sqrt{p_1^+p_2^+}}
       \int\frac{dp_1^{\prime +}dp_2^{\prime +}}{\sqrt{p_1^{\prime +} p_2^{\prime +}}} 
       \delta(p_1^+ + p_2^+-p_1^{\prime +}-p_2^{\prime +})
  a^\dagger(p_1^+) a^\dagger(p_2^+) a(p_1^{\prime +}) a(p_2^{\prime +}) .
\eea
The eigenstate with momentum $P^+$ is expanded as
\be 
|\psi(P^+)\rangle=\sum_n (P^+)^{\frac{n-1}{2}}\int\prod_i^n dy_i 
       \delta(1-\sum_i^n y_i)\psi_n(y_i)\frac{1}{\sqrt{n!}}\prod_{i=1}^n a^\dagger(y_iP^+)|0\rangle,
\ee
with the sum over $n$ restricted to odd or even numbers, because $\Pminus$
does not mix the two cases.  On substitution of this
Fock-state expansion and a dimensionless coupling
$g=\lambda/4\pi\mu^2$, the light-front Hamiltonian eigenvalue problem
${\cal P}^-|\psi(P^+)\rangle=\frac{M^2}{P^+}|\psi(P^+)\rangle$
becomes
\bea \label{eq:coupledsystem}
\lefteqn{\frac{n}{y_1}\psi_n(y_i)
+\frac{g}{4}\frac{n(n-1)}{\sqrt{y_1y_2}}
        \int\frac{dx_1 dx_2 }{\sqrt{x_1 x_2}}\delta(y_1+y_2-x_1-x_2) \psi_n(x_1,x_2,y_3,\ldots,y_n)}&& 
        \nonumber \\
& +\frac{g}{6}n\sqrt{(n+2)(n+1)}\int \frac{dx_1 dx_2 dx_3}{\sqrt{y_1 x_1 x_2 x_3}}
        \delta(y_1-x_1-x_2-x_3)\psi_{n+2}(x_1,x_2,x_3,y_2,\ldots,y_n) \nonumber \\
& +\frac{g}{6}\frac{(n-2)\sqrt{n(n-1)}}{\sqrt{y_1y_2y_3(y_1+y_2+y_3)}}
          \psi_{n-2}(y_1+y_2+y_3,y_4,\ldots,y_n)=\frac{M^2}{\mu^2}\psi_n(y_i).
\eea
This coupled system can then be solved for the mass $M$ and wave functions $\psi_n$.
The fully symmetric polynomials discussed in Sec.~\ref{sec:FnExpansion} were
designed for just this purpose; some results obtained in this way are discussed
below.  

\begin{figure}[ht]
\begin{center}
\epsfig{file=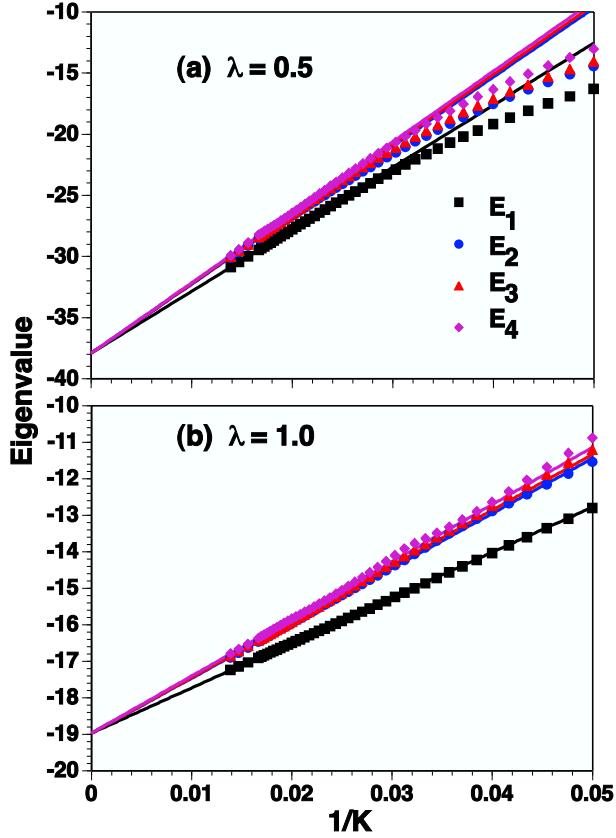,scale=0.75}
\caption{\label{fig:phi4HiRes} 
Mass-squared eigenvalues for odd and even states in the broken-symmetry phase, in units of $\mu^2$,
from DLCQ calculations with resolution $K$, from \protect\cite{Chakrabarti:2003tc}.
The values of the coupling constant $\lambda$ are quoted in units of $\mu^2$,
where $\mu$ is the bare mass of the constituents.  The solid lines are extrapolated 
fits to the numerical results.
}
\end{center}
\end{figure}

The earliest and highest resolution calculations have been done with DLCQ.
The resolution has been taken high enough to obtain degeneracy between
the odd and even sectors in the broken-symmetry case~\cite{Chakrabarti:2003tc},
following the work of Rozowsky and Thorn~\cite{RozowskyThorn} where degeneracy
was indicated but not achieved at their lower resolution.  Figure~\ref{fig:phi4HiRes}
contains two samples of the high-resolution DLCQ results, which show the degeneracy
between odd and even states in the infinite-resolution limit.
Low-level excitations at strong coupling can be associated with kink-antikink 
states~\cite{Chakrabarti:2003ha,Chakrabarti:2005zy,KrogerPauli}.  These may indicate formation of a kink 
condensate driving the transition to symmetry restoration for the negative
mass-squared case.
%

Inclusion of zero modes can improve the DLCQ calculations.
The solution of the DLCQ constraint equation for the zero mode can 
be used to study the critical coupling and critical exponents~\cite{phi4Pinsky}.
The solution can also be used to develop a controlled series of effective interactions
that improve the numerical convergence~\cite{ZeroModes}.

The LFCC method has also been applied to $\phi^4_2$ theory~\cite{LFCCphi4}.  For 
the odd case, the valence state is the one-particle state $a^\dagger(P^+)|0\rangle$.  
The leading contribution to the $T$ operator is
\be
T_2\equiv\int dp_1^+ dp_2^+ dp_3^+ t_2(p_1^+,p_2^+,p_3^+) 
    a^\dagger(p_1^+)a^\dagger(p_2^+)a^\dagger(p_3^+)a(p_1^++p_2^++p_3^+);
\ee
the function $t_2$ is symmetric in its arguments.  For $T$ truncated to $T_2$, the
projection $1-P_v$ is truncated to projection onto the three-particle state
$a^\dagger(p_1^+)a^\dagger(p_2^+)a^\dagger(p_3^+)|0\rangle$.  Zero modes
can also be included~\cite{LFCCZeroModes}.

The valence equation, the first in (\ref{eq:LFCCeqns}), can be reduced to
\be \label{eq:LFCCvalence}
1+g\int\frac{dx_1 dx_2}{\sqrt{x_1 x_2 x_3}}\tilde t_2(x_1,x_2,x_3)=M^2/\mu^2,
\ee
where $\tilde t_2$ is a rescaled function of longitudinal momentum 
fractions $x_i=p_i^+/P^+$,
\be \label{eq:tildet2}
\tilde t_2(x_1,x_2,x_3)\equiv P^+t_2(x_1P^+,x_2P^+,x_3P^+).
\ee
Given the definition of a dimensionless mass shift
\be \label{eq:Delta}
\Delta\equiv g\int\frac{dx_1 dx_2}{\sqrt{x_1 x_2 x_3}}\tilde t_2(x_1,x_2,x_3),
\ee
the valence equation can be written as simply $M^2=(1+\Delta)\mu^2$.

The reduced auxiliary equation, from the second equation in (\ref{eq:LFCCeqns}), is
\bea \label{eq:LFCCaux}
\lefteqn{\frac16\frac{g}{\sqrt{y_1 y_2 y_3}}
             +\frac{M^2}{\mu^2}\left(\frac{1}{y_1}+\frac{1}{y_2}+\frac{1}{y_3}-1\right)
                                    \tilde t_2(y_1,y_2,y_3)}&& \\
   &&  +\frac{g}{2}\left[\int_0^{1-y_1}dx_1
             \frac{\tilde t_2(y_1,x_1,1-y_1-x_1)}{\sqrt{x_1 y_2 y_3 (1-y_1-x_1)}} 
               + (y_1 \leftrightarrow y_2) + (y_1 \leftrightarrow y_3)\right] \nonumber \\
    &&  -\frac{\Delta}{2} \left(\frac{1}{y_1}+\frac{1}{y_2}+\frac{1}{y_3}\right) \tilde t_2(y_1,y_2,y_3)
              \nonumber \\
    && +\frac{3g}{2}\left\{\int_{y_1/(1-y_2)}^1 d\alpha_1 \int_0^{1-\alpha_1} d\alpha_2 
       \frac{\tilde t_2(y_1/\alpha_1,y_2,1-y_1/\alpha_1-y_2) \tilde t_2(\alpha_1,\alpha_2,\alpha_3)}
       {\sqrt{\alpha_1 \alpha_2 \alpha_3 y_3 (\alpha_1-y_1-\alpha_1 y_2)}}\right.      \nonumber \\
    && \rule{2.2in}{0mm} \left.  +(y_1\leftrightarrow y_2)+(y_1\leftrightarrow y_3)
                                               \rule{0mm}{0.15in} \right\} \nonumber \\
    && +\frac{3g}{2}\left\{ \left[ \int_{y_1+y_2}^1 d\alpha_1 \int_0^{1-\alpha_1} d\alpha_2
        \frac{\tilde t_2(y_1/\alpha_1,y_2/\alpha_1,1-(y_1+y_2)/\alpha_1)\tilde t_2(\alpha_1,\alpha_2,\alpha_3)}
        {\alpha_1 \sqrt{\alpha_2 \alpha_3 y_3 (\alpha_1-y_1-y_2)}} \right.  \right.   \nonumber \\
    && \rule{2.5in}{0mm} \left.  + (y_2 \leftrightarrow y_3)
                                                   \rule{0mm}{0.15in}\right]  \nonumber \\
     && \rule{1in}{0mm}  \left. +(y_1\leftrightarrow y_2)+(y_1\leftrightarrow y_3)
                                        \rule{0mm}{0.15in}\right\}=0, \nonumber
\eea
with $y_i=q_i^+/P^+$. 
For comparison, consider a Fock-state truncation that produces the same 
number of equations.  The truncated eigenstate
then contains only one and three-body contributions and the coupled
system of integral equations reduces to
\bea \label{eq:tildepsi1}
\lefteqn{1+g\int\frac{dx_1 dx_2}{\sqrt{x_1 x_2 x_3}}\tilde\psi_3(x_1,x_2,x_3)=M^2/\mu^2,} && \\
\label{eq:tildepsi3}
&& \frac16\frac{g}{\sqrt{y_1y_2y_3}}
   +\left(\frac{1}{y_1}+\frac{1}{y_2}+\frac{1}{y_3}-\frac{M^2}{\mu^2}\right)\tilde\psi_3(y_1,y_2,y_3) \\
     &&  +\frac{g}{2}\left[ \int_0^{1-y_1} dx_1
             \frac{\tilde\psi_3(x_1,y_1,1-y_1-x_1)}{\sqrt{x_1(1-y_1-x_1)y_2 y_3}} 
               + (y_1 \leftrightarrow y_2) + (y_1 \leftrightarrow y_3)\rule{0mm}{0.3in}\right]
                   =0,   \nonumber
\eea
with $\tilde\psi_3\equiv\psi_3/(\sqrt{6}\psi_1)$.

In each case, the first equation is of the same form; it provides for the self-energy
correction of the bare mass to yield the physical mass.  The second equations,
(\ref{eq:LFCCaux}) and (\ref{eq:tildepsi3}),
however, differ significantly.  The LFCC auxiliary equation includes the
physical mass in the three-body kinetic energy; the three-body equation 
of the Fock-truncation approach has only the bare mass and would
require sector-dependent renormalization~\cite{SecDep-Wilson,HillerBrodsky,Karmanov,SecDep}
to compensate. The fourth LFCC term is the nonperturbative
analog of the wave-function renormalization counterterm.  The last
two terms are partial resummations of higher-order loops.  These terms
do not appear in the truncated coupled system because the loops 
have intermediate states that are removed by the truncation.
     
\begin{figure}[ht]
\begin{center}
\epsfig{file=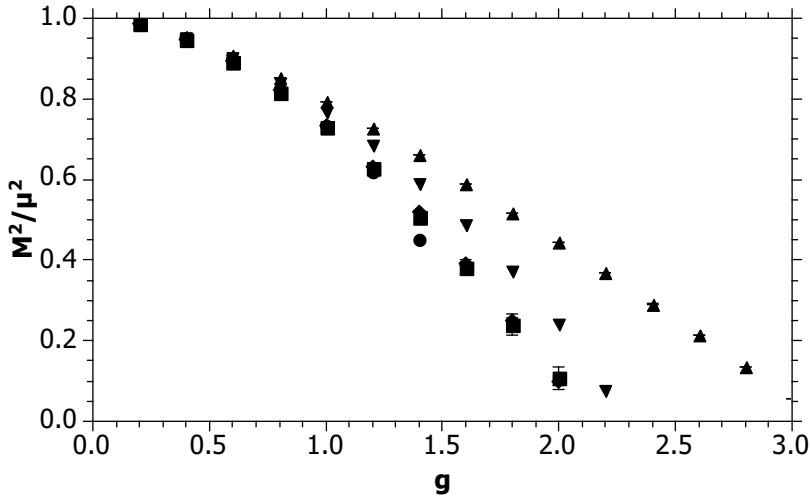,scale=0.5}
\caption{\label{fig:M2vsg} Mass-squared ratios $M^2/\mu^2$ versus
dimensionless coupling strength $g$ for the LFCC approximation (circles),
the Fock-space truncation to three (up triangles), five (squares),
and seven (diamonds) particles, 
and the Fock-space truncation to three
with sector-dependent masses (down triangles) for the odd eigenstate
of $\phi_{1+1}^4$.  The error bars reflect uncertainties in the
numerical extrapolations.  For small $g$, all of the results overlap;
for all $g$, the results for the five and seven-particle truncations
are nearly indistinguishable, indicating convergence of the Fock-state
expansion.
}
\end{center}
\end{figure}

These equations are solved numerically in \cite{LFCCphi4}, and more recently
for higher Fock-space truncation, with basis function expansions
that use the fully-symmetric multivariate polynomials~\cite{GenSymPolys}.
The converged results for the mass-squared eigenvalues are
shown in Fig.~\ref{fig:M2vsg}.  There is a distinct difference
between the LFCC approximation and the lowest-order Fock-space truncation.
This arises from two factors: the correct kinetic-energy mass
in each sector of the LFCC calculation and contributions from
higher Fock states.
When the lowest-order Fock-space truncation method is modified with 
sector-dependent masses~\cite{SecDep-Wilson,HillerBrodsky,Karmanov,SecDep},
the resulting mass values are intermediate between the these
two sets.  The higher-order Fock-space truncations, at five-body
and seven-body Fock sectors, are in agreement to within numerical
error, indicating convergence in the Fock-state expansion.  The
LFCC result is close to these, implying excellent representation
of contributions from the higher Fock states.
%

The Fock-space-converged results also allow estimation of the critical 
coupling of the theory.  In the plot, the mass gap vanishes at $g\simeq 2.1$.
This is consistent with a much older light-front (DLCQ) calculation
by Harindranath and Vary~\cite{VaryHari-phi4} but not with equal-time calculations,
as reported by Rychkov and Vitale~\cite{RychkovVitale}.  In their units,
the LFCC result for the critical coupling is $1.1$, and the DLCQ result is 1.38,
to compare with their result of 2.97.  However, these are values of
the ratio of bare parameters of the Lagrangian; as such, the different
quantizations need not yield the same values.  In fact, earlier work
by Burkardt~\cite{SineGordon} has shown that, while the bare coupling
$\lambda$ is unchanged, the renormalization of the bare mass $\mu$ 
is different in the two quantizations by a computable amount.
The bare mass is renormalized by tadpole contributions in equal-time quantization
but not in light-front quantization, and the two different masses
are related by~\cite{SineGordon}
\be
\mu_{\rm LF}^2=\mu_{\rm ET}^2
   +\lambda\left[\langle 0|:\frac{\phi^2}{2}:|0\rangle
       -\langle 0|:\frac{\phi^2}{2}:|0\rangle_{\rm free}\right].
\ee
The vacuum expectation values of $\phi^2$ resum the tadpole
contributions; the subscript {\em free} indicates the vev with
$\lambda=0$.  This distinction between bare masses in the
two quantizations implies that the dimensionless coupling $g=\lambda/4\pi\mu^2$
is also not the same.  Estimates of the critical coupling must
then be adjusted for the difference if they are to be compared.

\subsection{\it Yukawa Theory \label{sec:Yukawa}}

The standard Yukawa theory of fermions interacting with scalars has
received considerable attention.  A series of 
papers~\cite{DLCQYukawa,YukawaOneBoson,YukawaTwoBoson} investigated
the dressed fermion state on the light front, using PV regularization,
DLCQ, and Gauss--Legendre quadrature.  These built on preliminary work
with a soluble model~\cite{bhm1,bhm2} and considered various truncations
as well as different choices for the PV sector.  
A number of properties of the dressed-fermion eigenstate were extracted
from the Fock-state wave functions, including average numbers
of constituents and their average momenta, structure functions,
and the average radius.  

An analysis of the
exactly soluble case of PV-mass degeneracy was also done~\cite{ExactSoln},
in order to better understand the connection with equal-time quantization,
through the exact operator solution of the theory,
and to consider the possibility of developing a perturbation theory
based on mass differences; the eigenvalue problem in the Fock basis
is exactly soluble because it is triangular when the PV masses are degenerate
with the physical masses.

The one-loop fermion self-energy in light-front Yukawa theory
requires three PV scalars to subtract the quadratic and log divergences 
and to restore chiral symmetry in the massless limit~\cite{ChiralSymYukawa}.
In the preliminary work~\cite{bhm1}, a DLCQ approximation of this 
self-energy was shown to require a discrete set of PV Fock states that
was only 1.5 times the size of the set of ordinary Fock states; this
provided encouragement that the computational load associated with 
the introduction of the additional PV states was not too large.

The first calculation~\cite{DLCQYukawa}, with the DLCQ approach,
used these three scalars.  Subsequent work~\cite{YukawaOneBoson,YukawaTwoBoson}
used one PV scalar and one PV fermion, coupled in null combinations.  The advantage of the newer
PV scheme was in the absence of four-point instantaneous fermion interactions,
which are independent of the fermion mass and cancel between physical
and PV fermions.  Their absence significantly simplifies the matrix
representation of the Hamiltonian, which then contains only three-point
vertices.  This later work truncated the Fock space to include no
more than two bosons dressing the fermion and excluded pair production.
The one-boson truncation~\cite{YukawaOneBoson} is exactly solvable;
the two-boson truncation~\cite{YukawaTwoBoson} requires numerical
methods.

Closely following this work, there was a sequence of 
investigations~\cite{BrinetYukawa,KMSYukawaQED,KMSYukawa,KarmanovYukawa,KarmanovYukawa2}
within the context of covariant light-front dynamics and
eventually employing PV regularization and sector-dependent renormalization.
This work included computation of the anomalous magnetic
moment~\cite{KarmanovYukawa} and even the electromagnetic form 
factors~\cite{KarmanovYukawa2} of the fermion dressed by one or two bosons.
Dependence on the cutoff is eliminated to within numerical accuracy.
This is closely related to earlier work by Glazek and Perry~\cite{Glazek:1992aq}
on the sector-dependent approach to Yukawa theory; they show that triviality
imposes a limit on the cutoff. 

The primary gain from this work has been the experience of working with
fermions in nonperturbative calculations without the additional 
complications of QED.  For example, the form of PV regularization
which eliminates instantaneous fermion interactions was first developed
in Yukawa theory before being considered for QED.  Yukawa theory
also offers the opportunity to study an unquenched theory and the associated
renormalizations of the charge and scalar mass, something
for which the unstable scalar Yukawa theory cannot be used.  This aspect
is largely unexplored at present, although some preliminary work has been
done~\cite{DLCQYukawa,KarmanovYukawa2}.  Applications to modeling of
scalar meson exchange in nuclear physics are quite limited, given that
accurate phenomenology requires a more sophisticated interaction~\cite{Deuteron}.

The Yukawa-theory Lagrangian, regulated by one PV scalar and one PV fermion, is 
\bea
\lefteqn{{\cal L}=\frac{1}{2}(\partial_\mu\phi_0)^2-\frac{1}{2}\mu_0^2\phi_0^2
-\frac{1}{2}(\partial_\mu\phi_1)^2+\frac{1}{2}\mu_1^2\phi_1^2} \\
 &&+\frac{i}{2}\left(\psibar_0\g^\mu\partial_\mu-(\partial_\mu\psibar_0)\g^\mu\right)
     \psi_0
  -m_0\psibar_0\psi_0  \nonumber \\
&&-\frac{i}{2}\left(\psibar_1\g^\mu\partial_\mu-(\partial_\mu\psibar_1)\g^\mu\right)
      \psi_1+m_1\psibar_1\psi_1 
      -g(\phi_0 + \phi_1)(\psibar_0 + \psibar_1)(\psi_0 + \psi_1).
\nonumber
\eea
The subscript 0 indicates a physical field and 1, a PV field.  The fermion
masses are denoted by $m_i$, and the boson masses by $\mu_j$.
When fermion pairs are excluded, the resulting light-front
Hamiltonian is
\bea \label{eq:YukawaP-}
\lefteqn{\Pminus=
   \sum_{i,s}\int d\ub{p}
      \frac{m_i^2+p_\perp^2}{p^+}(-1)^i
          b_{i,s}^\dagger(\ub{p}) b_{i,s}(\ub{p})
    +\sum_{j}\int d\ub{q}
          \frac{\mu_j^2+q_\perp^2}{q^+}(-1)^j
              a_j^\dagger(\ub{q}) a_j(\ub{q})}  \nonumber \\
   && +\sum_{i,j,k,s}\int d\ub{p} d\ub{q}\left\{
      \left[ V_{-2s}^*(\ub{p},\ub{q})
             +V_{2s}(\ub{p}+\ub{q},\ub{q})\right]
                 b_{j,s}^\dagger(\ub{p})
                  a_k^\dagger(\ub{q})
                   b_{i,-s}(\ub{p}+\ub{q})\right.  \\
      &&\left.\rule{0.5in}{0in}
           +\left[U_j(\ub{p},\ub{q})
                    +U_i(\ub{p}+\ub{q},\ub{q})\right]
               b_{j,s}^\dagger(\ub{p})
                a_k^\dagger(\ub{q})b_{i,s}(\ub{p}+\ub{q})
                    + h.c.\right\},  \nonumber
\eea
where $a^\dagger$ creates a boson and $b^\dagger$ a fermion, and
\be
U_j(\ub{p},\ub{q})
   \equiv \frac{g}{\sqrt{16\pi^3}}\frac{m_j}{p^+\sqrt{q^+}},\;\;
V_{2s}(\ub{p},\ub{q})
   \equiv \frac{g}{\sqrt{8\pi^3}}
   \frac{{\vec\epsilon}_{2s}^{\,*}\cdot{\vec p}_\perp}{p^+\sqrt{q^+}},\;\;
{\vec \epsilon}_{2s}\equiv-\frac{1}{\sqrt{2}}(2s,i). 
\ee
The nonzero (anti)commutators are
\be
\left[a_i(\ub{q}),a_j^\dagger(\ub{q}')\right]
          =(-1)^i\delta_{ij}
            \delta(\ub{q}-\ub{q}'), \;\;
\left\{b_{i,s}(\ub{p}),b_{j,s'}^\dagger(\ub{p}')\right\}
     =(-1)^i\delta_{ij}   \delta_{s,s'}
            \delta(\ub{p}-\ub{p}').
\ee
The opposite signature of the PV fields is the reason that
no instantaneous fermion terms appear in $\Pminus$; they are
individually independent of the fermion mass and cancel
between instantaneous physical and PV fermions.

This scheme was explored in the
unphysical equal-mass limit where analytic solutions of
the field-theoretic eigenstate can be obtained without explicit
truncation~\cite{ExactSoln}.  The simplest such solution takes the form
\be
\beta_{+,\ub{k}}^\dagger |0\rangle + mg\int_0^{k^+} d\ub{l}
{U(\ub{k},\ub{l}) \over E_{1,0}(\ub{k}) - E_{1,1}(\ub{l},\ub{k} -
\ub{l})} b_{+,\ub{l}}^\dagger a_{\ub{k} - \ub{l}}^\dagger |0\rangle
+ g \int_0^{k^+} d\ub{l} {V(\ub{k},\ub{l}) \over
E_{1,0}(\ub{k}) - E_{1,1}(\ub{l},\ub{k} - \ub{l})} b_{-,\ub{l}}^\dagger
a_{\ub{k} - \ub{l}}^\dagger |0\rangle,
\label{df}
\ee 
where $\beta^\dagger$, $b^\dagger$, and $a^\dagger$ are creation
operators for null combinations of the fields, $m\equiv m_0=m_1$ 
is the fermion mass, and
\begin{eqnarray}
&&U(\ub{k},\ub{l}) \equiv {1 \over \sqrt{16\pi^3}}
          {1 \over \sqrt{k^+ -l^+}} 
       \left( {1 \over l^+} + {1 \over k^+}\right)\,, \\
&&V(\ub{k},\ub{l}) \equiv {1 \over \sqrt{16\pi^3}}
                      {1 \over \sqrt{k^+ -l^+}} 
    \left( {-l_1 - i l_2 \over l^+} + {k_1 + i k_2 \over k^+}\right)\,.
\end{eqnarray} 
The eigenvalue of the state is $(k_\perp^2+m^2)/k^+$.  
Such analytic solutions provide a
convenient check for numerical calculations.

The Fock-state expansion for the dressed-fermion state 
with $J_z=+1/2$ is
\bea
\lefteqn{|\psi_+(\ub{P})\rangle=\sum_i z_i b_{i+}^\dagger(\ub{P})|0\rangle
  +\sum_{ijs}\int d\ub{q} f_{ijs}(\ub{q})b_{is}^\dagger(\ub{P}-\ub{q})
                                       a_j^\dagger(\ub{q})|0\rangle}&& \\
 && +\sum_{ijks}\int d\ub{q}_1 d\ub{q}_2 f_{ijks}(\ub{q}_1,\ub{q}_2)
       \frac{1}{\sqrt{1+\delta_{jk}}}   b_{is}^\dagger(\ub{P}-\ub{q}_1-\ub{q}_2)
                 a_j^\dagger(\ub{q}_1)a_k^\dagger(\ub{q}_2)|0\rangle 
 +\ldots,       \nonumber
\eea
It is normalized by the requirement
$\langle\psi_{\sigma'}(\ub{P}')|\psi_\sigma(\ub{P})\rangle=\delta_{\sigma\sigma'}\delta(\ub{P}'-\ub{P})$.
The wave functions $f$ that define this state must satisfy the
usual coupled system of equations
The first three coupled equations are
\bea
m_i^2z_i&+& \sum_{i',j}(-1)^{i'+j} P^+ \int^{P^+} d\ub{q}
  \left\{ f_{i'j-}(\ub{q})[V_+(\ub{P}-\ub{q},\ub{q})+V_-^*(\ub{P},\ub{q})]
  \right. \nonumber \\
  &&\left.
   + f_{i'j+}(\ub{q})[U_{i'}(\ub{P}-\ub{q},\ub{q})+U_i(\ub{P},\ub{q})]\right\}
   = M^2z_i,
\eea
\bea \label{eq:oneboson}
\lefteqn{
\left[\frac{m_i^2+q_\perp^2}{1-y}+\frac{\mu_j^2+q_\perp^2}{y}\right]
  f_{ijs}(\ub{q}) 
  +\sum_{i'}(-1)^{i'}\left\{
    z_{i'}\delta_{s,-}[V_+^*(\ub{P}-\ub{q},\ub{q})+V_-(\ub{P},\ub{q})] 
                               \right.}\nonumber \\
  && \rule{2in}{0in}\left.
    +z_{i'}\delta_{s,+}[U_i(\ub{P}-\ub{q},\ub{q})+U_{i'}(\ub{P},\ub{q})]\right\}
     \\
    &&+2\sum_{i',k}\frac{(-1)^{i'+k}}{\sqrt{1+\delta_{jk}}}P^+ \int^{P^+-q^+}
    d\ub{q}' \left\{f_{i'jk,-s}(\ub{q},\ub{q}')
       [V_{2s}(\ub{P}-\ub{q}-\ub{q}',\ub{q}')\right. \nonumber \\
       && \rule{2in}{0in} +V_{-2s}^*(\ub{P}-\ub{q},\ub{q}')]
       \nonumber \\
           &&\left.+f_{i'jks}(\ub{q},\ub{q}')
             [U_{i'}(\ub{P}-\ub{q}-\ub{q}',\ub{q}')+U_i(\ub{P}-\ub{q},\ub{q}')]
             \right\} = M^2f_{ijs}(\ub{q}),\nonumber
\eea
and
\bea
\lefteqn{\left[\frac{m_i^2+(\vec{q}_{1\perp}+\vec{q}_{2\perp})^2}{1-y_1-y_2}
    +\frac{\mu_j^2+q_{1\perp}^2}{y_1}+\frac{\mu_k^2+q_{2\perp}^2}{y_2}\right]
        f_{ijks}(\ub{q_1},\ub{q_2})} \\
    &&+\sum_{i'}(-1)^{i'}\frac{\sqrt{1+\delta_{jk}}}{2}P^+ 
        \left\{f_{i'j,-s}(\ub{q_1})
[V_{-2s}^*(\ub{P}-\ub{q_1}-\ub{q_2},\ub{q_2})\right. \nonumber \\
  && \rule{2in}{0in} +V_{2s}(\ub{P}-\ub{q_1},\ub{q_2})] \nonumber \\
 &&+ f_{i'js}(\ub{q_1})
[U_i(\ub{P}-\ub{q_1}-\ub{q_2},\ub{q_2})+U_{i'}(\ub{P}-\ub{q_1},\ub{q_2})]
 \nonumber \\
 &&  + f_{i'k,-s}(\ub{q_2})
[V_{-2s}^*(\ub{P}-\ub{q_1}-\ub{q_2},\ub{q_1})+V_{2s}(\ub{P}-\ub{q_2},\ub{q_1})]
  \nonumber \\
 &&  \left. + f_{i'ks}(\ub{q_2})
[U_i(\ub{P}-\ub{q_1}-\ub{q_2},\ub{q_1})+U_{i'}(\ub{P}-\ub{q_2},\ub{q_1})]
\right\}+\ldots \nonumber \\  
   && = M^2f_{ijks}(\ub{q_1},\ub{q_2}). \nonumber
\eea

To best interpret the norms of different Fock sectors,
the physical wave functions were defined as the coefficients of
Fock states containing only positive-norm particles.
This reduction can be achieved by requiring all Fock states
to be expressed in terms of the positive-norm creation operators
$b_{0s}^\dagger$ and $a_0^\dagger$ and the zero-norm
combinations $b_s^\dagger\equiv b_{0s}^\dagger+b_{1s}^\dagger$
and $a^\dagger\equiv a_0^\dagger+a_1^\dagger$.
Any term containing a $b_s^\dagger$ or an $a^\dagger$,
which would be annihilated by the PV-generalized electromagnetic
current, is discarded, leaving the physical state
\bea
\lefteqn{|\psi_+\rangle_{\rm phys}=(z_0-z_1)b_{1+}^\dagger(\ub{P})|0\rangle
           + \sum_s \int d\ub{q}
\left(\sum_{ij}(-1)^{i+j}f_{ijs}(\ub{q})\right)
     b_{0s}^\dagger(\ub{P}-\ub{q})a_0^\dagger(\ub{q})
              |0\rangle } \nonumber \\
 && +\sum_s\int d\ub{q}_1 d\ub{q}_2 
 \left(\sum_{ijk}(-1)^{i+j+k}f_{ijks}(\ub{q}_1,\ub{q}_2)
       \frac{1}{\sqrt{1+\delta_{jk}}} \right)  \\
 &&\rule{2in}{0in}  \times   b_{0s}^\dagger(\ub{P}-\ub{q}_1-\ub{q}_2)
                 a_0^\dagger(\ub{q}_1)a_0^\dagger(\ub{q}_2)|0\rangle +\ldots,
                 \nonumber
\eea
From this state various physical quantities are extracted, including
the boson structure function for constituent-fermion helicity $s$
\bea \label{eq:fB}
\lefteqn{f_{Bs}(y)=\int d\ub{q} \delta(y-q^+/P^+)
   \left|\sum_{ij}(-1)^{i+j}f_{ijs}(\ub{q})\right|^2}&& \\
   &&+\int \prod_{n=1}^2 d\ub{q_n} \sum_{n=1}^2\delta(y-q_n^+/P^+)
   \left|\sum_{ijk}(-1)^{i+j+k}f_{ijks}(\ub{q_n})\right|^2+\ldots
   \nonumber
\eea

In traditional DLCQ, the lower bound of $1/K$ on longitudinal momentum
fractions provides a natural cutoff in the number of equations.
The range of the transverse integrations is cut off by imposing
$p_{i\perp}^2/x_i<\Lambda^2$ for each particle in a Fock state; this
reduces the DLCQ matrix problem to a finite size.  The transverse 
momentum indices $n_x$ and $n_y$ are limited by a transverse resolution
$N$.  For a reduced set of equations, alternative 
quadratures are more efficient. 
In particular, the transverse momentum $q_\perp$ can be mapped to
a finite range that compresses the wave function's tail to a
relatively small region, so that a Gauss--Legendre quadrature can
yield a good approximation~\cite{YukawaTwoBoson}. 

Given such a discretization and the consequent finite matrix
eigenvalue problem, one can compute mass eigenvalues and 
associated wave functions.  The bare parameters
$g$ and $m_0$ can be fixed by fitting ``physical'' constraints, such 
as specifying the dressed-fermion mass $M$ and its radius. In principle,
one then takes the infinite resolution and infinite momentum volume
limits, as well as the infinite PV-mass limits.

\subsection{\it Supersymmetric Yang--Mills theory \label{sec:SuperYM}}

A number of applications of SDLCQ to supersymmetric Yang--Mills (SYM) theories
have been carried out by the SDLCQ collaboration~\cite{SDLCQ,SYMspectrum,1+1correlator,%
subsequent,correlator2+1,Filippov,3DSYM,SYM-CS,BPSstates,lightstates,%
SYM-CSproperties,fundamentallight,fundamentalspectrum,N22Improved,%
N11FiniteTemp,ImprovedSYM,DirectEvidence}.  
The most recent work has been on the inclusion of fundamental
matter, i.e., supersymmetric QCD (SQCD) with a Chern--Simons (CS)
term~\cite{CSterm} in the large-$N_c$ 
approximation~\cite{fundamentallight,fundamentalspectrum,fundamentalmatter}.
Thermodynamic properties were of particular interest~\cite{FiniteTemp}.

The Chern--Simons term gives a mass to the adjoint
partons without breaking the supersymmetry; the adjoint
particles are less likely to form long strings, which
are difficult to approximate well and are not naturally
part of ordinary QCD.  Both a
dimensionally reduced model~\cite{SYM-CS,BPSstates} and the full (2+1)-dimensional 
theory~\cite{lightstates,SYM-CSproperties} have been investigated. 

A correlator $\langle T^{++}(x) T^{++}(y)\rangle$ of
the stress-energy tensor, has been computed~\cite{1+1correlator,subsequent}
as a test of a Maldacena conjecture~\cite{Maldacena1,Maldacena2}
of a correspondence between certain 
string theories and SYM theories at large-$N_c$.  These
can be tested directly if one is able to solve an SYM theory at
strong coupling, as SDLCQ is capable of doing.  The solution can 
be compared to a small-curvature supergravity approximation to 
the corresponding string theory. 

The particular calculation~\cite{Mtest} was done for of ${\cal N}=(8,8)$
SYM theory in two dimensions, which corresponds to a
type IIB string theory~\cite{Maldacena2}.  The supergravity
approximation and the opportunity for an SDLQ calculation
are discussed in \cite{JHEP} and \cite{subsequent}.  The
resolution needed to solve the SYM theory accurately
enough was reached with the work reported in \cite{Mtest}.
The method of calculation is discussed in Sec.~\ref{sec:SDLCQ}.

For comparison, the ${\cal N}$=(2,2) SYM theory was also
considered.  It is obtained through the dimensional reduction of
${\cal N}$=1 SYM theory from four to two dimensions~\cite{N22}.
The action is the same as for ${\cal N}$=(8,8), except that
the indices run from 1 to 2 instead of 1 to 8.
Because there are fewer dynamical fields, there are also fewer symmetries.
The smaller number of fields allowed calculations at
higher resolution, reaching $K=14$ for the $(2,2)$ theory
versus $K=11$ for the $(8,8)$ theory.  The important point of
the comparison is that there is no conjecture of correspondence
for the $(2,2)$ theory, and its correlator should behave 
differently.

\begin{figure}[ht]
\begin{center}
\begin{tabular}{cc}
\epsfig{file=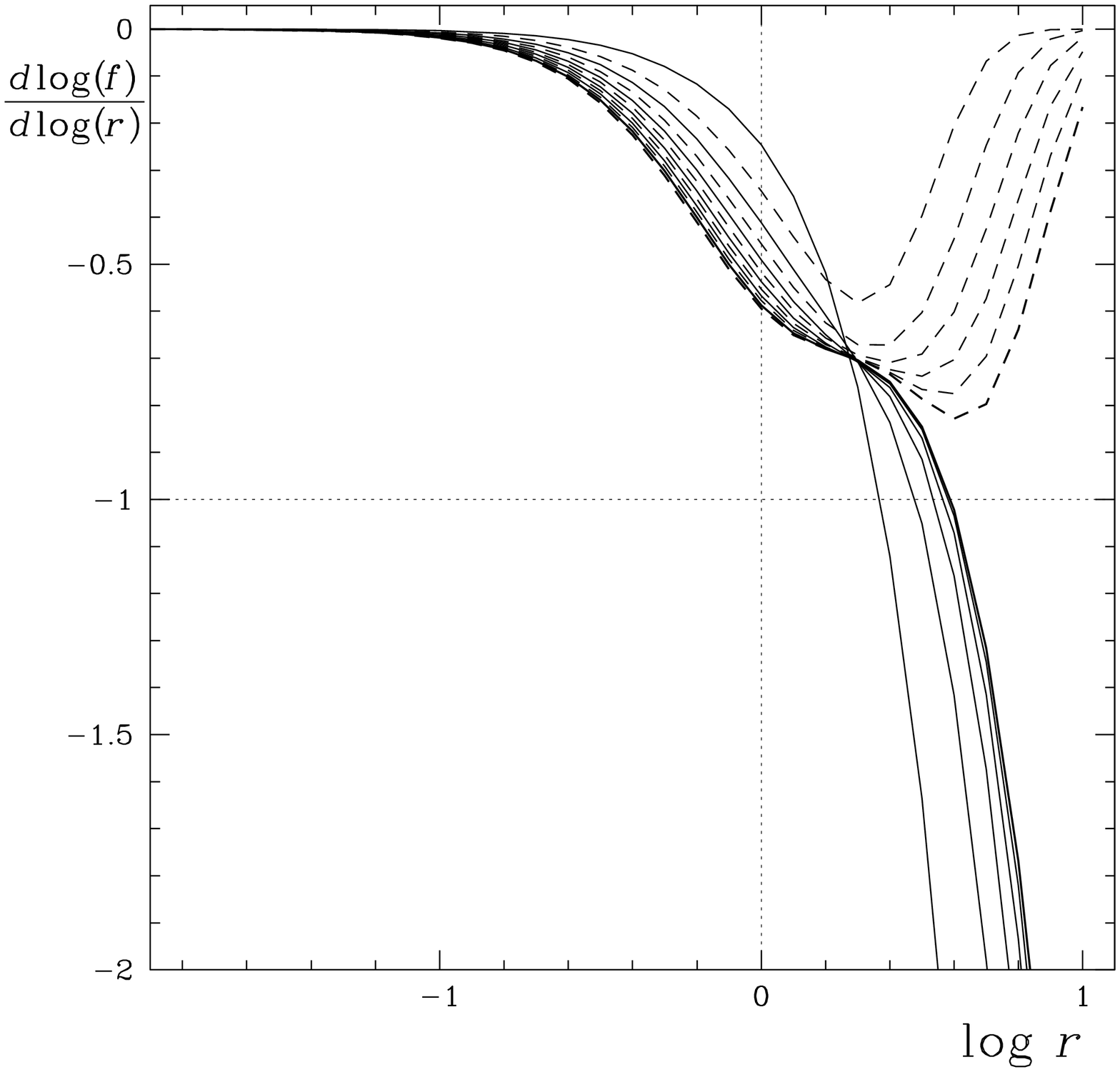,scale=0.43} &
\epsfig{file=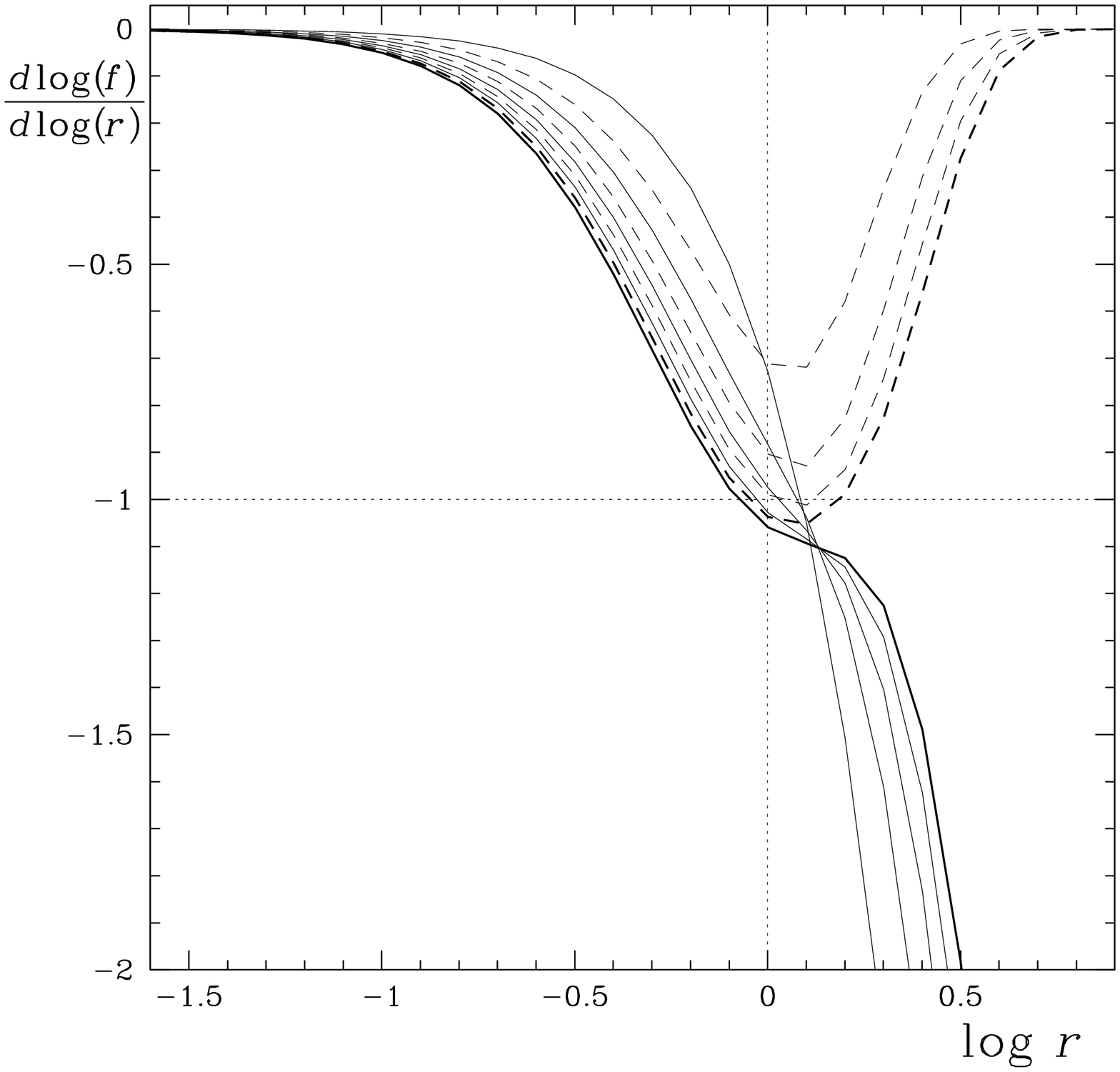,scale=0.43} \\
(a) & (b)
\end{tabular}
\caption{Plots of the log-log derivative of the
rescaled correlator $f$ for the (a) $\mathcal{N}=(2,2)$ and (b) $(8,8)$
SYM theories, from \protect\cite{Mtest}.  Each curve corresponds to a different resolution $K$,
with $K$ ranging from 3 to 14 in (a) and from 3 to 11 in (b).  For odd $K$
the curves are solid, and for even $K$ they are dashed.
The darker lines are for the highest resolutions; the lower resolutions converge
to these from above (odd) and below (even) for $\log r\geq 0.2$.
The separation $r\equiv\sqrt{2x^+x^-}$ is measured in units of $\sqrt{\pi/g^2N_c}$.}
\label{fig:loglog}
\end{center}
\end{figure}

Figure~\ref{fig:loglog} shows the
log-log derivative of the rescaled correlator $f$, as defined in (\ref{eq:rescaledF}),
for both SYM theories and for a range of resolution values.
For small $r\equiv\sqrt{2x^+x^-}$, the expected $1-1/K$ behavior, stated
in (\ref{eq:smallr}), appears for each resolution.
For large $r$, odd and even resolutions produce different behaviors; this
is because exactly massless states are missing from the odd case.  For
even $K$, there is a massless state that allows the correlator to
behave as $1/r^4$ at large $r$.  For odd $K$, the missing massless state
is recovered only in the limit of infinite $K$.

\begin{figure}[htbp]
\begin{center}
\begin{tabular}{cc}
\epsfig{file=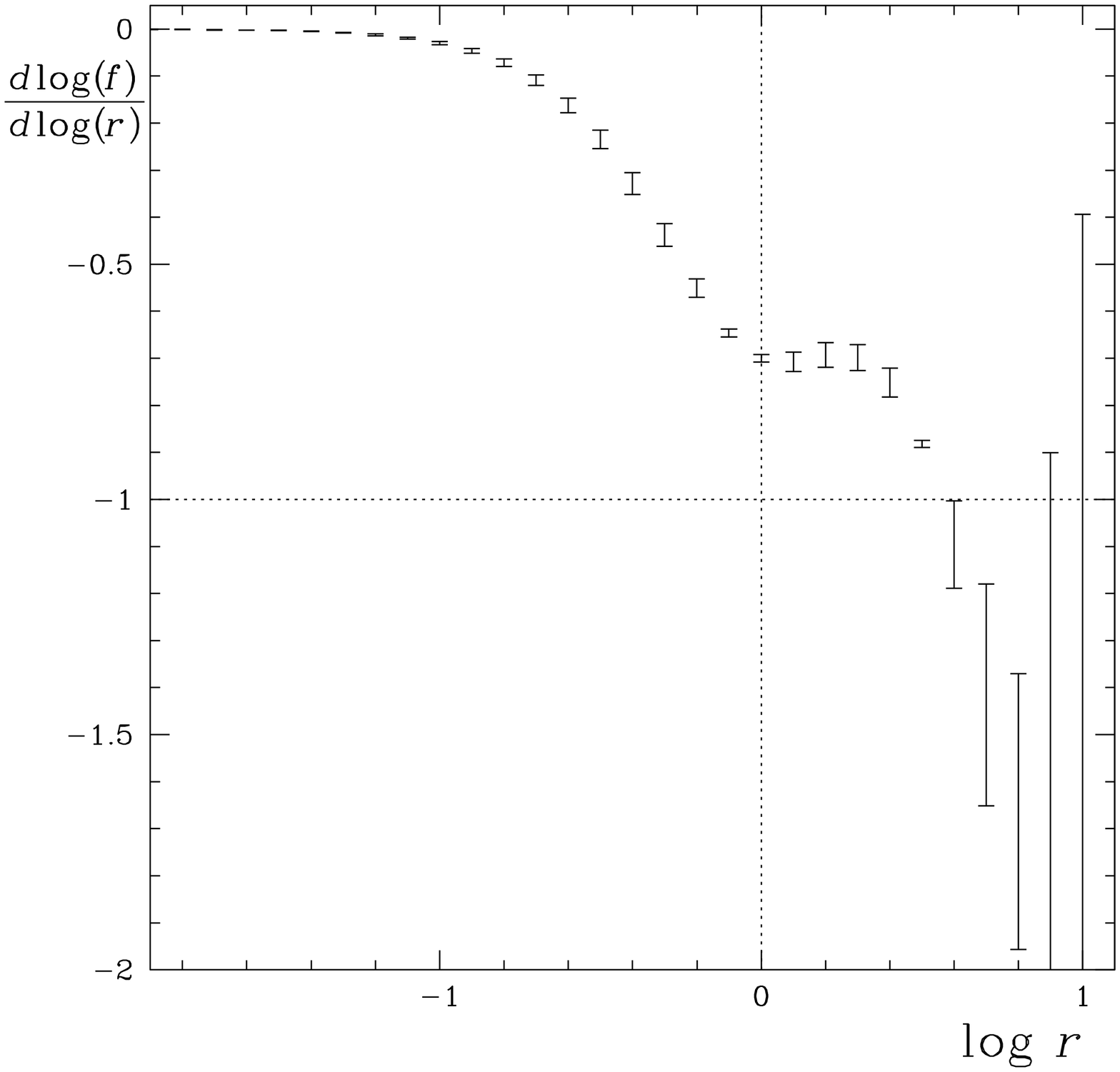,scale=0.43} &
\epsfig{file=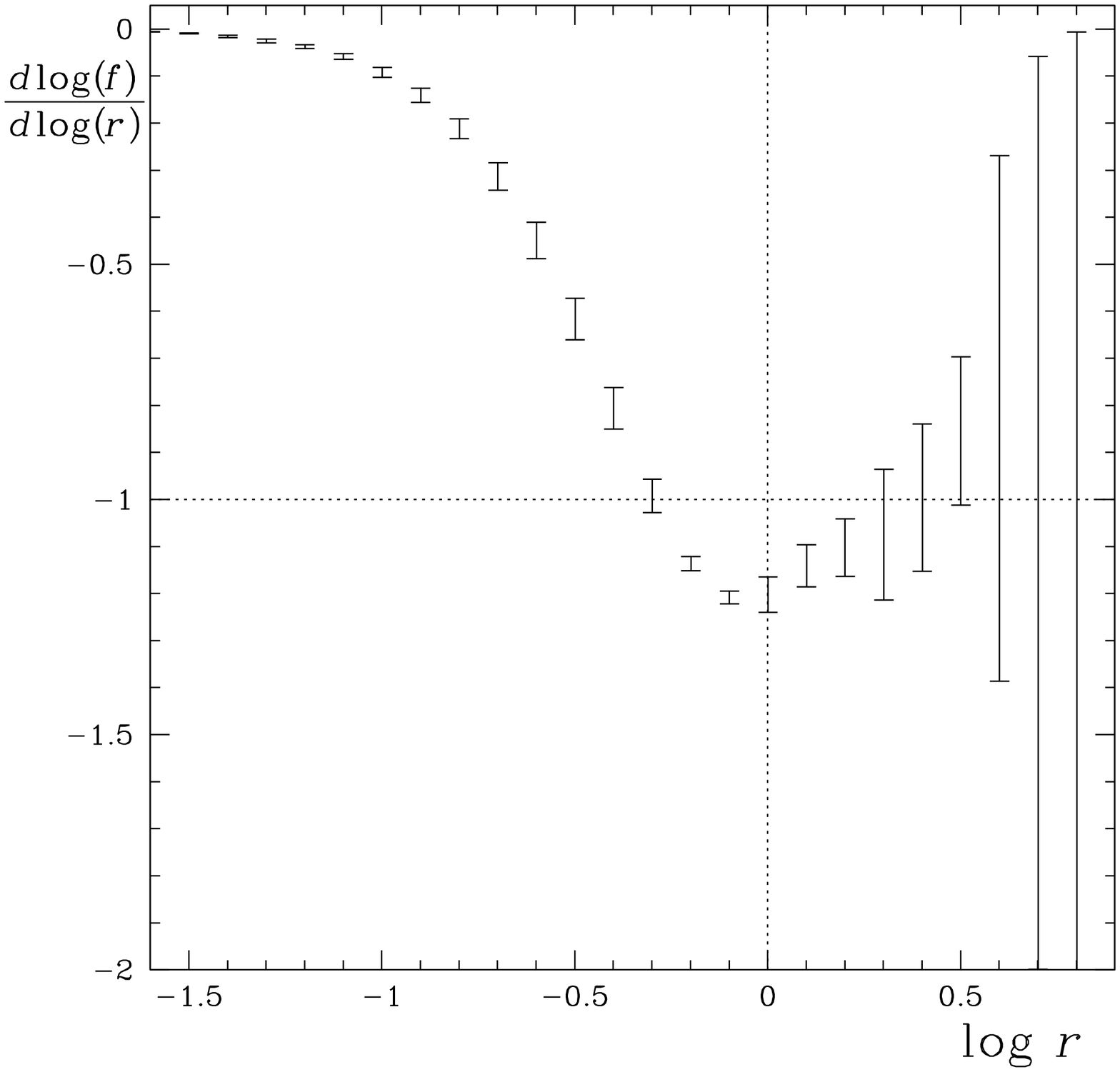,scale=0.43} \\
(a) & (b)
\end{tabular}
\caption{Summary of extrapolations to infinite resolution
for the (a) $\mathcal{N}=(2,2)$ and (b) $(8,8)$
SYM theories, from \protect\cite{Mtest}.  The 
vertical segments represent the intervals
obtained by various choices of extrapolations..}
\label{fig:extrap}
\end{center}
\end{figure}

For intermediate $r$, the expected behavior of the $(8,8)$ theory,
based on the correspondence, is $1/r^5$ and the log-log plot should
be near $-1$.  To investigate this region, the values of the 
scaled correlator were extrapolated to infinite resolution.
Estimates of the error in the extrapolation were made by
considering fits of different orders for the odd and even
resolutions separately.  The extrapolated results are shown
in Fig.~\ref{fig:extrap}.
The two theories are clearly different in their behavior for intermediate $r$.
Only the $(8,8)$ theory is consistent with the $1/r^5$
behavior predicted for it by the correspondence to the
supergravity approximation.

\subsection{\it Quantum Electrodynamics \label{sec:QED}}

Light-front QED has provided a number of useful tests of
various methods, particularly because it is the simplest
gauge theory.  Some work has been two-dimensional, where
QED is also known as the (massive) Schwinger 
model~\cite{Eller,Ma:1987wi,Mo:1992sv},
but the primary focus has been the full four-dimensional
theory.  The first DLCQ application was by Tang, Brodsky, and Pauli~\cite{Tang},
followed by a series of efforts to compute the states
of positronium~\cite{Krautgartner,Kaluza,TrittmannPauli}
with steadily improving methods, based primarily on
quadrature schemes and Fock-space truncations, culminating in
the current state-of-the-art calculations with the BLFQ
approach~\cite{BLFQpositronium}.  Related analytical 
calculations for level-splittings in positronium~\cite{Jones:1996vy}
focused on renormalization of the QED Hamiltonian.  There are also
calculations of fine and hyperfine structure~\cite{Brisudova:1996sv}
and the Lamb shift~\cite{Jones:1996pz} in hydrogen.

Another important QED observable is the anomalous magnetic 
moment, which has been computed in various ways.  Langnau
{\em et al.}~\cite{Langnau} investigated the DLCQ approximation
in the context of perturbation theory.
Hiller and Brodsky~\cite{HillerBrodsky} used DLCQ combined with
sector-dependent renormalization, an invariant-mass
cutoff, and a Fock space truncation to two photons.
More recent calculations~\cite{OnePhotonQED,TwoPhotonQED}
use PV regularization. The one-photon truncation of the
dressed electron state has also been computed by Karmanov 
{\em et al.}~\cite{KMSYukawaQED,Karmanov}.
The dressed-photon state has also been investigated~\cite{VacPol};
the appropriate PV regularization has been found to enforce a zero eigenmass
for the photon.

\subsubsection{\it anomalous moment of the electron}

The dressed-electron eigenstate of Feynman-gauge QED has been computed
in light-front quantization with a Fock-space truncation to 
include the two-photon/one-electron sector~\cite{ChiralLimit,SecDep,TwoPhotonQED,SSCthesis}.
Earlier work was limited to a one-photon/one-electron truncation~\cite{OnePhotonQED}.  
The theory is regulated by the inclusion of three massive Pauli--Villars (PV) particles,
one PV electron, and two PV photons.  In particular,
the chiral limit was investigated~\cite{ChiralLimit}, and the correct limit was found
to require two PV photon flavors, not just one as previously thought~\cite{OnePhotonQED}.
The renormalization and covariance of the electron current were
also analyzed~\cite{ChiralLimit}.  The plus component is well behaved
and is used in a spin-flip matrix element to compute the electron's
anomalous moment.  The dependence of the moment on the regulator
masses was shown to be slowly varying when the second PV photon flavor
is used to guarantee the correct chiral limit. However, in the
two-photon truncation, the chiral constraint
must be computed nonperturbatively~\cite{TwoPhotonQED}.  The nonperturbative constraint is
that the bare mass $m_0$ should be zero when the eigenmass $M$ is zero.

The motivation for the use of the plus component of the current
is that, for this component, additional renormalization is
not needed.  Because fermion-antifermion states are excluded,
there is no vacuum polarization. Thus, if the vertex and
wave function renormalizations cancel, there will be no
renormalization of the external coupling.
In covariant perturbation theory, this is a consequence of the Ward identity;
order by order, the wave function renormalization constant $Z_2$
is equal to the vertex renormalization $Z_1$.  
As discussed by Brodsky {\em et al}.~\cite{BRS},
this equality holds true more generally for nonperturbative bound-state 
calculations.  However, a Fock-space truncation can have the effect of
destroying the covariance of the electromagnetic current, so that some
components of the current require renormalization despite the
absence of vacuum polarization.  Also, the lack of fermion-antifermion 
vertices destroys covariance. However, in the particular case here, the 
couplings to the plus component are not renormalized~\cite{ChiralLimit}.

The violation of chiral symmetry in the massless electron limit
was not recognized in the earlier work on this 
particular PV regularization~\cite{OnePhotonQED}, because the
symmetry is restored in the limit of infinite PV electron mass,
but the need for three PV fields is quite
consistent with what has been found in different PV regularizations
of QED and Yukawa theory~\cite{ChiralSymYukawa}.
This symmetry is restored by the addition of a
second PV photon flavor, with its coupling strength and mass related
by a constraint.  For the one-photon truncation, this is the simple
condition~\cite{ChiralLimit}
\be
\sum_{i=0}^2(-1)^i\xi_i^2
   \frac{\mu_i^2/m_1^2}{1-\mu_i^2/m_1^2}\ln(\mu_i^2/m_1^2)=0.
\ee
For the two-photon
truncation, the coupling strength $\xi_2$ must be adjusted
nonperturbatively.  

\begin{figure}[ht]
\begin{center}
\epsfig{file=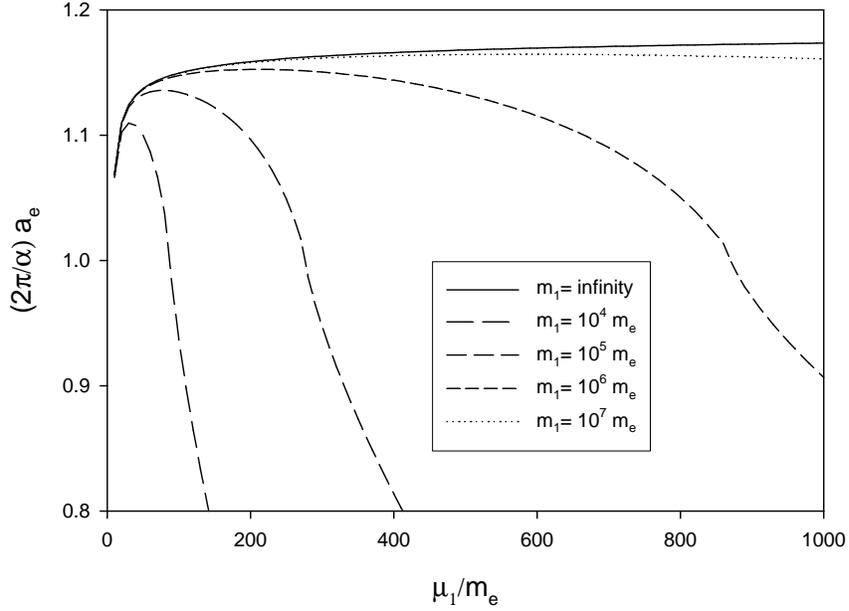,scale=0.75}
\caption{\label{fig:aeOnePhoton} The anomalous moment of the electron in
units of the Schwinger term (${\alpha/2\pi}$)~\protect\cite{Schwinger} plotted versus
the PV photon mass, $\mu_1$, for a few values of the PV electron mass,
$m_1$, from \protect\cite{ChiralLimit}. The second PV photon flavor 
is absent, and the chiral symmetry of the
massless limit is broken by the remaining regularization.  The Fock-state
expansion is truncated at one photon.}
\end{center}
\end{figure}

With the second PV photon flavor included, the electron's anomalous moment at finite
PV electron mass can be computed~\cite{ChiralLimit}.  Without the
second PV photon flavor, the anomalous moment has a strong 
dependence on the PV masses, as shown in Fig.~\ref{fig:aeOnePhoton}, 
but with the restoration of chiral symmetry in the limit of a massless
electron, the dependence on the PV mass is very mild.
%

From Fig.~\ref{fig:aeOnePhoton}, we see that, without the second
PV photon, the PV electron mass needs to be on the order of 
$10^7\,m_e$ before results for
the one-photon truncation approach the infinite-mass limit.
Thus, the PV electron mass must be at least this large
for a successful calculation with a two-photon Fock-space truncation, if only 
one PV photon flavor is included.  Unfortunately, such large mass values 
make numerical calculations difficult, because of contributions to 
integrals at momentum fractions of order $(m_e/m_1)^2\simeq 10^{-14}$, 
which are then subject to large round-off errors.  Therefore,
the two-photon calculation does require the second PV photon.

\begin{figure}[ht]
\begin{center}
\epsfig{file=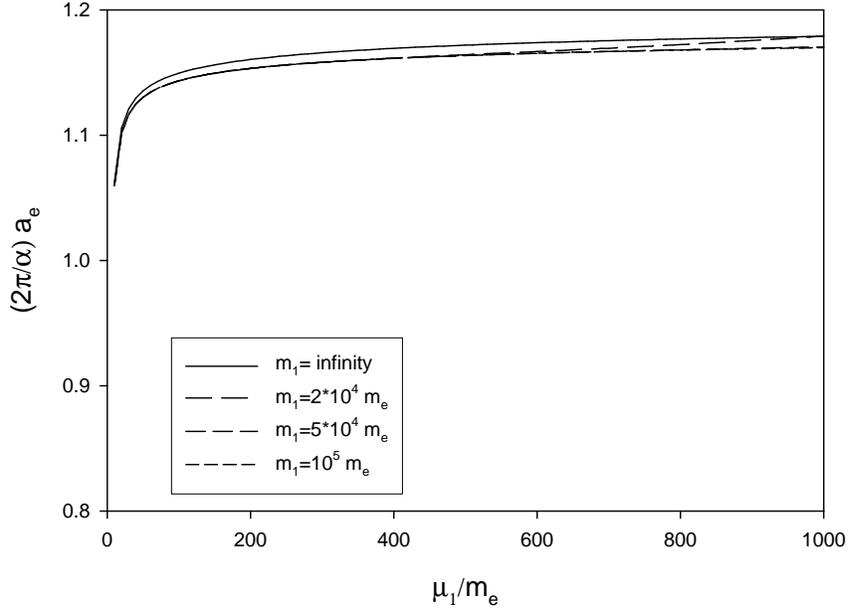,scale=0.75}
\caption{\label{fig:ae2OnePhoton} Same as Fig.~\ref{fig:aeOnePhoton},
and also from \protect\cite{ChiralLimit},
but with the second PV photon included, with a mass $\mu_2=\sqrt{2}\mu_1$,
and the chiral symmetry is restored.
The mass ratio is held fixed as $\mu_1$ and $\mu_2$ are varied.}
\end{center}
\end{figure}

When the anomalous moment is re-calculated in the
one-photon truncation with the second PV photon flavor included, the
result is given in Fig.~\ref{fig:ae2OnePhoton}, for PV masses
related by $\mu_2=\sqrt{2}\mu_1$.
Clearly, the dependence on the PV masses is greatly reduced.  

The value obtained for the anomalous moment differs from the leading-order
Schwinger result~\cite{Schwinger}, and thus from the physical value, by 17\%.  
This result is considerably improved if the self-energy contribution from 
the one-electron/two-photon Fock sector is included~\cite{SecDep}.
The alternative, sector-dependent renormalization method~\cite{HillerBrodsky,Karmanov}
accomplishes this by restricting the bare mass in the one-electron/one-photon
sector to being equal to the physical mass; however, this introduces
the usual infrared divergence of perturbation theory which requires a
nonzero photon mass and shifts the result for the anomalous moment
away from the standard result~\cite{HillerBrodsky}.

\begin{figure}[ht] 
\begin{center}
\epsfig{file=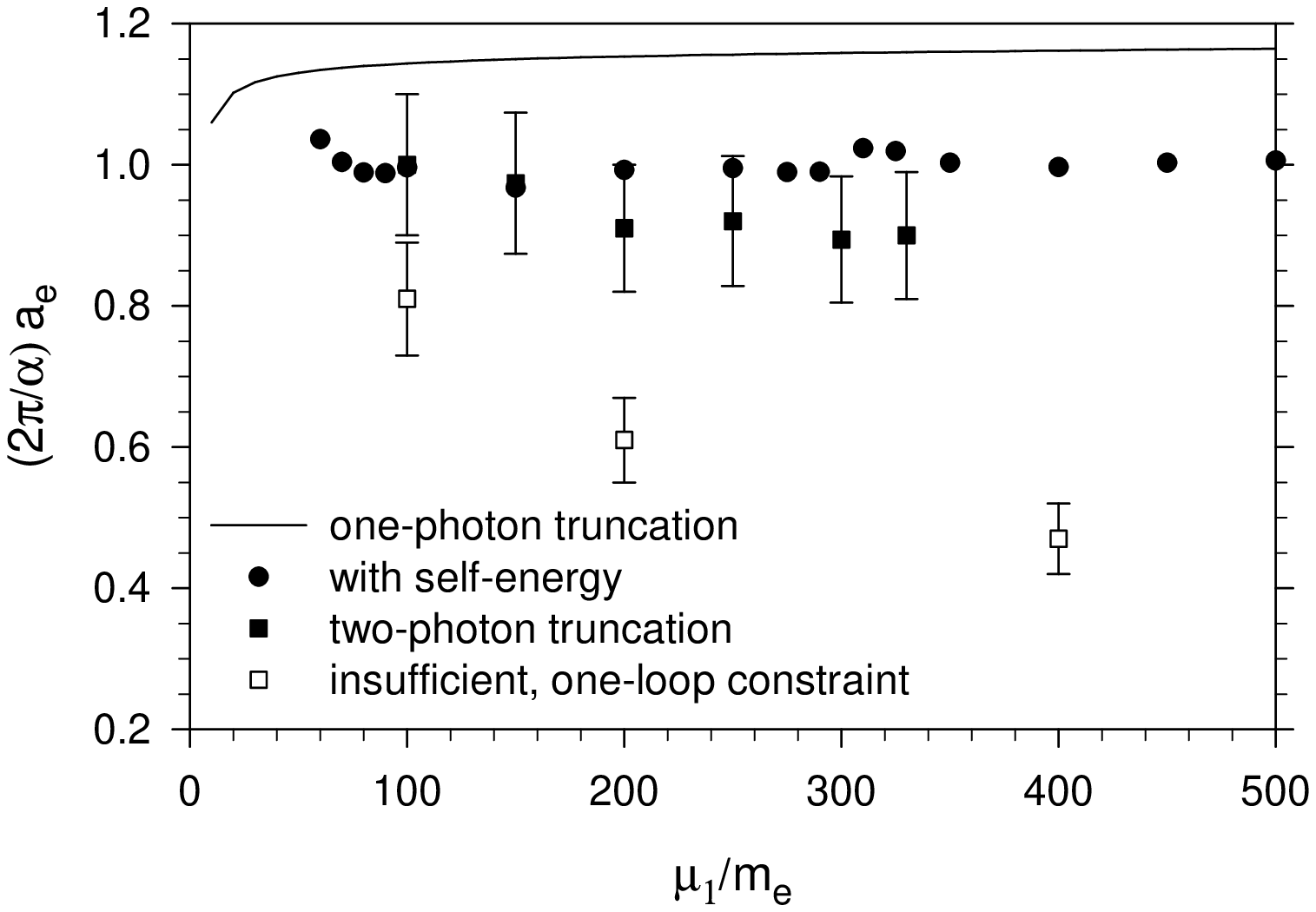,scale=0.75}
\caption{The anomalous moment of the electron in
units of the Schwinger term ($\alpha/2\pi$)~\protect\cite{Schwinger} plotted versus
the PV photon mass, $\mu_1$, with the second PV photon mass, 
$\mu_2$, set to $\sqrt{2}\mu_1$
and the PV electron mass $m_1$ equal to $2\cdot10^4\,m_e$, from \protect\cite{TwoPhotonQED}.
The solid squares are the result of the full two-photon truncation
with the correct, nonperturbative chiral constraint.  The open
squares come from use of a perturbative, one-loop constraint.
Results for the one-photon truncation~\protect\cite{ChiralLimit}
(solid line) and the one-photon truncation with the
two-photon self-energy contribution~\protect\cite{SecDep} (filled circles)
are also included.}
\label{fig:aeTwoPhoton}  
\end{center}
\end{figure}

The truncation of the dressed-electron
state was extended to include two photons, and the anomalous
moment was computed~\cite{TwoPhotonQED,SSCthesis}.  With the chiral symmetry properly controlled,
there is a plateau in the dependence on the 
PV mass, as shown in Fig.~\ref{fig:aeTwoPhoton}.
The main result for the two-photon truncation is in general
agreement with experiment, within numerical errors.  It is,
however, systematically slightly below, due to
the absence of two important contributions.  One is from the
Fock sector with an electron-positron loop, which contributes
at the same order as the two-photon sector in perturbation
theory, and the other is the three-photon self-energy contribution.

\begin{figure}[ht]
\begin{center}
\epsfig{file=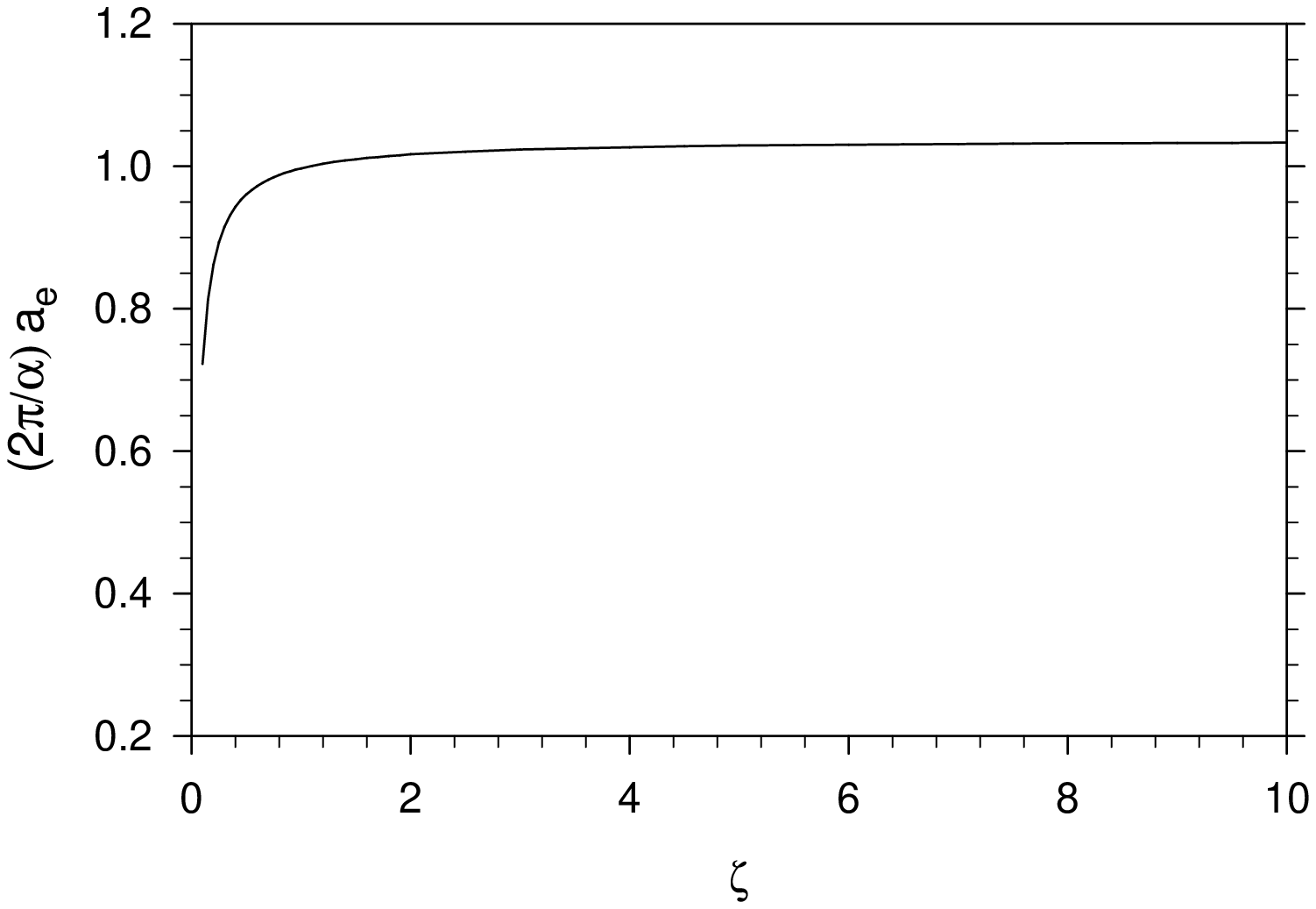,scale=0.75}
\caption{The anomalous magnetic moment $a_e$, relative to the Schwinger
term $\alpha/2\pi$~\protect\cite{Schwinger}, for the dressed-electron
state truncated to include at most the one-electron/one-photon
Fock states, as a function of the 
gauge parameter $\zeta$, with a bare photon mass of
$\mu_0=0.001\,m_e$ and PV photon mass of $\mu_1=200\,m_e$,
from \protect\cite{ArbGauge}.
}
\label{fig:aevszeta} 
\end{center}
\end{figure}

The work on QED has gone beyond Feynman gauge to include
arbitrary covariant gauges~\cite{ArbGauge}.  The gauge
dependence was checked for a one-photon/one-electron
truncation of the Fock expansion for the dressed electron.
The result for the anomalous moment is shown in 
Fig.~\ref{fig:aevszeta}.
The gauge parameter $\zeta$ is the coefficient of the gauge-fixing 
term in the Lagrangian; thus, $\zeta=0$ is a singular limit where
the gauge fixing is removed and the theory is undefined.  Except
for values of $\zeta$ near 0, the anomalous moment varies only
slightly.  This variation is due to Fock-space
truncation errors.

The one-photon truncation has also been solved with the BLFQ 
method~\cite{BLFQelectron,BLFQelectron2},
both in a transverse cavity and as free (modeled as a weak cavity field).
The dressed electron problem was formulated in light-cone gauge;
however, the instantaneous interactions were neglected, making the
coupled system for the wave functions very similar in structure to the 
Feynman-gauge system.  The ultraviolet divergences were
regulated by the basis choice and by the basis truncation.
Sector-dependent mass renormalization was also used, with the physical
electron mass assigned in the $|e\gamma\rangle$ sector
and the bare mass in the $|e\rangle$ sector adjusted to yield the
correct eigenmass for the dressed state.  Without sector-dependent
coupling renormalization, to absorb the uncanceled divergence from
the broken Ward identity, the norm of the wave function diverges
in the infinite basis limit, and results had to be extracted
from a range of finite cutoffs.  Later work argues for an
explicit renormalization of the matrix element for the
anomalous moment~\cite{XZhao}.

\begin{figure}[ht]
\begin{center}
\epsfig{file=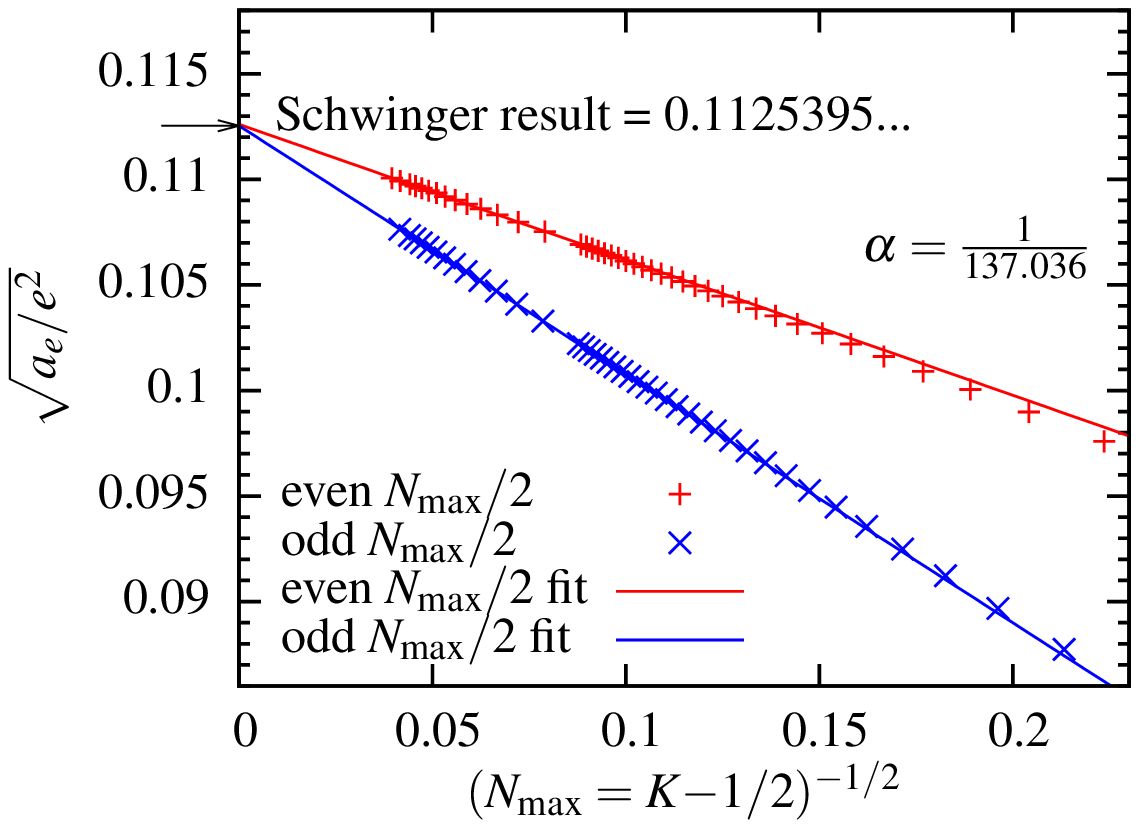,scale=1.0}
\caption{The square-root of the anomalous magnetic moment $a_e$, relative to 
the electron charge $e$, as a function of the BLFQ basis-size parameter 
$N_{\rm max}$, with resolution $K$ set to $N_{\rm max}+1/2$ and the
oscillator-basis energy scale $b=m_e$, from \protect\cite{BLFQelectron2}.  
The lines are from linear fits for $N_{\rm max}>150$.
}
\label{fig:aeBLFQ} 
\end{center}
\end{figure}

These investigations included careful studies
of convergence with respect to both the basis size $N_{\rm max}$ and the
oscillator energy scale $\Omega$.  In the free case, convergence
is slow because the harmonic oscillator basis functions are not a good
match to the power-law behavior of the electron's wave functions.
However, the harmonic oscillator functions should be a good approximation
for the confined quarks of QCD, and this work serves as test of the method.
The results are comparable to the known perturbative result and the
other nonperturbative calculations, as illustrated in Fig.~\ref{fig:aeBLFQ}.

As another illustration of the LFCC method, the dressed-electron state
can be investigated~\cite{LFCCqed}.  To simplify the calculations, electron-positron
pairs are excluded; however, an infinite number of photons is retained.
The anomalous moment can then be computed from a
spin-flip matrix element of the current.
The valence state is the bare electron, and
the $T$ operator is truncated to just simple photon emission from
an electron:
\bea
T&=&\sum_{ijls\sigma\lambda}\int dy d\vec{k}_\perp 
   \int\frac{d\ub{p}}{\sqrt{16\pi^3}}\sqrt{p^+}\, t_{ijl}^{\sigma s\lambda}(y,\vec{k}_\perp) \\
&& \rule{1in}{0mm} \times a_{l\lambda}^\dagger(yp^+,y\vec{p}_\perp+\vec{k}_\perp)
   b_{js}^\dagger((1-y)p^+,(1-y)\vec{p}_\perp-\vec{k}_\perp)b_{i\sigma}(\ub{p}) .
   \nonumber
\eea
This includes as much physics as the two-photon Fock-space truncation considered
earlier~\cite{TwoPhotonQED}.

\begin{figure}[ht]
\begin{center}
\epsfig{file=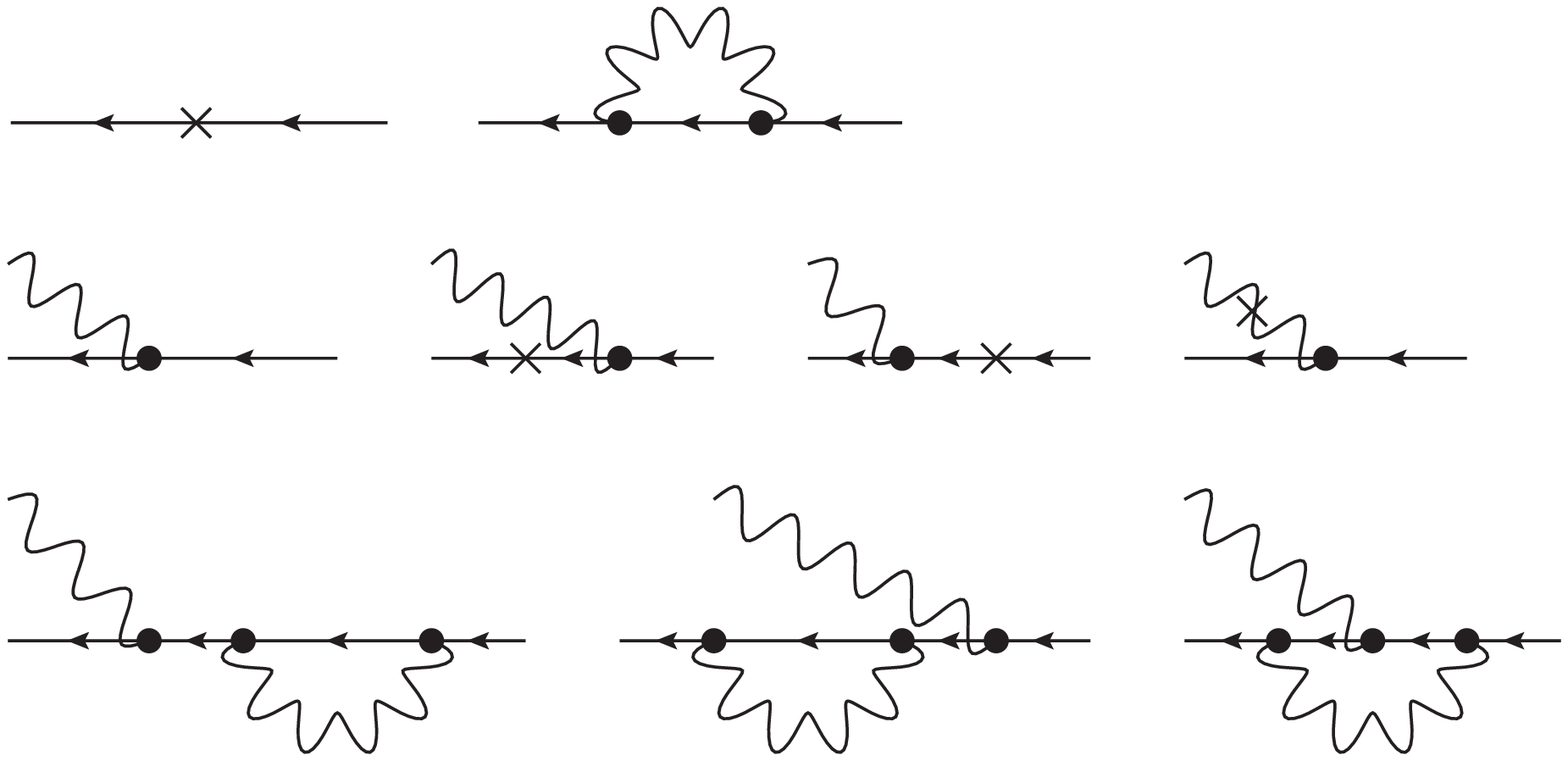,scale=0.75}
\caption{Graphical representation of the terms of the LFCC effective 
QED Hamiltonian.  Each graph represents an operator that
annihilates an electron and creates either a single electron or an electron
and photon.  The crosses indicate light-front kinetic-energy contributions.}
\label{fig:effPminus}
\end{center}
\end{figure}

A graphical representation of the effective Hamiltonian $\ob{\Pminus}$, 
excluding terms that annihilate the valence state, is
given in Fig.~\ref{fig:effPminus}.
The self-energy loop is the same
in each contribution, without the sector and spectator dependence found in
calculations with Fock-space truncation.  Another consequence of using the
LFCC method is that terms corresponding to all three graphs for the
Ward identity in Fig.~\ref{fig:WardID} appear in the effective Hamiltonian,
as indicated by the last row of Fig.~\ref{fig:effPminus}, which prevents the
uncanceled divergence encountered in a Fock-space truncation.  Also,
in the equations for the functions $t_{ijl}^{\sigma s\lambda}(y,\vec{k}_\perp)$,
the physical mass of the eigenstate replaces the bare mass in
the kinetic energy term, without use of sector-dependent 
renormalization~\cite{SecDep-Wilson,HillerBrodsky,Karmanov,SecDep}.

The anomalous moment $a_e=F_2(0)$ is obtained from the spin-flip
matrix element of the current 
$J^+=\overline{\psi}\gamma^+\psi$~\cite{BrodskyDrell}, 
coupled to a photon of momentum $q$
in the Drell--Yan ($q^+=0$) frame~\cite{DrellYan}, as given in Eq.~(\ref{eq:Jplus}).
The formalism for a projected expectation value~\cite{LFCCqed},
with the projection $P_s$ being the projection onto
the physical subspace with the two transverse polarizations $\lambda=1,2$,
yields for the current matrix element
\be 
\langle\psi_a^\sigma(\ub{P}+\ub{q})|J^+(0)|\psi_a^\pm(\ub{P})\rangle
=\frac{\langle\widetilde\psi_a^\sigma(\ub{P}+\ub{q})|
    e^{-T}P_s^\dagger J^+(0) P_s e^T|\phi_a^\pm(\ub{P})\rangle}
 {\int d\ub{P}' \langle\widetilde\psi_a^\pm(\ub{P}')|
                       e^{-T}P_s^\dagger P_s e^T|\phi_a^\pm(\ub{P})\rangle},
\ee
with $\langle\widetilde\psi_a^\sigma(\ub{P}+\ub{q})|$ 
the left-hand eigenstate of $\ob{\Pminus}$
\bea
\lefteqn{|\widetilde\psi_a^\pm(\ub{P})\rangle=|\tilde\phi_a^\pm(\ub{P})\rangle}&&\\
&&    + \sum_{jls\lambda}\int dy d\veck \sqrt{\frac{P^+}{16\pi^3}}
       l_{ajl}^{\pm s\lambda}(y,\veck)
  a_{l\lambda}^\dagger(yP^+,y\vec{P}_\perp+\veck)
       b_{js}^\dagger((1-y)P^+,(1-y)\vec{P}_\perp-\veck)|0\rangle , \nonumber
\eea
The truncation to $|e\rangle+|e\gamma\rangle$ is consistent with the truncation
of $T$.  The form factors can then be extracted.

This can be checked by considering the perturbative solutions for the right and
left-hand wave functions; substitution into the expression for $a_e$ gives
immediately the Schwinger result~\cite{Schwinger} of $\alpha/2\pi$,
in the limit of zero photon mass, for any covariant gauge~\cite{LFCCqed}.
A complete calculation requires a numerical solution of
the eigenvalue problems.  This will yield all contributions to $a_e$
of order $\alpha^2$, except those with electron-positron pairs,
and a partial summation of higher orders.

\subsubsection{\it positronium}

The first application of light-front techniques to positronium in
light-cone gauge was by
Tang {\em et al.}~\cite{Tang}.  An improved treatment of the Coulomb
singularity led to better convergence~\cite{Krautgartner}.  Both
of these attempts considered an effective interaction in the $|e^+e^-\rangle$
sector, with the $|e^+e^-\gamma\rangle$ sector integrated out and all other
sectors neglected.  A direct diagonalization of the light-front QED
Hamiltonian was attempted by Kalu\v{z}a and Pauli~\cite{Kaluza}.
This work culminated in the more exhaustive study by Trittmann and
Pauli~\cite{TrittmannPauli}, which included the annihilation channel
and numerical convergence to restoration of rotational invariance;
the light-front equations were discretized by Gauss-Legendre quadrature.

More recently, the BLFQ method has been used to study 
positronium~\cite{BLFQpositronium}, again in light-cone gauge,
with a Fock-space truncation to the $|e^+e^-\rangle$ and $|e^+e^-\gamma\rangle$
sectors.  The annihilation channel, with coupling to the $|\gamma\rangle$
sector, is not included.  With use of the Bloch formalism ~\cite{Bloch},
the higher sector is eliminated to yield an effective 
one-photon-exchange interaction in the lower sector, and the self-energy
contributions neglected.  This is equivalent to the truncations made
for the earlier work~\cite{Krautgartner,TrittmannPauli}, 
but the numerical approximation is quite different.
More important, it will serve as a point of reference for future calculations
with self-energies and sector-dependent renormalization.

If the full Hamiltonian were used instead of the Bloch-reduced Hamiltonian,
the matrix representation would be much larger but much more sparse.
However, the self-energy terms would be automatically included with
no simple way to exclude them.  Of course, eventually this is
exactly what needs to be considered, to approximate the full
many-body problem.  This will require a better understanding of
the renormalization generated by the self-energy corrections.

The calculation is done at $\alpha=0.3$, in order to make higher-order
effects visible numerically.  The results compare quite favorably with those from the
Schr\"odinger equation with (perturbative) relativistic corrections,
to order $\alpha^4$.
The dependence on the regulators and the rate of convergence
were thoroughly studied.  The convergence rate with respect to the 
basis cutoff $N_{\rm max}$ was found to be strongly affected by
the choice of oscillator energy scale $b$; the convergence rate
with respect to the longitudinal DLCQ resolution $K$ was 
affected by the photon mass $\mu$.

The convergence is again slow, because harmonic oscillator functions are
not the best approximation to the exponential functions of the 
Schr\"odinger-Coulomb problem.  However, as preparation for the
calculation of meson states, the use of oscillator functions
is worthwhile.  Similarly, though the two-body sector lends itself
to solution in terms of relative coordinates, the single-particle
basis set of BLFQ was retained, as preparation for calculations that
include more Fock sectors explicitly.

The instantaneous-photon interaction is singular for small 
longitudinal photon momentum.  In a perturbative calculation,
this is canceled by a contribution from regular photon exchange,
which can be matched term by term.  In the nonperturbative
calculation, the cancellation is not complete and requires
a counterterm.  Similar counterterms were used in earlier
work~\cite{Krautgartner,TrittmannPauli}, where extensive
discussion of this counterterm can be found.

The integrable Coulomb singularity, while controlled in
principle, causes difficulties numerically.  A photon mass
is introduced to regulate it.

There is also a logarithmic divergence from the one-photon-exchange
kernel~\cite{Krautgartner,Brinet}  that requires a counterterm.
This takes into account effects of the (missing) $|e^+e^-\gamma\gamma\rangle$
sector.  Unfortunately, the counterterm worsens the breaking
of rotational symmetry.

The Fock-sector truncation and the asymmetry between the longitudinal
and transverse discretizations destroy the rotational invariance
of the Hamiltonian.  Thus, the $M_J$ multiplets are not degenerate.
However, the individual projections are typically close enough
to allow assignment of $J$ values.

\subsubsection{\it true muonium}

The general approach used for positronium can be naturally extended
to analyze true muonium, the bound states of the $\mu^+$-$\mu^-$ 
system.  These states are metastable, with lifetimes
short compared to the muon lifetime.  The methods of \cite{TrittmannPauli}
have been extended to true muonium by Lamm and Lebed~\cite{LebedLamm}.
They include $|\mu^+\mu^-\rangle$ and $|e^+e^-\rangle$ sectors and
effective interactions obtained by integrating out the one-photon 
annihilation channel and the higher sectors with an added photon.
The impact of the $|e^+e^-\rangle$ component is investigated by
allowing the electron mass to be comparable to the muon mass.
The coupling strength is again held at $\alpha=0.3$.

The numerical methods were based on Clenshaw--Curtis~\cite{ClenshawCurtis}
and Gauss-Chebyshev quadratures.  The Coulomb singularity
required different counterterms in each sector.  In general,
the cutoff dependence and choice of counterterms are studied
carefully, leading to good agreement with nonrelativistic 
calculations that include relativistic effects perturbatively.
The agreement provides confirmation that the light-front calculations
are done correctly.  However, one must keep in mind that these
and other tests of light-front methods in QED are just that,
tests of the method, and will never be competitive with equal-time
calculations.  The true home of light-front methods is in
strongly interacting theories where perturbative corrections
of some zero-order (non)relativistic approximation cannot be used effectively.

\subsection{\it Quantum Chromodynamics \label{sec:QCD}}

Two-dimensional QCD has been studied quite extensively in
the DLCQ approximation~\cite{2Dqcd}.
The earliest application to four-dimensional QCD is a 
DLCQ calculation by Hollenberg {\em et al.}~\cite{Hollenberg};
however, such four-dimensional applications have lacked a
consistent regularization.  The various methods developed since,
such as PV regularization~\cite{BRSTPVQCD,Paston} and 
sector-dependent renormalization
should provide a path forward.  The primary purpose of this
review is to facilitate and encourage progress along that path.

As an alternative to direct diagonalization of the QCD Hamiltonian,
one can use light-front techniques to construct and analyze
quark models, particularly in parallel with the AdS/QCD approach~\cite{AdSQCD}.
The BLFQ method has been modified to use relative coordinates
and study heavy quarkonium in exactly this way~\cite{BLFQquarkonium}.
This allows the calculation of masses and decay constants.
McCartor and Dalley~\cite{McCartorDalley} have derived
an effective Hamiltonian operator to induce chiral symmetry breaking;
this could be incorporated into a study of mesons in QCD.
There is also a construction of a confining interaction
based on the requirement of restoration for rotational
symmetry~\cite{Brisudova:1997rv,Brisudova:1996vw,Brisudova:1995hv}.

Glueball states have been studied~\cite{Allen:1999kx} with a renormalized light-front
Hamiltonian, constructed from recursion relations that restore
symmetries and eliminate cutoff dependence.  The Fock space
was truncated to the two-gluon sector, and the Hamiltonian
provided an effective interaction to bind the gluons.  The 
renormalization scheme was developed in the context of
scalar theories~\cite{Kylin:1998wr,Allen:1998bk} and
builds on the fundamental ideas of Wilson {\em et al.}~\cite{Wilson:1994fk,Perry:1993mn}
for a derivation of the constituent quark model from light-front QCD.
Along similar lines, asymptotic freedom in pure-glue QCD has been
investigated with the related RGPEP~\cite{RGPEPgluons}.

\section{Future Challenges} \label{sec:Summary}

The preceding sections have provided an in-depth review of
the nonperturbative methods applied to light-front calculations
in a variety of quantum field theories.  In each case, there
remain many interesting questions to pursue, not the least
of which, of course, is the solution of QCD itself, and
there are many other applications not yet considered.

For quenched scalar Yukawa theory, perhaps the most 
important open question is the convergence of the
LFCC method.  This theory is the simplest theory for
such a test, because the $T$ operator has the simplest
possible form, and the theory is not very far removed
from the simple model where the leading LFCC approximation
gives the exact solution~\cite{LFCC}.  Another important
question to be understood is why sector-dependent
renormalization works so well for calculations with
Fock-space truncation~\cite{QSY}, seemingly avoiding the
pitfalls found in Yukawa theory and QED~\cite{SecDep}.
Is the boson nature of the theory the key?  In which case,
does this bode well for the gluons of QCD?

This leads to the more general question of whether
sector-dependent renormalization can be interpreted in 
some way to allow meaningful calculations.  The triviality
limit on cutoffs~\cite{Glazek:1992aq,SecDep} must somehow
be absorbed into the rules for extraction of physical
observables.  Avoidance of Fock-space truncation, via
the LFCC method, may be the best route forward, but the
work on quenched scalar Yukawa theory~\cite{QSY} indicates
that Fock-space truncation and sector-dependent renormalization
may also provide a useful set of rules for calculation.

In $\phi^4_{1+1}$ theory, the standard bearer for theories
with broken symmetry, there is still much to consider, if
only to show that light-front methods can do as good a job
as any other approach.  Specifically, the critical couplings
and critical exponents should be calculable for the 
restoration of symmetry in the negative-mass-squared case
as well as for the dynamical symmetry breaking in both
the positive and negative mass-squared cases.  However,
any comparison with equal-time calculations that is done 
in terms of bare parameters must take into account the
different renormalizations~\cite{SineGordon}.  Near the critical
coupling, where the mass of the dressed state is close to
zero and the probabilities of higher Fock states should be
large, use of sector-dependent mass renormalization or
the LFCC method is probably crucial.

Zero modes can be useful in the analysis of $\phi^4$ theory
but are not necessarily
required~\cite{RozowskyThorn}.  In any case, much hinges
on the interpretation of the Fock vacuum $|0\rangle$.
Critical couplings are typically signaled by degeneracy of
a dressed eigenstate with the (massless) vacuum.  An increase
of the coupling beyond this point drives the ``massive''
eigenstate below zero and therefore below the alleged
vacuum, a nonsensical outcome.  This implies that the
quantization has been done incorrectly; the Fock vacuum
should always be the lowest state.  A correct quantization
should either shift the field explicitly~\cite{Hornbostel}
or introduce zero modes that give some structure to the
vacuum.  Either way, the field then develops the
necessary vacuum expectation value.

The solution of Yukawa theory should be carried out with
an additional PV scalar, to nonperturbatively restore
the chiral symmetry of the massless limit.  The work on
Yukawa theory was done~\cite{YukawaTwoBoson,KarmanovYukawa}
before the understanding of the
symmetry restoration was achieved in the context of
PV-regulated QED.

The work on supersymmetric Yang--Mills theory should
be extended to 3+1 dimensions.  This would bring the
possibility of analyzing QCD as a limit of supersymmetric QCD,
with the superpartners replacing PV fields as the regulators.
Obviously, this also entails supersymmetry breaking, to
give the superpartners a large mass.  However, this
is distinct from supersymmetric extensions of the
Standard Model, which are theories in their own right.
Instead, the idea is to use supersymmetry as a regularization
tool, with no physical interpretation given to the 
superpartners, which would be removed from the calculation
in an infinite-mass limit.  The limitations on the
form of supersymmetry breaking, that come from the
appearance of unobserved processes such as flavor-changing
decays, would not necessarily be applicable, which
broadens the options for the mechanism of the breaking.

For both Yukawa theory and QED, very little has been
done on the inclusion of vacuum polarization~\cite{VacPol}.  Typically,
the Fock space is truncated to exclude additional fermion-antifermion
pairs.  This leaves open the question of nonperturbative
charge renormalization.  The charge renormalization done
in sector-dependent renormalization is `only' an artifact
of absorbing wave function and vertex renormalization, leftover
from a broken Ward identity. Of course, the LFCC method
sidesteps issues of Fock-space truncation, but needs to
be tested in both Yukawa theory and QED.

There is a similar renormalization issue in the treatment
of self-energy corrections for multiparticle bound states.
Calculations have typically divided into two types.  One
type focuses on the dressing of a single particle; here
self-energies are the center of the piece.  The other type
focuses on binding two constituents, usually by an 
effective interaction obtained by integrating out other
Fock sectors; here self-energies are explicitly neglected,
to understandably simplify the calculation.  In the
future, the knowledge gained from the first type needs to 
be merged with the second.  This is particularly important
before moving on to less severe Fock-space truncations
for which a reduction to a single effective equation
and identification of self-energy contributions cannot
be done.  The LFCC method should be useful because there
the self-energy corrections tend to be spectator independent.

Simply by measuring the length of the section on 
applications to QCD in comparison with the length
of the other sections, one can see that there is much
to be done.  However, many if not all of the tools
necessary for the job have been established.  Nonperturbative
calculations of simpler systems such as glueballs and
heavy mesons are within reach.  The potential for
an explanation of confinement is particularly interesting in these
systems, perhaps based on RGPEP~\cite{RGPEP} or the LFCC method;
in the latter method, an infinite number of gluons can be
included.

There is also much formal work to be done.  A proof of
renormalizability is needed for the proposed PV regularization
scheme~\cite{BRSTPVQCD}.  A small step in this direction is
the introduction of Nakanishi--Lautrup fields~\cite{NL}
to make the BRST transformation nilpotent off-shell.
However, there are more than just the usual complications
of perturbative renormalization, because, after all,
the objective is use in a nonperturbative calculation.
The philosophy employed so far has been that a regularization
adequate for perturbation theory is then sufficient for
nonperturbative calculations, implicitly assuming that
renormalizability carries over as well.  This may not
be the case; for example, in any gauge-fixed theory,
one must contend with the
existence of Gribov copies~\cite{GribovCopies}.

Nevertheless, the time is ripe for tremendous progress in light-front
solutions for systems in QCD.  Consider the progress that
lattice gauge theory has made in the twenty-five years
since its originator expressed so much pessimism~\cite{Wilsonremarks}.
This progress was accomplished through the concerted
and combined efforts of a large community.  The challenge
of the future is for light-front methods to be carried
forward in just such a manner.

\section*{Acknowledgements}
Some of the reported work was done by the author,
in collaboration with several researchers, including
S.J. Brodsky, S.S. Chabysheva, 
G. McCartor, S.S. Pinsky, and U. Trittmann.  This work
was supported in part by the US Department of
Energy through Contract No.\ DE-FG02-98ER41087
and in part by the Minnesota Supercomputing Institute
of the University of Minnesota with grants of computing
resources.


\end{document}